\documentclass[fleqn,usenatbib]{mnras}

\usepackage{newtxtext,newtxmath}

\usepackage[T1]{fontenc}
\usepackage{float}
\usepackage{lscape}
\usepackage{chngpage}
\DeclareRobustCommand{\VAN}[3]{#2}
\let\VANthebibliography\thebibliography
\def\thebibliography{\DeclareRobustCommand{\VAN}[3]{##3}\VANthebibliography}

\usepackage{gensymb}
\usepackage{graphicx}	
\usepackage{amsmath}	
\usepackage{ulem}
\newcommand{\Vrot}{v_\textrm{rot}}
\newcommand{\Wmx}{W_\textrm{mx}}
\newcommand{\Wmxc}{W_\textrm{mx}^c}
\newcommand{\parTF}{\theta_{\rm TF}}
\newcommand{\parPV}{\theta_{\rm PV}}
\def\sun{\hbox{$\odot$}}

\def\la{\mathrel{\hbox{\rlap{\hbox{\lower4pt\hbox{$\sim$}}}\hbox{$<$}}}}
\def\ga{\mathrel{\hbox{\rlap{\hbox{\lower4pt\hbox{$\sim$}}}\hbox{$>$}}}}

\newcommand{\kms}{{\,km\,s$^{-1}$}}

\newcommand{\HI}{\mbox{\normalsize H\thinspace\footnotesize I}}

\def\apj    {{ApJ }}
\def\apjl   {{ApJL }}
\def\apjs   {{ApJS }}

\def\mnras  {{MNRAS }}
\def\nat    {{Nature }}

\title[Forward-modelling Tully-Fisher peculiar velocities]{Large-scale motions and growth rate from forward-modelling Tully-Fisher peculiar velocities}

\author[P.\ Boubel et al.]{
Paula Boubel,$^{1}$\thanks{E-mail: paula.boubel@anu.edu.au}
Matthew Colless,$^{1}$
Khaled Said,$^{2}$
and Lister Staveley-Smith$^{3}$
\\
$^{1}$The Australian National University, Mount Stromlo Observatory, Cotter Road, Canberra, ACT 2611, Australia\\
$^{2}$School of Mathematics and Physics, University of Queensland, Brisbane, QLD 4072, Australia\\
$^{3}$International Centre for Radio Astronomy Research (ICRAR), University of Western Australia, 35 Stirling Hwy, Crawley, WA 6009, Australia\\
}

\date{Accepted XXX. Received YYY; in original form ZZZ}

\pubyear{2024}

\begin{document}
\label{firstpage}
\pagerange{\pageref{firstpage}--\pageref{lastpage}}
\maketitle

\begin{abstract}
Peculiar velocities are an important probe of the mass distribution in the Universe and the growth rate of structure, directly measuring the effects of gravity on the largest scales and providing a test for theories of gravity. Comparing peculiar velocities predicted from the density field mapped by a galaxy redshift survey with peculiar velocities measured using a distance estimator such as the Tully-Fisher relation yields the growth factor for large-scale structure. We present a method for forward-modelling a sample of galaxy magnitudes and velocity widths that simultaneously determines the parameters of the Tully-Fisher relation and the peculiar velocity field. We apply this to the Cosmicflows-4 (CF4) Tully-Fisher dataset, using the peculiar velocities predicted from the 2M++ redshift survey. After validating the method on mock surveys, we measure the product of the growth rate and mass fluctuation amplitude to be $f\!\sigma_8$\,=\,0.35\,$\pm$\,0.03 at an effective redshift of $z$\,=\,0.017. This is consistent at 3$\sigma$ with the Planck CMB prediction, even though the uncertainty does not fully account for all sources of sample variance. We find the residual bulk flow from gravitational influences outside the 2M++ survey volume to be $|V|$\,=\,227\,$\pm$\,11\kms, $(l,b)$\,=\,(303$^\circ$,$-$1$^\circ$) in Galactic polar coordinates and the CMB frame. Using simulations, we show that applying our methodology to the large new sample of Tully-Fisher peculiar velocities expected from the WALLABY \HI\ survey of the southern sky can improve the constraints on the growth rate by a factor of 2--3.
\end{abstract}

\begin{keywords}
galaxies: distances and redshifts -- cosmology: cosmological parameters -- cosmology: large-scale structure of Universe
\end{keywords}


\section{Introduction}
\label{intro}

The continuous action of gravity over the history of structure growth in the Universe is reflected by the present-day velocities of galaxies. Peculiar velocities, the individual motions of galaxies separate from the overall expansion of the Universe, thus offer a powerful method to measure the amount and distribution of matter and test General Relativity (GR) via the time evolution of the growth rate of structure. This probe is sensitive to small deviations from GR because small differences in galaxy acceleration are amplified over time \citep{Strauss_1995}. However, more precise measurements of the growth rate than currently exist are required to distinguish GR from plausible alternatives.

In the linear regime, the peculiar velocity field is directly related to the density field of galaxies through the peculiar acceleration \citep{Peebles1980,Peebles1993},
\begin{equation}
\mathbf{v}(\mathbf{r}) = \frac{H_{0}\beta}{4\pi} \int \textrm{d}^{3} \mathbf{r'} \frac{\delta_{g}(\mathbf{r'})(\mathbf{r'}-\mathbf{r})}{|\mathbf{r'}-\mathbf{r}|^{3}}
\label{eq:velocity_beta}
\end{equation}
where $\beta$\,$\equiv$$f$/$b$ is the cosmology- and bias-dependent velocity scaling relating the velocities and densities through the growth rate of structure, $f$\,$\approx$\,$\Omega_m(z)^\gamma$, and the linear galaxy bias parameter, $b$\,=\,$\sigma_{8,g}/\sigma_8$ \citep{Strauss_1995}. Comparing the measured peculiar velocity field to the predicted peculiar velocity field $\mathbf{v}(\mathbf{r})$ derived using Equation~\ref{eq:velocity_beta} from the galaxy density field $\delta_{g}(\mathbf{r})$ (obtained in a redshift survey) thus yields an estimate for $\beta$.

Peculiar velocities can be measured statistically from galaxy clustering correlations, such as redshift-space distortions \citep[RSD;][]{Kaiser_1987}. These methods enable velocity-density comparisons which have been used to estimate the growth rate of structure at low-to-intermediate redshift \citep{Turner_2023, Huterer_2017, Qin_2019, Pezzotta_2017, Howlett_2016}. In this paper, the focus is on a complementary technique: \textit{direct} measurements of peculiar velocities at low redshift using distance-indicator relations.

\citet{Carrick_2015} applied this method by comparing the reconstructed velocity field from the 2M++ galaxy redshift survey with peculiar velocities from the SFI++ Tully-Fisher dataset \citep{Springob_2009} and the First Amendment supernova dataset \citep{Turnbull_2011}, obtaining $\beta$\,=\,0.431\,$\pm$\,0.021. The strength of the approach adopted by \citet{Carrick_2015} lies in the precise determination of the relative amplitudes of peculiar velocities from the 2M++ density field; its weakness lies in the constraint on the overall scaling of the velocity field (i.e.\ $\beta$) from limited peculiar velocity datasets. Here we assume the peculiar velocities predicted from the 2M++ density field are correct up to scaling by $\beta$, and obtain a more precise and independent estimate of $\beta$ by comparing Carrick et~al.'s velocity field reconstruction to the peculiar velocities we derive from the Cosmicflows-4 (CF4) Tully-Fisher dataset \citep{2020cf4}. We focus on this dataset in order to refine the methodology for Tully-Fisher \HI\ peculiar velocity measurements in preparation for the WALLABY \HI\ Tully-Fisher survey \citep{Koribalski_2020, Westmeier_2022}, which is expected to greatly increase the number of galaxies for which such measurements can be made.

To study peculiar velocities directly, both redshifts and measures of galaxy distance that are independent of redshift are needed. These usually involve scaling relations between distance-dependent quantities and distance-independent quantities. Measurement of a distance-independent quantity leads, through such a scaling relation, to a prediction of a distance-dependent quantity based on redshift. The difference with the observed value gives an estimate of the peculiar velocity and the comoving distance. Such methods include the Tully-Fisher relation between luminosity and rotation velocity for late-type galaxies \citep{Tully_1977} and the Fundamental Plane relation between size, surface brightness and velocity dispersion for early-type galaxies \citep{Dressler_1987, Djorgovski_1987}. 

Conventionally, the parameters of the scaling relation are determined in a first step, either by using a subset of galaxies with independent distance measurements or by assuming the galaxies' distances are given by their redshifts (i.e.\ no peculiar velocities). Distances are then estimated in a second step based on the calibrated scaling relation. Calibration assuming no peculiar velocities works because redshifts are relatively precise indicators of distance at sufficiently large distances that the Hubble velocity is much greater than typical peculiar velocities. However, this approach inflates the scatter around the distance-indicator relation by ignoring peculiar velocities. The alternative, self-consistent, approach advocated here recovers the parameters of the scaling relation and the linear-regime peculiar velocity field simultaneously and jointly in a single step \citep[cf.][]{Said_2020,Dam_2020}, which in principle offers greater precision.

This method assumes parametric models for the velocity field and the Tully-Fisher relation. Given values of the model parameters, we can predict the observable quantities for the galaxy sample. Combining (weak) priors on the model parameters with the likelihood of observing the actual dataset given the predicted dataset for these values of the model parameters, we can compute the posterior probability distribution for the model parameters given the observed data, the assumed model, and the priors. This forward modelling approach is derived from \citet{Saglia_2001} and \citet{Colless_2001}, who used 3D Gaussians to fit the Fundamental Plane distance indicator for early-type galaxies; \citet{Said_2020} and \citet{Howlett_2022} developed this approach further. \citet{Willick_1997b} used a forward-modelling approach for Tully-Fisher data, but with limited success due to the smaller datasets available at the time. Here, we combine the work of \citet{Willick_1997b} and \citet{Said_2020}, to provide a single-step forward-modelling maximum likelihood method for simultaneously fitting the Tully-Fisher relation and the peculiar velocity field. We apply this to the largest available Tully-Fisher sample to date, that collated in the CF4 dataset \citep{2020cf4}, to obtain an improved constraint on the velocity scale parameter $\beta$ and hence on the growth rate of structure. 

The paper is organised as follows. In Section~\ref{methodology} we describe the method used to simultaneously fit the Tully-Fisher relation and the peculiar velocity model; we also show how to optimally estimate distances and peculiar velocities for individual galaxies. In Section~\ref{TFmodel} we describe the CF4 Tully-Fisher data and the model used to fit and simulate it, which includes a non-linear Tully-Fisher relation, a linearly variable scatter, the sample selection criteria, and the treatment of outliers. In Section~\ref{selection} we describe the selection function for the CF4 Tully-Fisher data. In Section~\ref{mocks} we summarise our simulation framework, describing the models used and the procedure for generating mocks. In Section~\ref{results} we present the results of our analysis applied to the CF4 Tully-Fisher catalogue, giving estimates of the Tully-Fisher relation parameters, the velocity field parameters, and the growth rate of structure, as well as comparing this growth rate estimate to previous results to results in the literature. We provide estimates of the distances and peculiar velocities for individual galaxies, both from the Tully-Fisher relation alone and in combination with the best-fit velocity field model. We also discuss the differences between our Tully-Fisher relation and that of \citep{2020cf4}. In Section~\ref{WALLABY} we discuss prospects for future applications of this method to the WALLABY Tully-Fisher survey currently under way, forecasting the expected constraints that can be obtained on the growth rate of structure. In Section~\ref{conclusions} we state our conclusions.

Unless otherwise noted, we assume a flat $\Lambda$CDM cosmology with $\Omega_m$\,=\,0.315, as favoured by the Planck collaboration \citep{2018planck}, and use $H_0$\,=\,100$h$\kms\,Mpc$^{-1}$ (including when calculating distance moduli and absolute magnitudes).

\section{Method}
\label{methodology}

We describe a forward-modelling method (i.e.\ one comparing observed quantities to the same quantities predicted from a model) in which the parameters of the distance-indicator relation and the peculiar velocity field are inferred jointly in one step via maximum likelihood. The general form of this methodology was originally developed by \citet{Willick_1997b} for comparing Tully-Fisher observations of spiral galaxies to the peculiar velocity and density fields predicted from redshift surveys. A variation was used by \citet{Saglia_2001} and \citet{Colless_2001} for comparing Fundamental Plane observations of early-type galaxies with peculiar velocity models, then extended and improved by \citet{Springob_2014}, \citet{Said_2020} and \citet{Howlett_2022} for application to the large 6dFGS and SDSS surveys. Here we slightly modify Willick et~al.'s original method for estimating Tully-Fisher distances to apply in redshift space. This avoids the effects of Malmquist bias and requires only a model for the peculiar velocity field and not the density field. We use the `forward' form of the Tully-Fisher relation to minimise dependence on the intrinsic distribution of observables and conditional probabilities to reduce uncertainties due to selection effects.

\subsection{Forward model of conditional probabilities}
\label{forwardmodel}

Following \citet{Willick_1997b}, we develop an expression for the conditional probability of a galaxy with an observed magnitude $m$, an observed velocity width $w \equiv \log{\Wmxc}-2.5$, an observed 21\,cm (velocity-integrated\footnote{In this work, all 21\,cm fluxes are {\it velocity} integrated flux densities as defined in \citet{2017meyer}, and thus have units Jy\kms.}) flux density $s \equiv \log{S_{21}}$, and an observed redshift $z$, given a model for the peculiar velocity field. We start from the joint probability distribution of the observables and the galaxy's position $\mathbf{r}$, which can be expressed in terms of its cosmological redshift $z_c$ and sky coordinates $(\alpha,\delta)$. The only non-observable quantity is the cosmological redshift $z_c$, which we will subsequently marginalise over. The joint probability is 
\begin{align}
P(&m,w,s,z,\mathbf{r}\,|\,\parTF,\parPV) 
 = P(m,w,s,z,z_c,\alpha,\delta\,|\,\parTF,\parPV) \nonumber \\
 &= P(m,w,f\,|\,z,z_c,\alpha,\delta,\parTF,\parPV)\,P(z,z_c,\alpha,\delta\,|\,\parTF,\parPV) \nonumber \\
 &= P(m,w,f\,|\,z,\alpha,\delta,\parTF,\parPV)\,P(z_c\,|\,z,\alpha,\delta,\parPV)\,P(z,\alpha,\delta)
\label{eq:jointprob}
\end{align}
where $\parTF$ represents the parameters of the Tully-Fisher relation model and $\parPV$ represents the parameters of the peculiar velocity field model (note that the peculiar velocity model requires the galaxy's sky position and either its redshift $z$ or its distance $z_c$). In the first factor on the final line, we explicitly model the Tully-Fisher observables as depending on the galaxy's observed redshift, the parameters of the Tully-Fisher relation, and the galaxy's true distance through a combination of the redshift and the parameters of the peculiar velocity model (see below). We exclude any dependence on its sky position, which is appropriate for the sample studied here but not if the selection function varied over the sky. In the second factor on the final line, we choose to factor $P(z,z_c,\ldots)$ by conditioning on $z$ rather than $z_c$. If we had factored this as $P(z\,|\,z_c,\dots)P(z_c,\ldots)$, then evaluating the first of these factors would require the peculiar velocity model and the second would require the (related) real-space density model. With our chosen factorisation, the peculiar velocity model is required for the first factor, $P(z_c\,|\,z,\alpha,\delta,\parPV)$, but not for the second factor, since $P(z,\alpha,\delta)$ is the redshift-space distribution, an observable quantity. In either case this factor does not depend on the parameters of the Tully-Fisher relation.

Before we can turn this probability factorisation into a model, we first have to establish the relation between observed and cosmological redshifts given a model for the peculiar velocity field. In general, observed redshift $z$ is related to cosmological redshift $z_c$ and peculiar redshift $z_p \equiv 1 + v_{\!p}/c$ (where $v_{\!p}$ is the peculiar velocity) by
\begin{equation}
(1+z) = (1+z_c)(1+z_p) ~.
\label{eq:redshifts}
\end{equation}
So, given a galaxy at $(z,\alpha,\delta)$ and a model that predicts the line-of-sight peculiar velocity at each location in redshift space, $v_{\!p}^\prime = cz_p^\prime = u(z,\alpha,\delta)$, the predicted cosmological redshift is
\begin{equation}
z_c^\prime(z,\alpha,\delta,\parPV) = \frac{z-z_p^\prime}{1+z_p^\prime} 
 = \frac{z-u(z,\alpha,\delta,\parPV)/c}{1+u(z,\alpha,\delta,\parPV)/c} ~.
\label{eq:zcpred}
\end{equation}

The peculiar velocity field model used here has four parameters, $\parPV = \{ \beta, V_x, V_y, V_z \}$; these are the velocity scaling $\beta$ and the components of the residual bulk flow $\mathbf{V}_\textrm{ext} = (V_x, V_y, V_z)$. The model takes the form
\begin{equation}
u(z,\alpha,\delta,\parPV) = \beta V_{\textrm{pred}}(z,\alpha,\delta) + \mathbf{V}_\textrm{ext} \cdot {\mathbf{\hat{r}}(z,\alpha,\delta)}
\label{eq:velocitymodel}
\end{equation}
where $V_\textrm{pred}$ is the predicted radial peculiar velocity given by the 2M++ model of \citet{Carrick_2015}, normalised so that $\beta=1$ and $\mathbf{V}_\textrm{ext}=0$. The velocity scaling is a combination of the mean density of the universe $\Omega_m$ and the linear bias $b$ of the galaxy sample, $\beta = \Omega_m^\gamma/b$, where $\gamma \approx 0.55$ for General Relativity. The residual bulk flow accounts approximately for the effect of the unknown density distribution outside the 2M++ redshift survey volume (i.e.\ at distances beyond about 200\,$h^{-1}$\,Mpc).

To model the Tully-Fisher relation, we also require the relation between these redshifts and luminosity distance. Apparent magnitude $m$ and absolute magnitude $M$ are related by the distance modulus $\mu$, which is a function of both observed redshift $z$ and cosmological redshift $z_c$ \citep[as explained by][]{Calcino_2017}:
\begin{align}
m = M + \mu(z,z_c) &= M + 25 + 5\log D_L(z,z_c) \nonumber \\
 &= M + 25 + 5\log(1+z) + 5\log D_C(z_c)
\label{eq:distancemodulus}
\end{align}
where $D_L(z,z_c) = (1+z)D_C(z_c)$ is the luminosity distance and $D_C(z_c)$ is the comoving distance. Given the above peculiar velocity model, the predicted apparent magnitude can be expressed as
\begin{align}
m^\prime(w,z,\alpha,\delta,\parPV,&\parTF) = \nonumber \\
&M^\prime(w) + 25 + 5\log(1+z) + 5\log D_C(z_c^\prime)
\label{eq:mpred}
\end{align}
where $M^\prime(w)$ is the absolute magnitude predicted from the velocity width by the Tully-Fisher relation and $z_c^\prime(z,\alpha,\delta,\parPV)$ is given by Equation~\ref{eq:zcpred}. Note that by using $z_c^\prime$ rather than $z_c$ we are using the best estimate from the velocity field model, thereby making this expression more sensitive to the parameters of the model while at the same time avoiding the need to include this quantity in consideration when marginalising over $z_c$. In doing so we are departing from the procedure introduced by \citet{Willick_1997b}, but this approach simplifies the method and reduces the computational cost with negligible effect on the precision of the fitted parameters, since $z_c$ for any individual galaxy is much more strongly constrained by the velocity field model than by its offset from the Tully-Fisher relation (see below).

We can now generate a model for each of the factors in Equation~\ref{eq:jointprob}. As shown by \citet{Willick_1994}, the probability of measuring a particular apparent magnitude given a galaxy's \HI\ velocity width and observed redshift (the first factor on the right-hand side of Equation~\ref{eq:jointprob}) can be represented by a Gaussian distribution of the form
\begin{align}
P(m,w,f\,|\,z,\alpha,\delta,&\parTF,\parPV) \nonumber \\
 &\propto \frac{\phi(w)F(m,w,s,z)}{\sigma_\textrm{TF}} 
 \exp \left[ -\frac{(m - m^\prime)^{2}}{2\sigma_\textrm{TF}^2} \right]
\label{eq:TFmodel}
\end{align}
where $\phi(w)$ is the intrinsic velocity width distribution, $F(m,w,s,z)$ is the sample selection function, and $m^\prime = m^\prime(w,z,\alpha,\delta,\parPV)$ is given by Equation~\ref{eq:mpred}. We allow the selection function to depend on magnitude, velocity width, flux density and redshift, but {\it not} on the galaxy's comoving distance nor on its sky position (which may not hold for heterogeneous samples). This `forward' model predicts the galaxy's apparent magnitude in terms of its distance-independent \HI\ velocity width and its observed redshift (given a peculiar velocity field model), and assumes there is a Gaussian scatter about the Tully-Fisher relation, varying linearly with velocity width, of $\sigma_\textrm{TF}(w) = a_\sigma w + b_\sigma$ magnitudes.

The second factor on the right-hand side of Equation~\ref{eq:jointprob} gives the probability of a galaxy having a particular cosmological redshift given its redshift-space location, and encapsulates the peculiar velocity model in the form
\begin{equation}
P(z_c\,|\,z,\alpha,\delta,\parPV) \propto \frac{1}{\sigma_{v}} 
 \exp \left[ -\frac{(cz_c - cz_c^\prime)^{2}}{2\sigma^{2}_{v}} \right]
\label{eq:PVmodel}
\end{equation}
where $z_c^\prime(z,\alpha,\delta,\parPV)$ is the predicted cosmological redshift given by the peculiar velocity model at the galaxy's location (see Equation~\ref{eq:zcpred}). We assume that the cosmological redshift has a Gaussian distribution around the value predicted by the velocity model, with a scatter $\sigma_{v}$. This scatter parametrises our ignorance of the fine details of the velocity field, and accounts for the effects of (small) observational errors in the redshift measurements, uncertainties in the reconstruction of the linear velocity model, and additional errors in the velocity field due to non-linearities. This velocity `noise' is expected to be about 150\kms \citep{Carrick_2015}, although it cannot be globally constant, as systematics in the reconstruction of the linear velocity field will vary with location and non-linearities in the velocity field will be greater in denser regions. A conservative estimate is $\sigma_v$\,=\,250\kms, approximately the dispersion in the peculiar velocities predicted by the 2M++ velocity field model.

The third and final factor on the right-hand side of Equation~\ref{eq:jointprob}, $P(z,\alpha,\delta)$, is the observed redshift-space distribution of galaxies along a given line of sight. 

Having constructed models for all the factors of the joint probability given in Equation~\ref{eq:jointprob}, we marginalise over the sole non-observable quantity, the comoving redshift $z_c$, to make the joint probability a function only of the observables and the model parameters
\begin{align}
P&(m,w,s,z,\alpha,\delta\,|\,\parTF,\parPV) \nonumber \\
 &= \int\!\!\!P(m,w,s,z,z_c,\alpha,\delta\,|\,\parTF,\parPV)\,dz_c \nonumber \\
 &= P(m,w,f\,|\,z,\alpha,\delta,\parTF,\parPV)\,.\!\!
 \int\!\!\!P(z_c\,|\,z,\alpha,\delta,\parPV)\,dz_c\,.\, P(z,\alpha,\delta) \nonumber \\
 &\propto \frac{\phi(w)F(m,w,s,z)}{\sigma_\textrm{TF}} 
 \exp \left[ -\frac{(m - m^\prime)^{2}}{2\sigma_\textrm{TF}^2} \right]
 P(z,\alpha,\delta)
\label{eq:jointprobobs}
\end{align}
where (here and below) integration is over the whole domain of the variable. Since $z_c$ only appears in $P(z_c\,|\,z,\alpha,\delta,\parPV)$, which integrates to unity, we obtain a result that only depends on the model parameters $\parTF$ and $\parPV$ through the predicted magnitude $m^\prime$ (set by the Tully-Fisher relation and the peculiar velocity field model; see Equation~\ref{eq:mpred}) and the Tully-Fisher relation scatter, $\sigma_{\rm TF}$ (which combines the intrinsic scatter in the Tully-Fisher relation and scatter due to systematic errors in the peculiar velocity model).

However, as noted by \citet{Willick_1997b}, using the joint probability to estimate the parameters of the Tully-Fisher relation and the peculiar velocity field is not optimal because of its sensitivity to the velocity width distribution, $\phi(w)$, the selection function $S(m,w,z,f)$, and the line-of-sight redshift-space distribution $P(z,\alpha,\delta)$, none of which are of direct interest and all of which are less precisely known than is desirable. It is therefore better to focus on a conditional probability function, which for the forward Tully-Fisher relation used here means the conditional probability of observing a given apparent magnitude $m$ given all the other observables and the Tully-Fisher and peculiar velocity models. Noting that
\begin{align}
P&(m,w,s,z,\alpha,\delta\,|\,\parTF,\parPV) \nonumber \\
 &= P(m\,|\,w,s,z,\alpha,\delta,\parTF,\parPV)\,P(w,s,z,\alpha,\delta\,|\,\parTF,\parPV) \nonumber \\
 &= P(m\,|\,w,s,z,\alpha,\delta,\parTF,\parPV) \nonumber \\
 &\quad\quad .\!\int\!\!\!P(m,w,s,z,\alpha,\delta\,|\,\parTF,\parPV)\,dm 
\end{align}
we have
\begin{align}
P(m\,|\,w,s,z,\alpha,\delta,&\parTF,\parPV) 
 = \frac{P(m,w,s,z,\alpha,\delta\,|\,\parTF,\parPV)}{\int\!P(m,w,s,z,\alpha,\delta\,|\,\parTF,\parPV)\,dm} \nonumber \\
 &= \frac{F(m,w,s,z)\exp\left[-\frac{(m-m^\prime)^{2}}{2\sigma_\textrm{TF}^2}\right]}{\int F(m,w,s,z)\exp\left[-\frac{(m-m^\prime)^{2}}{2\sigma_\textrm{TF}^2}\right]\,dm}
\label{eq:mcondprob}
\end{align}
where the $\phi(w)$ and $P(z,\alpha,\delta)$ factors have cancelled in the numerator and denominator. In the simplest case, where there is no magnitude selection because the sample is defined by other parameters, independent of magnitude, the selection function cancels in the numerator and denominator and the conditional probability then reduces to
\begin{equation}
P(m\,|\,w,z,\alpha,\delta,\parTF,\parPV) = \frac{1}{\sqrt{2\pi}\sigma_{\rm TF}} \exp\left[-\frac{(m-m^\prime)^{2}}{2\sigma_\textrm{TF}^2}\right]
 \label{eq:mcondprobfinal}
\end{equation}
where the flux has been dropped as it is no longer relevant. The next simplest case is when the magnitude selection is a simple hard limit, $m_{\rm lim}$, giving
\begin{equation}
P(m\,|\,w,s,z,\alpha,\delta,\parTF,\parPV) 
= \frac{\exp\left[-\frac{(m-m^\prime)^{2}}{2\sigma_\textrm{TF}^2}\right]}{\int_{-\infty}^{m_{\rm lim}} \exp\left[-\frac{(m-m^\prime)^{2}}{2\sigma_\textrm{TF}^2}\right]\,dm} 
\label{eq:mcondprobmaglim}
\end{equation}
where the integral in the denominator accounts for the part of the distribution not observed due to the magnitude limit.

Finally, given fully specified models for the Tully-Fisher relation, the peculiar velocity field and the sample selection function, we estimate the model parameters by maximising the (logarithm of) the likelihood constructed as the product (over all sample galaxies $i$) of the conditional probability computed from Equation~\ref{eq:mcondprob}, \ref{eq:mcondprobfinal} or \ref{eq:mcondprobmaglim}. For example, with no magnitude selection, the likelihood is simply
\begin{align}
\ln \mathcal{L} &= \sum_i \ln P(m_i\,|\,w_i,z_i,\alpha_i,\delta_i,\parTF,\parPV) \nonumber \\
 &= \sum_i -\ln\left(\sqrt{2\pi}\sigma_{\rm TF}\right)-\left[\frac{(m-m^\prime)^2}{2\sigma_{\rm TF}^2}\right] ~.
\label{eq:lnlike}
\end{align}
We use Markov chain Monte Carlo (MCMC) to sample this likelihood. The specific algorithm and its implementation is described in \citet{emcee}.

The key points in the derivation of Equation~\ref{eq:mcondprob} are: (i)~The factorisation of the probability distribution in Equation~\ref{eq:jointprob}, using observed redshift rather than cosmological redshift (i.e.\ redshift space rather than real space), simplifies the method and means only the (redshift-space) peculiar velocity model is required, and not the (real-space) density model as well. (ii)~Using the best estimate of the peculiar velocity model to estimate the predicted magnitude (Equation~\ref{eq:mpred}), rather than marginalising over $z_c$, departs from the procedure of \citet{Willick_1997b} but simplifies the method and reduces computational cost. It has negligible effect on the precision of the fitted parameters, since $z_c$ for any individual galaxy is much more strongly constrained by the peculiar velocity model (Equation~\ref{eq:PVmodel}) than by its offset from the Tully-Fisher relation (Equation~\ref{eq:TFmodel}), since in velocity units $\sigma_v$$\ll$\,$\sigma_{\rm TF}$. (iii)~The use of the conditional probability for the magnitudes removes the dependence on the intrinsic distribution of velocity line widths and on the observed line-of-sight redshift distribution. (iv)~Combining the forward Tully-Fisher relation (i.e.\ making magnitude the dependent variable) and the conditional probability distribution means selection function factors independent of magnitude cancel; to the extent they are independent of magnitude, the selection functions on velocity width, flux, and redshift are irrelevant. (v)~The magnitude selection criterion for the sample needs to be determined and modelled. Equation~\ref{eq:mcondprob} reduces to Equation~\ref{eq:mcondprobfinal} for no magnitude selection or to Equation~\ref{eq:mcondprobmaglim} for a fixed magnitude limit. (vi)~The models for the Tully-Fisher relation and the peculiar velocity field affect the likelihood only through the predicted magnitude $m^\prime$ (Equation~\ref{eq:mpred}) and the Tully-Fisher scatter $\sigma_{\rm TF}$ (which incorporates both the intrinsic scatter about the relation in the absence of peculiar velocities and the scatter due to non-linearities in the velocity field or errors in the velocity field model).

\begin{figure*}\centering
\includegraphics[width=\textwidth]{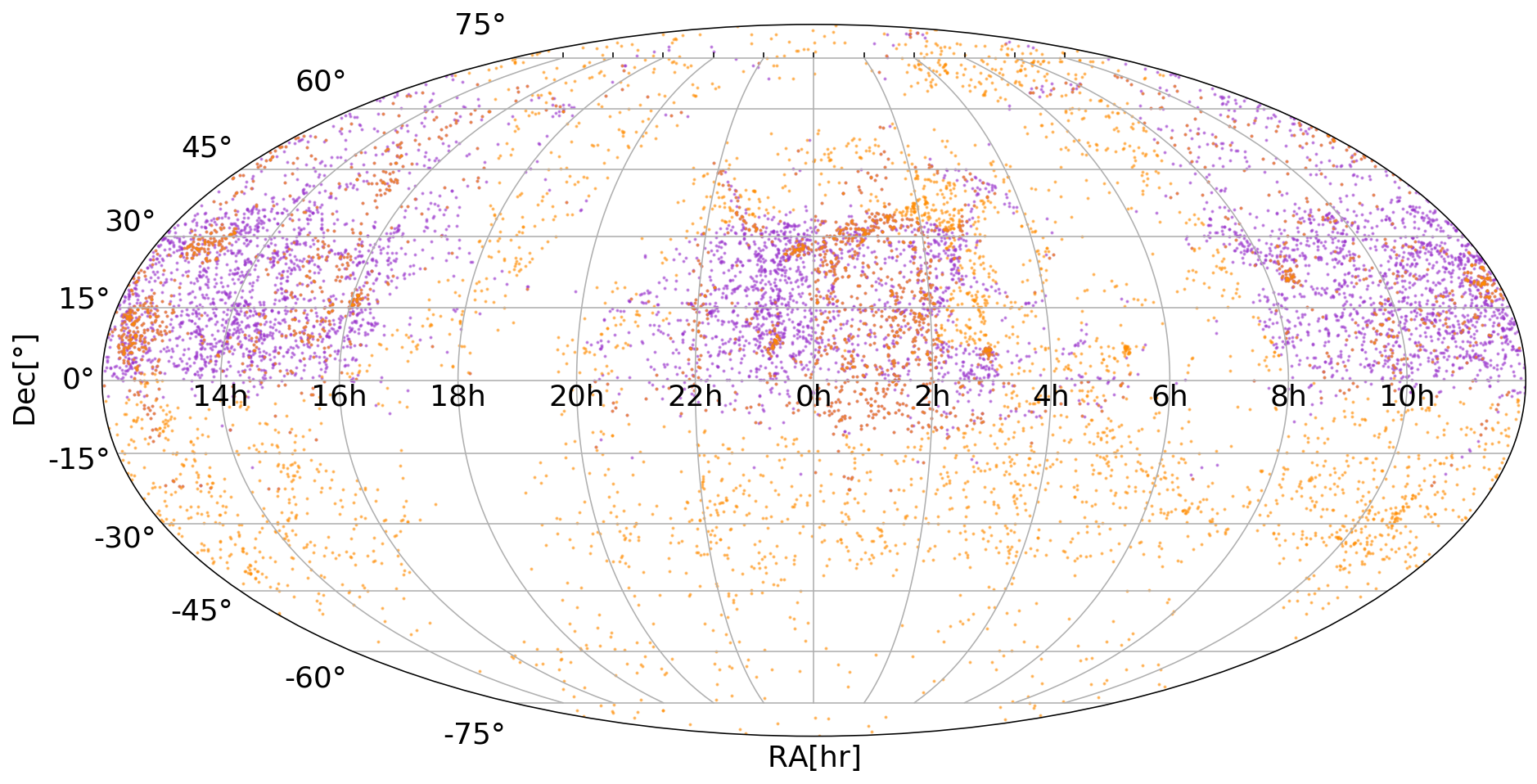}
\caption{Sky coverage of the CF4 Tully-Fisher galaxies with photometry provided by WISE $W1$-band magnitudes (orange) and SDSS $i$-band magnitudes (purple). While WISE covers the whole sky, SDSS is limited to less than half the northern sky.}
\label{fig:footprint}
\end{figure*}

\subsection{Estimating distances and peculiar velocities}
\label{estdistpv}

Given the Tully-Fisher and peculiar velocity field parameters estimated from the entire sample of galaxies, we now want to estimate distances and peculiar velocities for individual galaxies. We choose to fix the parameters of the Tully-Fisher relation $\parTF$ and peculiar velocity model $\parPV$ at their best-fit values, although in principle we could marginalise over these parameters. We can then write the conditional probability of a galaxy having cosmological redshift $z_c$ given the observables and the fixed model parameters as follows:
\begin{align}
P(z_c\,&|\,m, w,z, \alpha,\delta,\parTF,\parPV) \nonumber \\
 &= \frac{P(m\,|\,w,z,\parTF,z_c)\,P(z_c\,|\,z,\alpha,\delta,\parPV)}{P(m\,|\,w,z,\alpha,\delta,\parPV,\parTF)} \nonumber \\
 &= \frac{P(m\,|\,w,z,\parTF,z_c)\,P(z_c\,|\,z,\alpha,\delta,\parPV)}{\int\!P(z_c,m\,|\,w,z,\alpha,\delta,\parPV,\parTF)\,dz_c} \nonumber \\
 &\propto \frac{1}{2\pi\sigma_\textrm{TF}\sigma_{v}}\exp\left[-\frac{(m-m^\prime)^{2}}{2\sigma_\textrm{TF}^2}\right]\exp \left[ -\frac{(cz_c - cz_c^\prime)^{2}}{2\sigma^{2}_{v}} \right]
\label{eq:zccondprob}
\end{align}
where we omit irrelevant parameters in the conditional probabilities and, in the final step, use Equation~\ref{eq:PVmodel} for the conditional probability of the comoving redshift and Equation~\ref{eq:mcondprobfinal} for the conditional probability of the apparent magnitude (the latter could be replaced by Equation~\ref{eq:mcondprob} or~\ref{eq:mcondprobmaglim}, as appropriate).

This estimate for each galaxy's distance combines two constraints on its comoving redshift: one derived from the Tully-Fisher relation, obtained by solving $m^\prime(z_c)$\,=\,$m$ (Equation~\ref{eq:mpred}), and one derived from the velocity field model by setting $z_c$\,=\,$z^\prime$ (Equation~\ref{eq:zcpred}). The relative weights determining the best overall estimate are set by the scatters of the Tully-Fisher relation, $\sigma_{\rm TF}$, and the velocity field model, $\sigma_v$.

\section{Tully-Fisher relation}\label{TFmodel}

At large radii, spiral galaxy rotation curves reach a maximum rotation velocity $\Vrot$. The tight correlation observed between $\Vrot$ and intrinsic luminosity $L$ is called the Tully-Fisher relation \citep{Tully_1977}: simply stated, brighter spirals rotate faster. This relation is well-modelled by a power law, $L \propto \Vrot^\alpha$ \citep[see, e.g.,][]{Strauss_1995}. Writing the Tully-Fisher relation in terms of the absolute magnitude, $M = \textrm{const} - 2.5\log{L}$, and the velocity width of the rotation curve, $2\Vrot$, gives
\begin{equation}
M = a(\log{2\Vrot}-2.5) + b
\end{equation} 
where $a$ and $b$ are the slope and zero-point of the Tully-Fisher relation. The distance-dependent observed quantity is the apparent magnitude $m$, which is related to the absolute magnitude by the distance modulus (see Equation~\ref{eq:distancemodulus}).

Rotation velocities are closely related to the inclination-corrected broadening of the emitted neutral hydrogen (\HI) line, in particular the \HI\ line width, $W_\textrm{m50}$, at the level corresponding to 50\% of the average flux within the range covering 90\% of the total \HI\ flux, as explained by \citet{Courtois_2009}. We adopt their use of $W_\textrm{mx}$ for the quantity approximating $2\Vrot\sin{i}$, where $i$ is the galaxy inclination; throughout this paper, we will use $\Wmxc = W_\textrm{mx}/\sin{i}$ in place of $2\Vrot$ as the independent variable of the Tully-Fisher relation, and we will call $\Wmxc$ the `velocity width'.

The Cosmicflows-4 (CF4) catalogue \citep{2020cf4} is currently the largest full-sky catalogue of galaxies with Tully-Fisher distances and peculiar velocities. It is derived from heterogeneous datasets and contains 10737 galaxies with \HI\ redshifts and line widths, together with optical or infrared photometry. The \HI\ data is taken primarily from the All Digital \HI\ (ADHI) catalogue \citep{Courtois_2009}, which itself is composed mainly of good-quality \HI\ data from the ALFALFA survey \citep{2018alf}. Optical photometry in the AB system is provided by the Sloan Digital Sky Survey \citep[SDSS;][]{York_2000} in the $u$, $g$, $r$, $i$ and $z$ optical bands for most of these galaxies. Infrared $W1$ and $W2$ magnitudes from the Wide-field Infrared Satellite Explorer \citep[WISE;][]{Wright_2010} are used for galaxies outside the SDSS survey region, since WISE provides photometry for the entire sky. However, not every galaxy in the CF4 catalogue has a WISE magnitude due to the difficulty of measuring WISE photometry and its relatively poorer quality (see Section~2.2.2 and Section~8 of \citet{2020cf4}). WISE magnitudes are in the Vega system, so fixed Vega-to-AB offsets\footnote{https://wise2.ipac.caltech.edu/docs/release/allsky/expsup/sec4\_4h.html} of 2.699 were applied in \citet{Kourkchi_2019}. In general, a homogeneous dataset is preferred because it has a uniform magnitude selection function. However, the separate magnitude limits of the datasets that comprise CF4 are all subdominant to the \HI\ flux limit. As discussed in Section~\ref{selection}, this simplification of the effective selection function obviates the need to account for dataset-dependent selection effects.

\begin{figure}\centering 
\includegraphics[width=1.0\columnwidth]{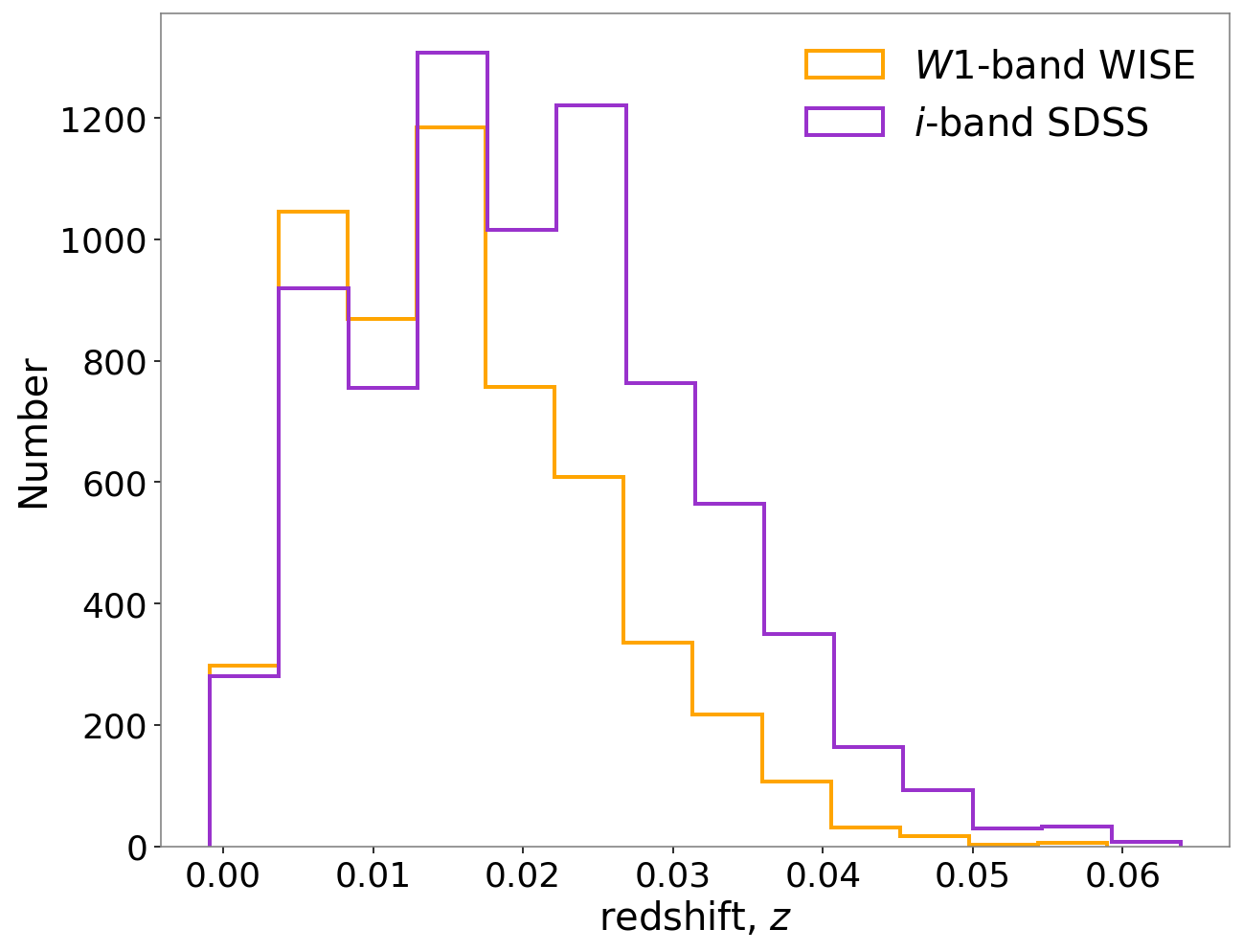}
\caption{Redshift distributions of the CF4 Tully-Fisher galaxies with WISE $W1$-band magnitudes (orange) and SDSS $i$-band magnitudes (purple).}
\label{fig:redshift}
\end{figure}

There are 7,502 targets with SDSS $i$-band magnitudes, 5479 targets with WISE $W1$-band magnitudes, and 2,244 targets with both, giving a total of 10737 unique galaxies. The positions of the CF4 Tully-Fisher galaxies on the sky are shown in Figure~\ref{fig:footprint}, while their redshift distributions are shown in Figure~\ref{fig:redshift}; the galaxies with SDSS and WISE photometry are shown in both samples.

For the rest of this section, as we develop and fit a model for the Tully-Fisher relation, we use the method described in the previous section but fix the peculiar velocity field, so that we are only fitting (and comparing) the Tully-Fisher relations. Ultimately, however, we will fit the Tully-Fisher model and the peculiar velocity model simultaneously---see Section~\ref{results}. The peculiar velocity for each galaxy adopted here is from \citet{Carr2022}, using the peculiar velocity model of \citet{Carrick_2015} based on the 2M++ redshift compilation\footnote{https://github.com/KSaid-1/pvhub}. We convert the measured apparent magnitudes to absolute magnitudes using these peculiar-velocity-corrected distance estimates and $H_0$\,=\,100$h$\kms\,Mpc$^{-1}$. The 2M++ velocity field reconstruction was chosen because previous studies have found it to be at least as good as other representations. For example, \citet{Peterson_2022} found it produced the smallest uncertainties in cosmological parameters such as $H_0$ when making peculiar velocity corrections to redshifts.

\begin{figure*}\centering
\includegraphics[width=2\columnwidth]{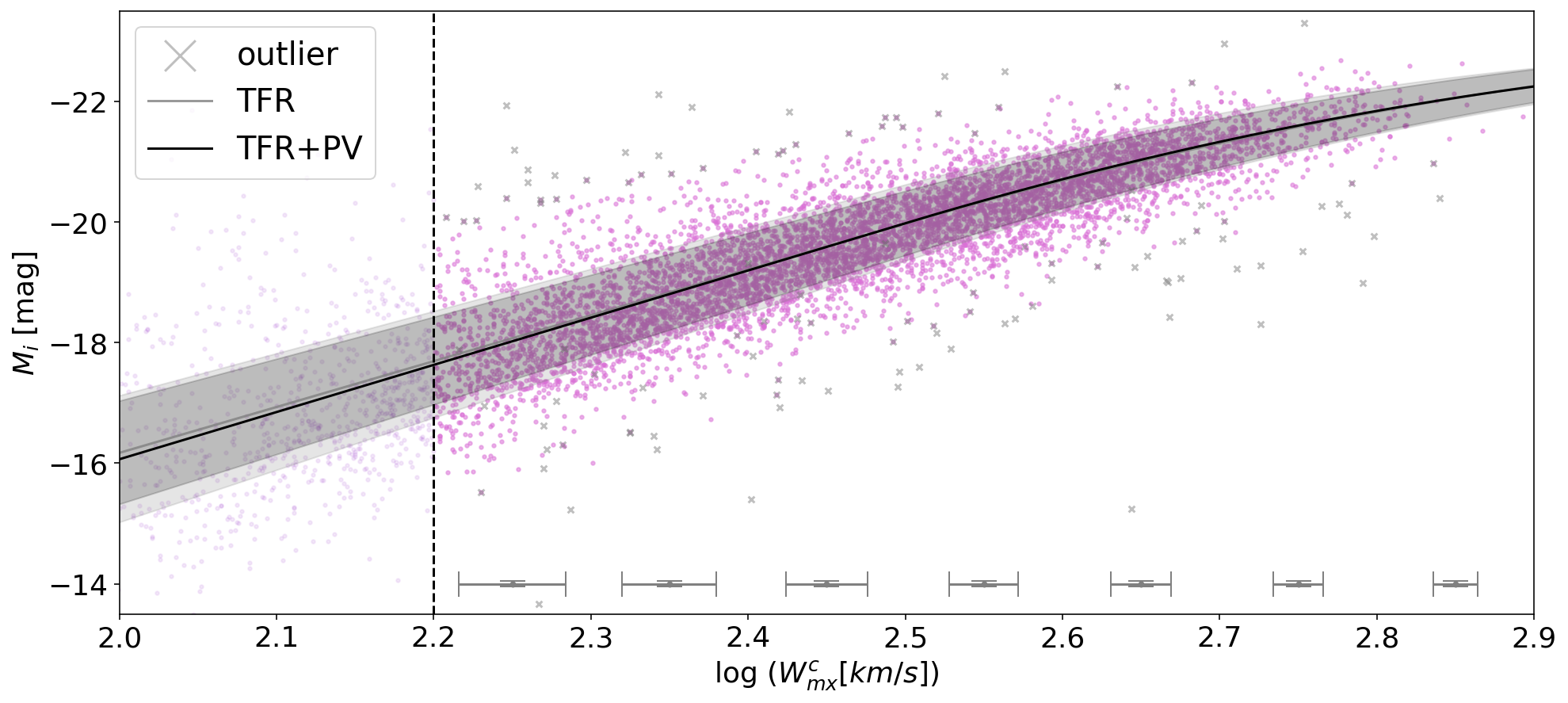}
\includegraphics[width=2\columnwidth]{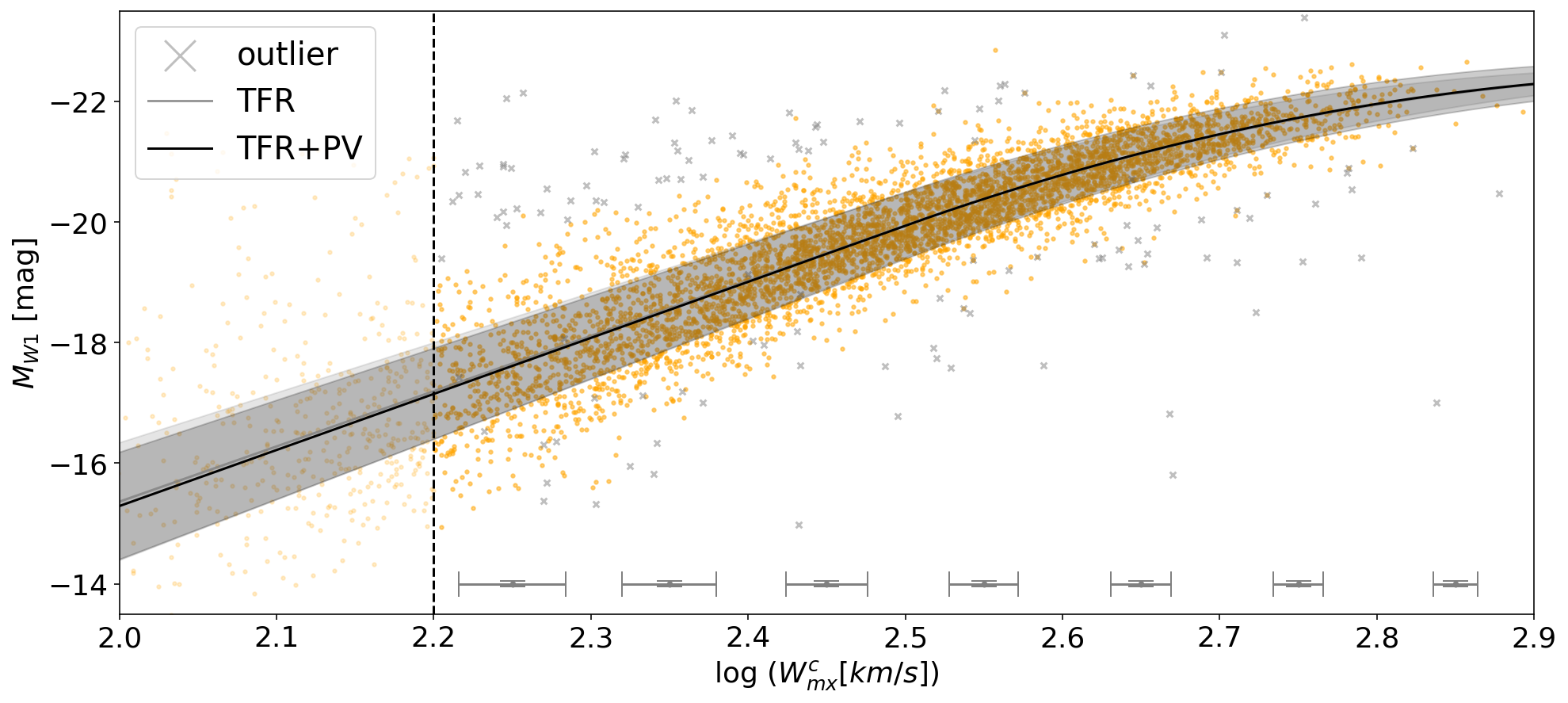}
\caption{Fits to the Tully-Fisher relation between absolute magnitude and velocity width for galaxies in CF4. Galaxy SDSS $i$-band magnitudes are shown in the top panel and WISE $W1$-band magnitudes are shown in the bottom panel. The Tully-Fisher relation fits (see Section~\ref{nonlinearity}) are shown as solid lines, and the shaded regions indicate the fitted scatter (see Section~\ref{scattermodel}). Two fits are shown in each panel: (i)~the light grey line and shading show the initial Tully-Fisher fits, where the distances used to compute the galaxies’ absolute magnitudes were obtained using their observed redshifts corrected for the peculiar velocities predicted by the 2M++ peculiar velocity model (see Table~\ref{tab:tfr_fit} for parameter values) and $H_0$\,=\,100$h$\kms\,Mpc$^{-1}$; and (ii)~the dark grey line and shading show the combined Tully-Fisher and peculiar velocity fits (see Table~\ref{tab:paramsfinal} for parameter values). Galaxies are excluded from the fit of the Tully-Fisher relation if they have velocity widths less than $\log{\Wmxc}=2.2$ (the vertical dashed line; see Section~\ref{cuts}) or if they lie more than 3$\sigma$ from the initial fit and have problematic magnitudes based on visual inspection (see Section~\ref{outliers}); the latter are shown as grey crosses. The error bars at the bottom show the median uncertainties in the velocity widths and magnitudes for the included galaxies in bins with $\Delta\log\Wmxc = 0.1$\,dex, plotted at the $\log{\Wmxc}$ bin centres.}
\label{fig:cf4data}
\end{figure*}

The following subsections describe the choices made in identifying and removing outliers, in setting the velocity width lower limit for the sample, and in determining the most appropriate models for the Tully-Fisher relation and the scatter about this relation. We present these choices in a logical linear sequence, although in practice they were arrived at through a self-consistent and convergent iterative process. The resulting $i$-band and $W1$-band Tully-Fisher relations (correcting for the 2M++ peculiar velocities) are shown in Figure~\ref{fig:cf4data}, together with the best fits to these relations obtained using the method of Section~\ref{methodology}; the parameters of these fits and their uncertainties are listed in Table~\ref{tab:tfr_fit}. After removing outliers (Section~\ref{outliers}) and applying a velocity width cut (Section~\ref{cuts}), we are left with 6,224 galaxies with $i$-band magnitudes and 4,723 galaxies with $W1$-band magnitudes (including 1896 galaxies with both $i$ and $W1$ magnitudes).

\subsection{Outliers}
\label{outliers}

One important assumption of our method is that, at any point along the Tully-Fisher relation, the residuals in absolute magnitude have a Gaussian distribution. Checking this assumption by examining the distribution of the residuals in slices of absolute magnitude, we find they have extended wings and are not pure Gaussians. Applying a 3$\sigma$-clipping criterion, 179 galaxies with $\log{\Wmxc}>2.2$ (116 with SDSS magnitudes and 118 with WISE magnitudes, with 55 having both) were identified as outliers from the Tully-Fisher relation. If these outliers were valid, they would break the assumption of Gaussian scatter and warrant a more complex model. 

\begin{figure}\centering 
\includegraphics[width=\columnwidth]{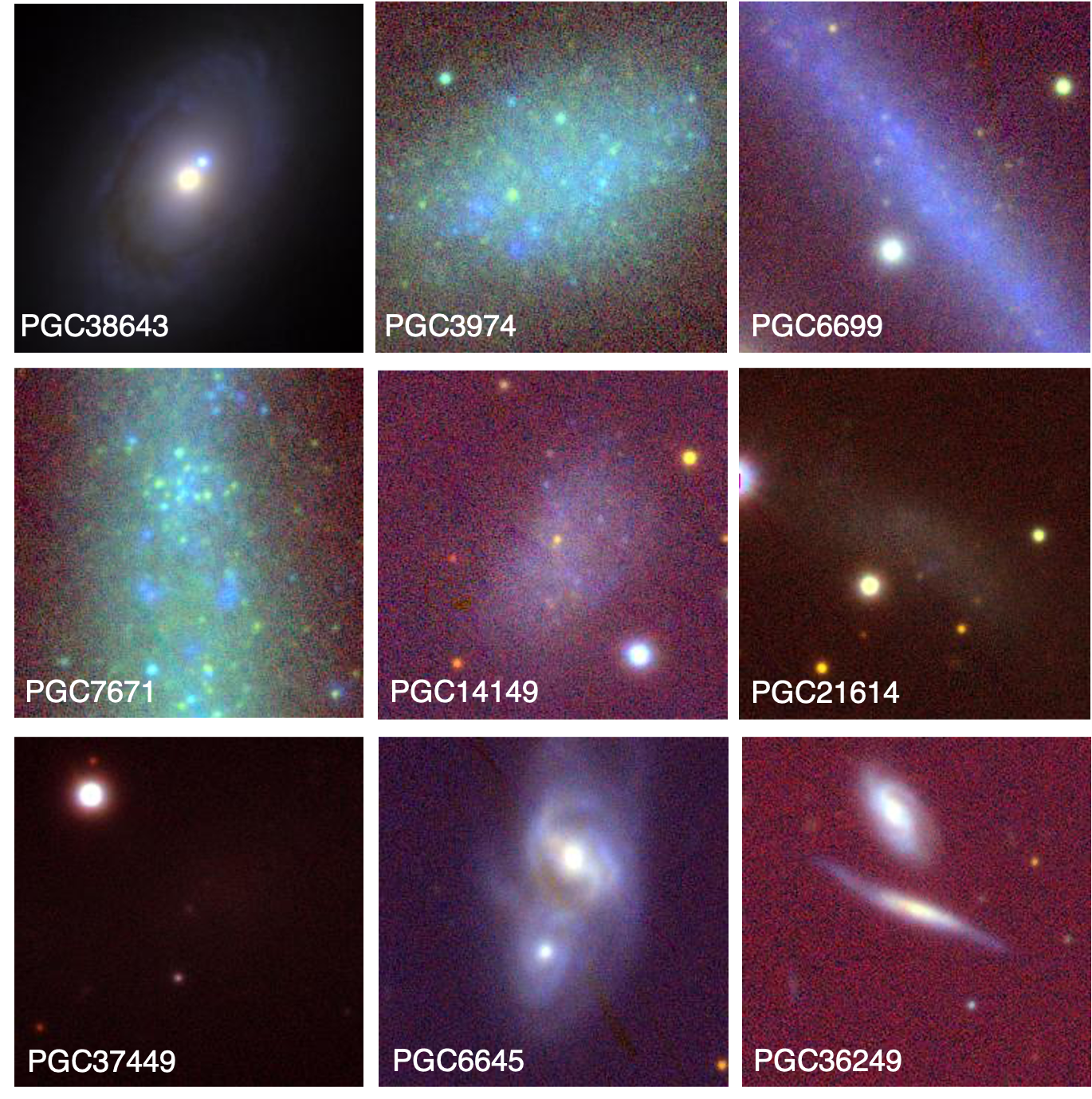}
\caption{Pan-STARRS images of example outlier galaxies (combined $y$+$i$+$g$ images from \url{https://ps1images.stsci.edu/cgi-bin/ps1cutouts}).}
\label{fig:outliers}
\end{figure}

We therefore visually inspected the Pan-STARRS, WISE and SDSS images (where available) at each source's coordinates and identified several categories of problem: (i)~Galaxies that appear clearly inconsistent with their assigned inclination. Galaxies in the CF4 samples are nominally limited to those with inclinations greater than 45 degrees, so anything appearing to be completely face-on is excluded (note that inclinations for the CF4 sample were assigned by eye). (ii)~Galaxies contaminated by very bright nearby sources. (iii)~Galaxy blends/mergers. (iv)~Images containing only faint blurs or resolved dwarf galaxies. (v)~Images containing no evident galaxies. Galaxies with these obvious issues were excluded from the analysis, resulting in the removal of 168 outliers out of 179 possibles (95\%). Figure~\ref{fig:outliers} shows examples of the galaxies that were removed. 

\begin{figure}\centering 
\includegraphics[width=0.99\columnwidth]{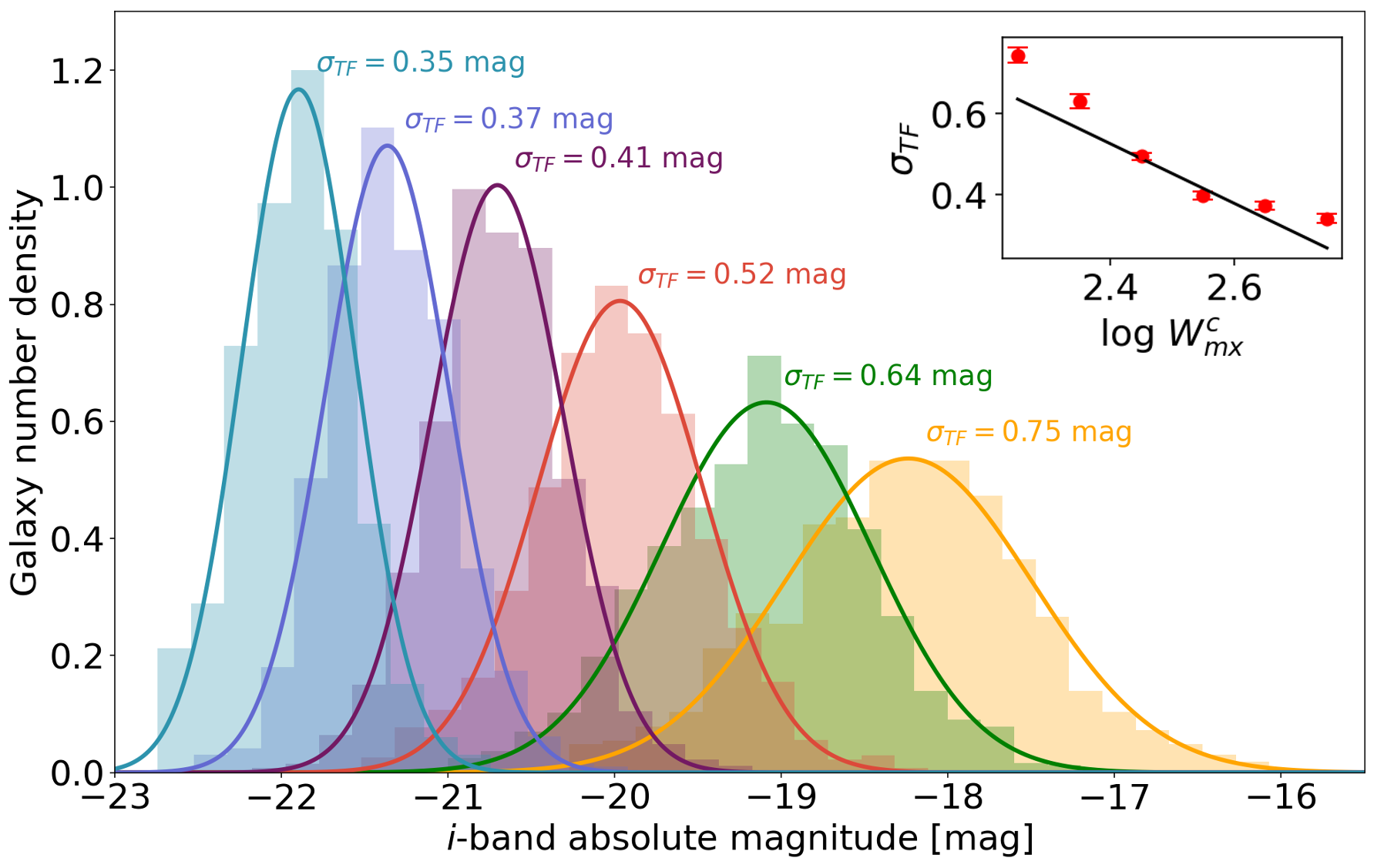} \vspace*{6pt} \\
\includegraphics[width=0.99\columnwidth]{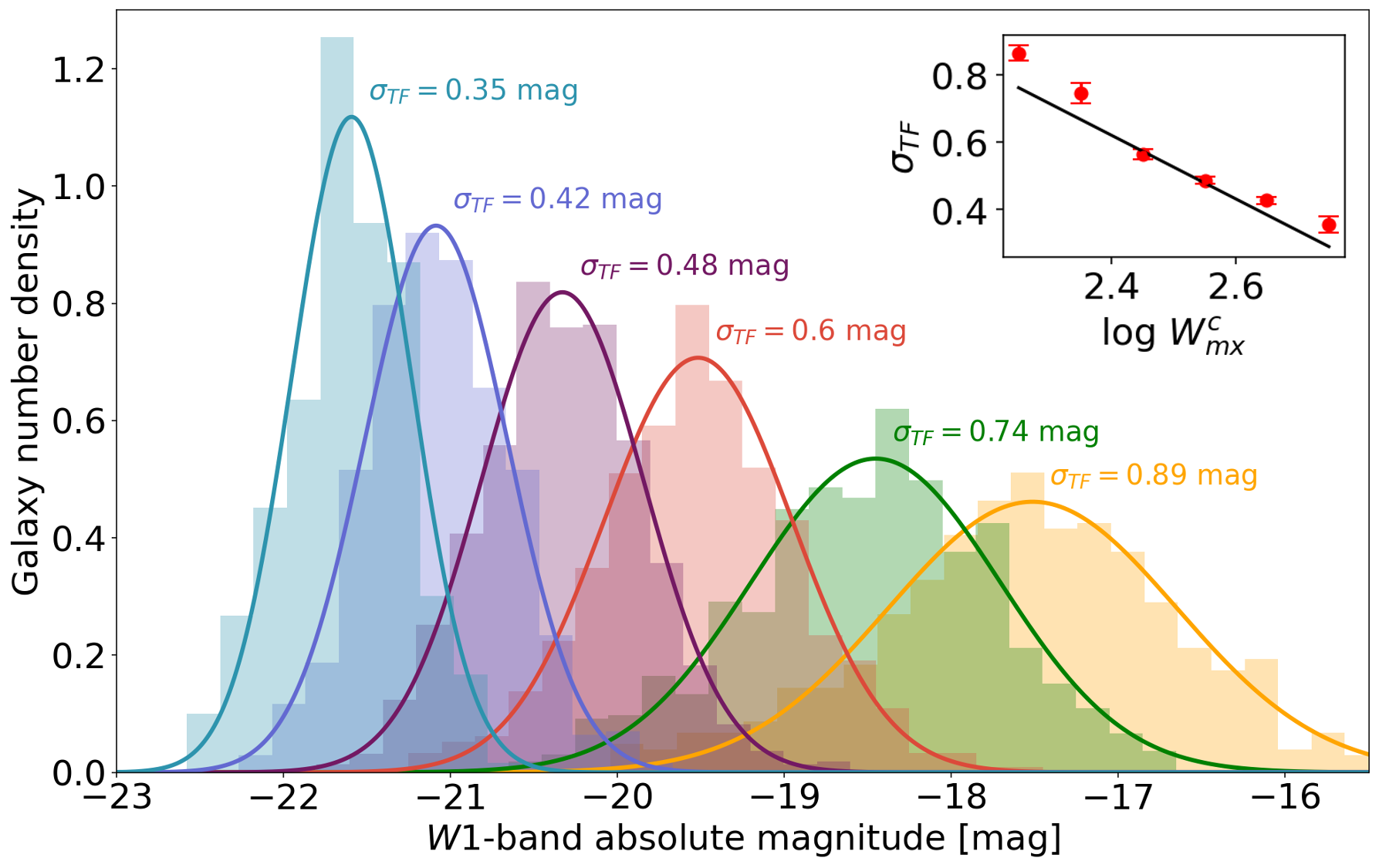}
\caption{Histograms of absolute SDSS $i$-band magnitudes (top) and WISE $W1$-band magnitudes (bottom) in 0.1\,dex velocity bins over $2.25 < \log{\Wmxc} < 2.85$, with their best-fitting Gaussians. The insets show the fit to the linear scatter model (black line) compared to the scatter measured in each $\log{\Wmxc}$ bin (red points).}
\label{fig:mag_scatter}
\end{figure}

Once the galaxies with these obvious issues are removed from the sample, we find the residuals are satisfactorily described by Gaussian distributions. This is demonstrated in Figure~\ref{fig:mag_scatter}, which shows Gaussian fits to the magnitude distributions for various slices along the Tully-Fisher relation after the removal of visually-confirmed outliers.

\begin{table}
\centering
\caption{Parameters of the Tully-Fisher relation model for CF4 data corrected for the peculiar velocities predicted by the 2M++ velocity model, fitted to 6,224 galaxies with SDSS $i$ magnitudes and 4723 galaxies with WISE $W1$ magnitudes.}
\label{tab:tfr_fit}
\begin{tabular}{ccc}
\hline\hline
Parameter & SDSS $i$-band & WISE $W1$-band \\
\hline
$a_0$ & $-$19.972\,$\pm$\,0.008 & $-$20.631\,$\pm$\,0.009 \\
$a_1$ & $-$7.60\,$\pm$\,0.07 & $-$9.39\,$\pm$\,0.08 \\
$a_2$ & 4.5\,$\pm$\,0.4 & 8.6\,$\pm$\,0.4 \\
$\epsilon_{0}$ & 2.15\,$\pm$\,0.06 & 2.77\,$\pm$\,0.07 \\
$\epsilon_{1}$ & $-$0.65\,$\pm$\,0.04 & $-$0.90\,$\pm$\,0.03 \\
\hline
\end{tabular}
\end{table}

\subsection{Lower limit on velocity width}
\label{cuts}

Figure~\ref{fig:cf4data} shows that galaxies scatter more about the Tully-Fisher relation at fainter magnitudes and lower velocity widths. Including all these galaxies gives poorer fits and larger parameter uncertainties. \citet{K2020} ameliorated this by imposing a low-luminosity limit at $M_i = -17$ in fitting the inverse Tully-Fisher relation. Here we fit the forward Tully-Fisher relation, minimising residuals in magnitude, so a cut in velocity width is more appropriate, as it will not result in biased estimates for the Tully-Fisher relation parameters (see Section~\ref{methodology}).

\begin{figure}\centering 
\includegraphics[width=1\columnwidth]{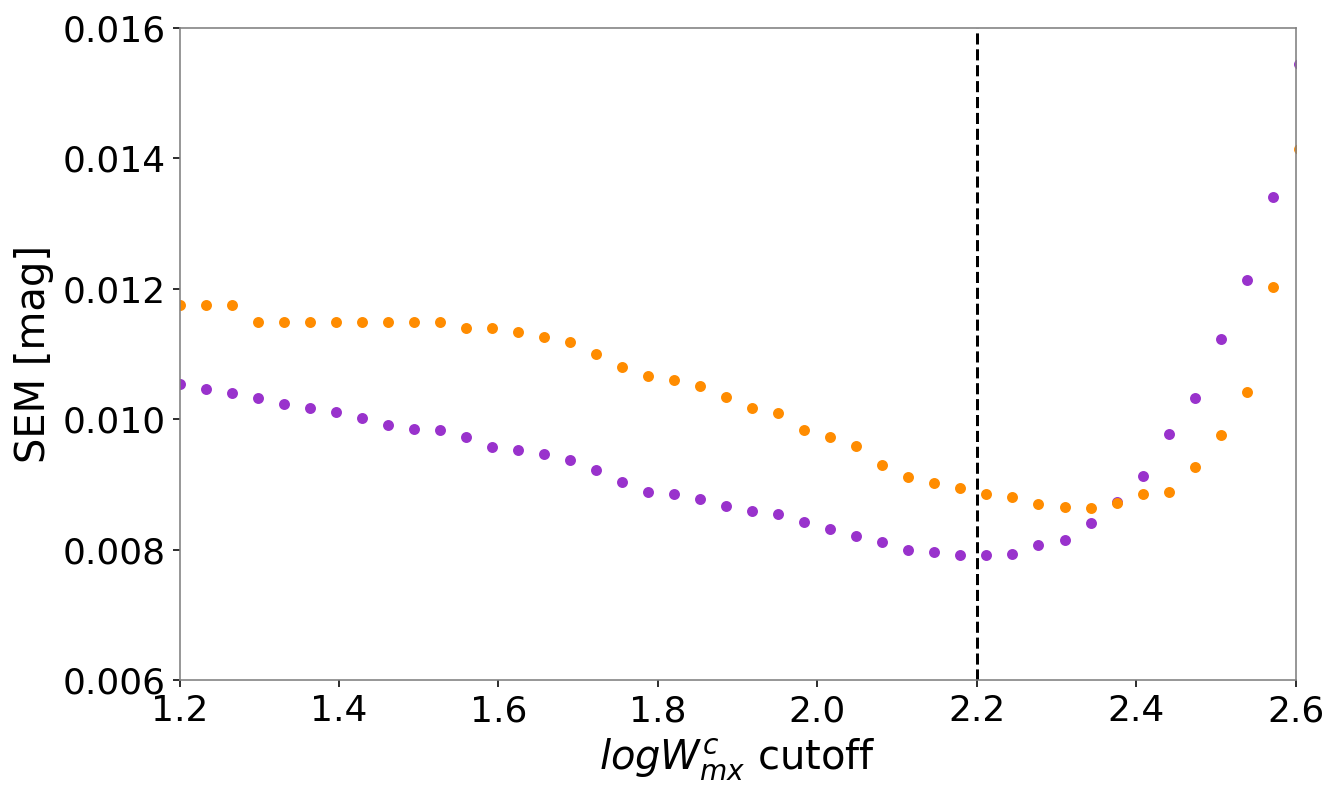}
\caption{Standard error of the mean (SEM) for the Tully-Fisher model fit to the CF4 data with SDSS $i$-band magnitudes (purple) and WISE $W1$-band magnitudes (orange) as a function of the $\log{\Wmxc}$ cutoff applied to the sample. The adopted lower limit of $\log{\Wmxc}=2.2$ is indicated by the vertical line.}
\label{fig:residuals}
\end{figure}

The placement of this cut should strike a balance between improving the quality of the fit (minimising the scatter for individual galaxies) and keeping as much data as possible (maximising the sample size). A suitable figure of merit to minimise is the RMS scatter about the fit divided by the Poisson gain from increased sample size (the square root of the sample size); this corresponds to minimising the standard error in the mean (SEM) of the fit. Figure~\ref{fig:residuals} shows the SEM of the fit as a function of the lower limit on the velocity width. The minimum occurs around $\log{\Wmxc}=2.2$ for the SDSS $i$-band Tully-Fisher relation and at slightly higher $\log{\Wmxc}$ for the WISE $W1$-band Tully-Fisher relation. We adopt $\log{\Wmxc}=2.2$ as the cutoff, as it is fairly close to the minimum in both bands and slightly more data-preserving. This approximately optimal lower limit on the velocity width removes about 16\% and 11\% of the $i$-band and $W1$-band samples respectively.

\subsection{Tully-Fisher relation non-linear model}
\label{nonlinearity}

Conventionally, the Tully-Fisher relation is a linear relation linking absolute magnitude and velocity width, and having a constant RMS scatter in absolute magnitude. However \citet{K2020}, in their analysis of the CF4 peculiar velocities, added a quadratic term to the Tully-Fisher relation for $\log{W_\textrm{m50}^c} > 2.5$ to allow for a flattening of the relation at the bright end. This curvature of the Tully-Fisher relation is not explained by selection effects known to exist within the relevant data sets, as it only affects galaxies that are well outside the selection limits; thus it appears to be a physical change in the Tully-Fisher relation for bright, high-mass galaxies. Kourkchi et~al.\ chose to fix the break-point from the linear relation at $\log{W_{m50}} = 2.5$. We initially allowed the break-point to be a free parameter, but it was under-constrained in our model, and we found that fixing it at $\log{W_{m50}} = 2.5$ gave satisfactory results. Our adopted model for the CF4 Tully-Fisher relation thus has the form
\begin{equation} 
M =
\begin{cases}
a_0 + a_1w & (w < 0) \\
a_0 + a_1w + a_2w^2 & (w \geq 0) \\
\end{cases}
\label{eq:TFreln}
\end{equation}
where $w \equiv \log{\Wmxc} - 2.5$ and $w = 0$ is the break-point above which the relation shows curvature.

We check whether a linear model gives as good a fit as this curved model to the observed CF4 Tully-Fisher relation. We find the additional free parameter of the curved model produces a marginally better representation of the data, in the sense that it gives a lower reduced $\chi^{2}$ (i.e.\ $\chi^{2}/\nu$ where $\nu$ is the number of degress of freedom) for the SDSS $i$-band Tully-Fisher relation: 0.98 with $\nu$\,=\,6,219 compared to 1.02 with $\nu$\,=\,6,220 for the linear model. For the $W1$-band Tully-Fisher relation we find a reduced 0.97 with $\nu$\,=\,4,718 for the curved model and 0.97 with $\nu$\,=\,4,719 for the linear model. In these $\chi^{2}$ calculations the data points are assumed to be independent, since any correlations between points due to the velocity field are expected to be dominated by the uncorrelated errors in the distance estimates from the Tully-Fisher relation.

\subsection{Tully-Fisher relation intrinsic scatter model}
\label{scattermodel}

In the calibration of the Tully-Fisher relation of the CF4 galaxies by \citet{K2020}, a non-linear model of the Tully-Fisher scatter was adopted (see their Figure~9). They found that the RMS scatter in absolute magnitude could be approximated by a quadratic relation in absolute magnitude. This model was determined using only the calibrator galaxies, a much smaller sample than the full CF4 Tully-Fisher dataset and encompassing a narrower range of absolute magnitudes. We find that model cannot be applied here, as it gives excessive scatter for the brightest galaxies in our sample (the quadratic approximation reaches a minimum and increases again before the end of our magnitude range). We thus seek a new model to approximate the Tully-Fisher scatter. 

The inset panels in Figure~\ref{fig:mag_scatter} show the trend of decreasing scatter in magnitude about the Tully-Fisher relation as velocity width increases (the scatter is calculated in magnitude rather than velocity width because we are using the forward Tully-Fisher relation), and suggest a weakly parabolic rather than linear fit. Adding a quadratic term does not result in a noteworthy improvement to the reduced $\chi^{2}$ (in the $i$-band, 0.998 with $\nu=6,219$ decreases to 0.982 with $\nu=6,218$). We therefore adopt a linear model for the scatter in magnitude about the Tully-Fisher relation as a function of velocity width
\begin{equation}
\sigma_\textrm{TF} = \epsilon_{0} + \epsilon_{1} (w + 2.5) 
\label{eq:TFscatter}
\end{equation} 
where $\sigma_\textrm{TF}$ accounts for both scatter intrinsic to the Tully-Fisher relation and scatter due to unmodelled peculiar velocities (errors in the linear velocity field model or non-linear peculiar motions).

\section{Sample selection function}
\label{selection}

\begin{figure}\centering 
\includegraphics[width=\columnwidth]{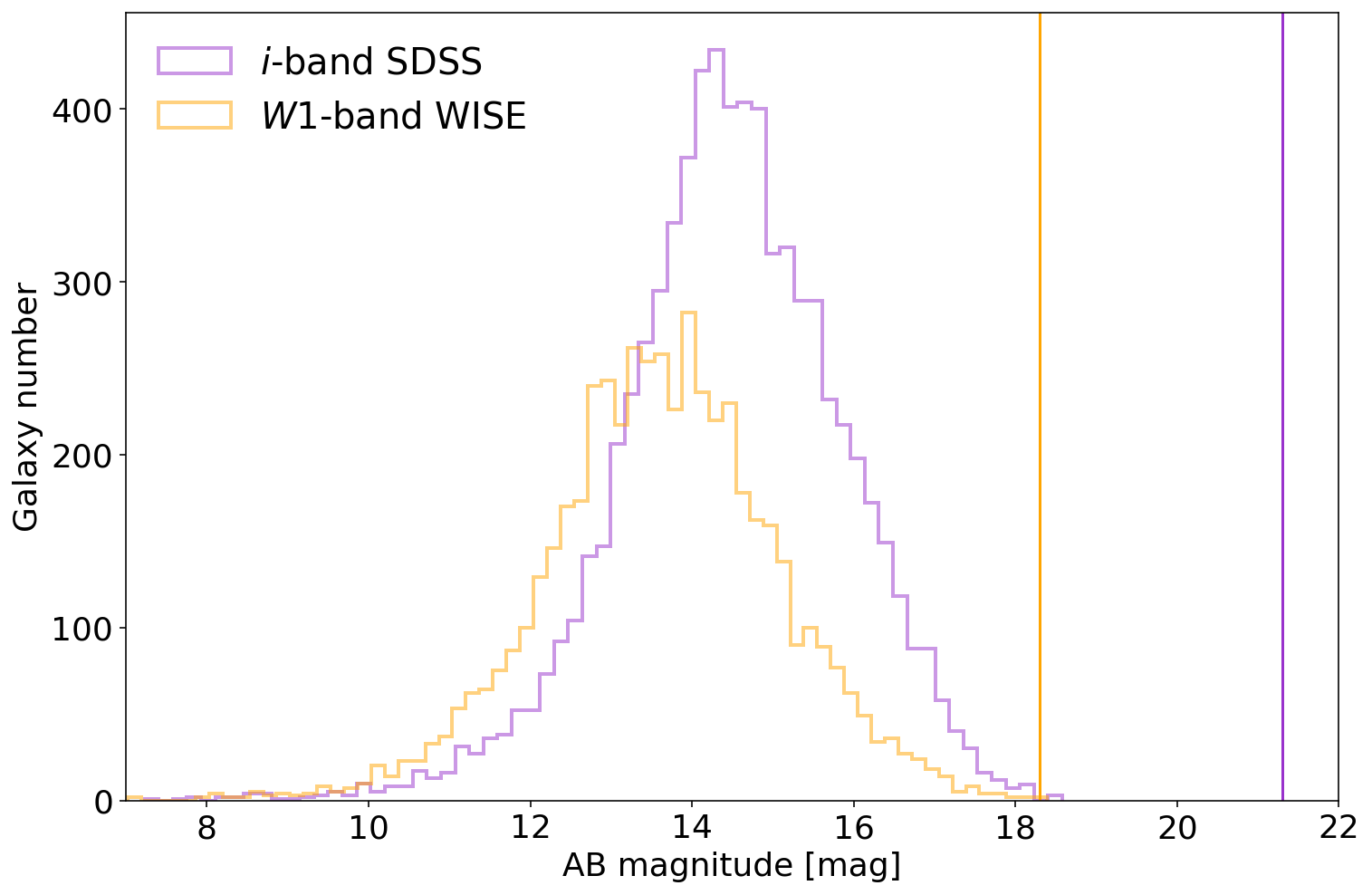}
\caption{Distribution of SDSS $i$-band magnitudes (purple) and WISE $W1$-band magnitudes (orange) from the CF4 Tully-Fisher sample (above the velocity width limit and excluding outliers). Vertical lines correspond to the 95\% completeness limit for SDSS (purple) and the conservative S/N=20 magnitude limit for WISE (orange).}
\label{fig:ABmags}
\end{figure}

In this section we describe the sample selection criteria for each of the observed quantities. As emphasised above, we must understand the dependence of the sample selection function on magnitude, and whether direct or indirect magnitude selection has a material impact. It emerges that the magnitude distributions of the Cosmicflows-4 Tully-Fisher sample are determined by the \HI\ flux measurement limits and the imposed velocity width limit, and not by the magnitude limits of the optical/infrared photometry \citep{K2020}.

For the SDSS $i$-band magnitudes, we can estimate the effective limiting magnitude from the number counts of galaxies. The SDSS number counts in the $r$-band have been compared to those from COMBO-17 \citep{Wolf_2003}, a much deeper survey\footnote{https://live-sdss4org-dr12.pantheonsite.io/imaging/other\_info/}. This comparison suggests a 95\% completeness limit at $r$\,$\approx$\,21. The mean colour of CF4 galaxies is $\langle r-i \rangle = 0.3$, which implies the $i$-band 95\% completeness limit is $i$\,$\approx$\,20.7. This is significantly fainter than the faintest $i$ magnitude in CF4 ($i$\,=\,19.94) and much fainter than the faintest $i$ magnitude in the fitted subsample once outliers are excluded and the lower limit in velocity width is imposed ($i$\,=\,18.58)---see Figure~\ref{fig:ABmags}. We can therefore be confident that the SDSS optical magnitude limit is irrelevant to the CF4 Tully-Fisher sample selection. 

For the WISE $W1$-band magnitudes, Section VI.5.b of the WISE Explanatory Supplement\footnote{https://wise2.ipac.caltech.edu/docs/release/allsky/expsup/sec6\_5.html} shows internal and external tests of the WISE Source Catalog completeness. From internal tests, they have determined that a signal-to-noise ratio of S/N=20 is achieved at a Vega magnitude of 15.6\,mag, corresponding to $W1$\,=\,18.3 in the AB magnitude system used by CF4. (The actual completeness depends somewhat on sky position, so this is the `typical' completeness corresponding to the median sky coverage). This conservative estimate of the completeness limit is comparable to the faintest $W1$ magnitude in CF4 ($W1$\,=\,18.39; this is also the faintest magnitude in the fitted subsample). Thus, as Figure~\ref{fig:ABmags} shows, the WISE limiting magnitude is clearly not significant in the selection of the sample.

The observed redshifts of the galaxies in the CF4 Tully-Fisher catalogue range from $-$0.0009 to 0.0639 in the CMB frame (see Figure~\ref{fig:redshift}). We impose a lower limit of $z_{\rm min}$\,=\,0.002 (i.e.\ 600\kms) to avoid negative redshifts and galaxies with peculiar velocities comparable to their distances; this removes 128 galaxies from the sample. We do not impose an upper redshift limit. For the fitted Tully-Fisher samples, the mean redshifts are $\langle z \rangle$\,=\,0.021, 0.016 and 0.019 (6210, 4820 and 5760\kms) for the combined, SDSS and WISE samples respectively. The effective redshifts for peculiar velocities (the mean redshifts weighted by the inverse square of the peculiar velocity uncertainties) are $\langle z_{\rm eff} \rangle$\,=\,0.0172 and 0.0165 (5135 and 4936\kms) for the SDSS and WISE samples.

The velocity width selection limit $w_{\rm lim}$ is a minimum velocity width chosen to exclude galaxies in the dwarf regime that depart significantly from the linear Tully-Fisher relation and have larger scatter. As discussed in Section~\ref{cuts}, the optimal velocity width limit for the CF4 dataset was found to be $\log{\Wmxc}=2.2$, corresponding to $w_{\rm lim}=\log{\Wmxc}-2.5=-0.3$. This is higher than the limit imposed by the minimum measurable \HI\ line width for the ALFALFA observations.

\begin{figure}
\centering 
\includegraphics[width=\columnwidth]{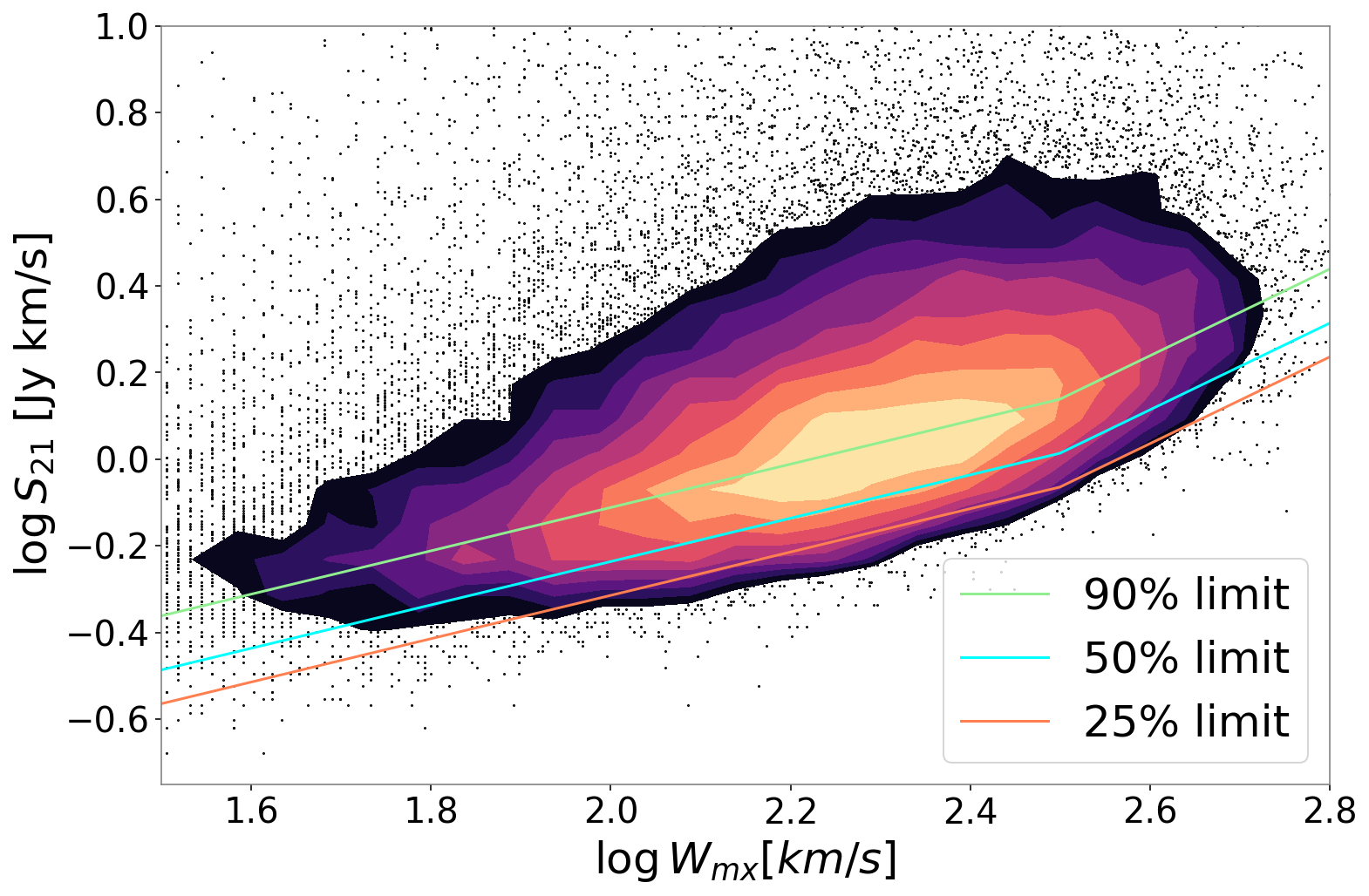}
\caption{ALFALFA flux density as a function of velocity width, showing the 25\%, 50\%, and 90\% completeness limits. The contour represents data from the ALFALFA survey where we have set a threshold of 50 points per 2D bin at which to begin drawing contours.}
\label{fig:completeness}
\end{figure}

Flux has the most complicated selection function. Spectroscopic \HI\ surveys like ALFALFA are not simply flux-limited; at the same flux, they are also less sensitive to broader line widths than to narrower ones. The ALFALFA survey is thus flux-limited in a manner that depends on the velocity width. Moreover, the flux limit itself is not a hard cutoff; there is a gradual decrease in the probability of a source being detected at fainter fluxes. \citet{2011alf} determined the completeness of the ALFALFA survey as the fraction of galaxies of a given flux density that are detected and included in the survey. They give the logarithm of the flux density limit, $\log{S_{21,\textrm{lim}}}$, corresponding to completenesses of 25\%, 50\% and 90\% as linear functions of $\log{W_{m50}} = \log{\Wmx + 6}$ \citep[for ALFALFA line widths; see][]{K2020}. These completeness limits are shown in Figure~\ref{fig:completeness}. In exploring the effects of flux selection through mocks (see next section) we use the 50\% completeness limit given by
\begin{equation} 
\log{S_{21,\textrm{lim}}} =
\begin{cases}
 0.5\log{W_{m50}} - 1.30 & (\log{W_{m50}} < 2.5) \\
 \log{W_{m50}} - 2.55 & (\log{W_{m50}} \geq 2.5) ~. \\
\end{cases}
\label{eq:complim}
\end{equation}

Despite the CF4 sample not being directly affected by the magnitude limits of the SDSS or WISE photometry, the \HI\ flux limit can, through its correlation with optical/infrared luminosity, impose an effective apparent magnitude selection function. The relationship between \HI\ flux and optical magnitude is complex and difficult to model, so instead we determine an empirical magnitude selection function, $F_m(m)$, by comparing the observed distribution of magnitudes in the sample with the expected distribution in the absence of selection effects. Figure~\ref{fig:ABmags} shows that there is magnitude-dependent selection affecting both the $i$-band and $W1$-band magnitudes, as the observed numbers of galaxies peak near $m_1$\,$\approx$\,14 then decrease roughly linearly to zero around $m_2$\,$\approx$\,18. A simple model for the observed number of galaxies in this magnitude range is thus $N_\textrm{obs}(m) = N(m_1)(m_2-m)/(m_2-m_1)$. However, since $log N(m) \propto 0.6m$ in Euclidean space \citep{Hubble_1926}, at these low redshifts we expect the true number of galaxies to increase with magnitude approximately as $N_{\rm true}(m) = N(m_1)10^{0.6(m-m_1)}$. The effective selection function for $m$ can therefore be approximated as
\begin{align} 
F_m(m) &= N_{\rm obs}(m)/N_{\rm true}(m) \nonumber \\
       &=
\begin{cases}
1 & m \leq m_1 \\
\frac{m_2-m}{m_2-m_1}10^{0.6(m_1-m)} & m_1 < m \leq m_2 \\
0 & m > m_2
\end{cases}
\label{eq:maglimsoft}
\end{align}
where in this case we adopt $m_1=14$ and $m_2=18$. This is the selection function we insert in Equation~\ref{eq:mcondprob} to compute the conditional probability of the observed apparent magnitude.

\section{Mock test}
\label{mocks}

\begin{figure*}\centering
\includegraphics[width=2\columnwidth]{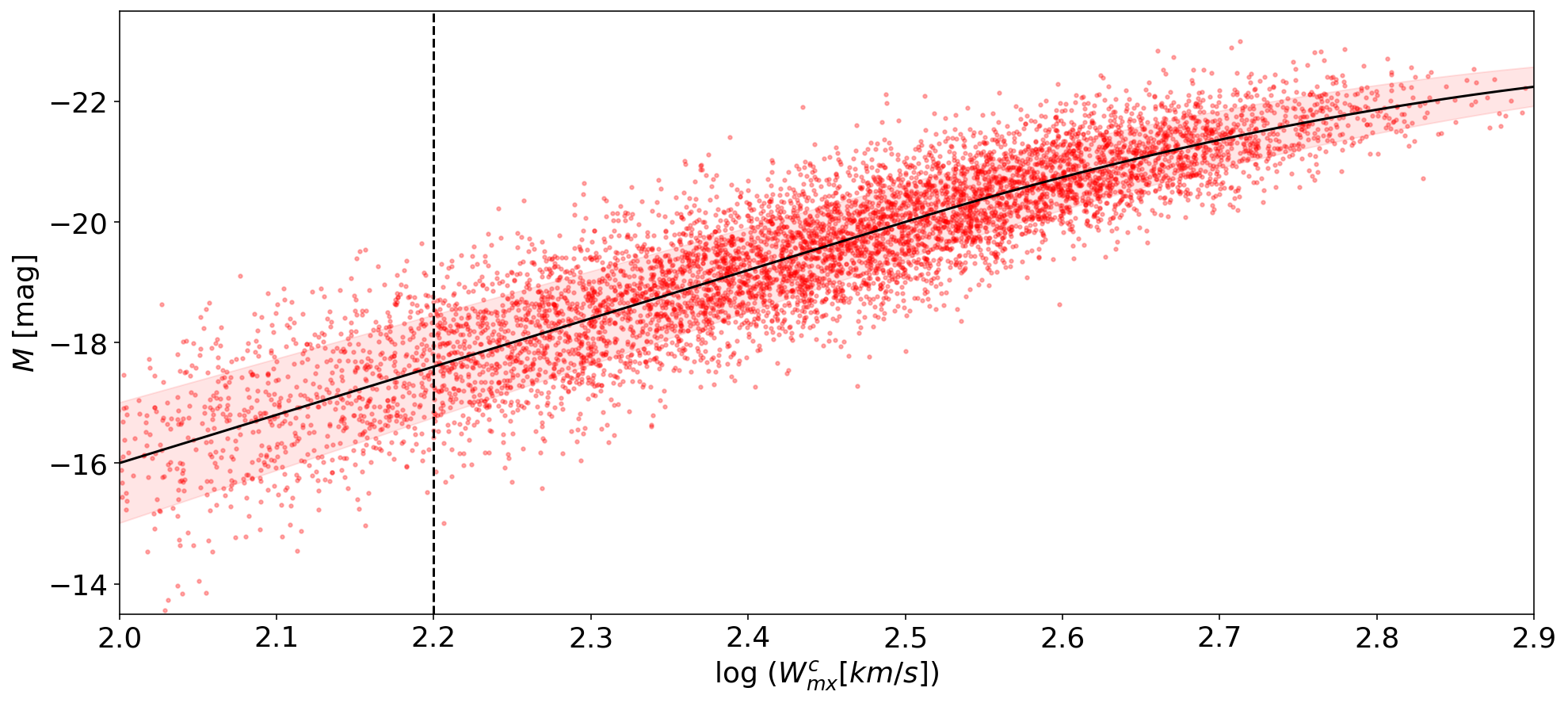}
\caption{The Tully-Fisher relation between absolute magnitude and velocity width for a 10,000-galaxy mock of the CF4 Tully-Fisher dataset. The black curve shows the input Tully-Fisher model and the shaded region shows the input Tully-Fisher 1$\sigma$ scatter in absolute magnitude.}
\label{fig:mockTFR}
\end{figure*}

\begin{figure}\centering 
\includegraphics[width=\columnwidth]{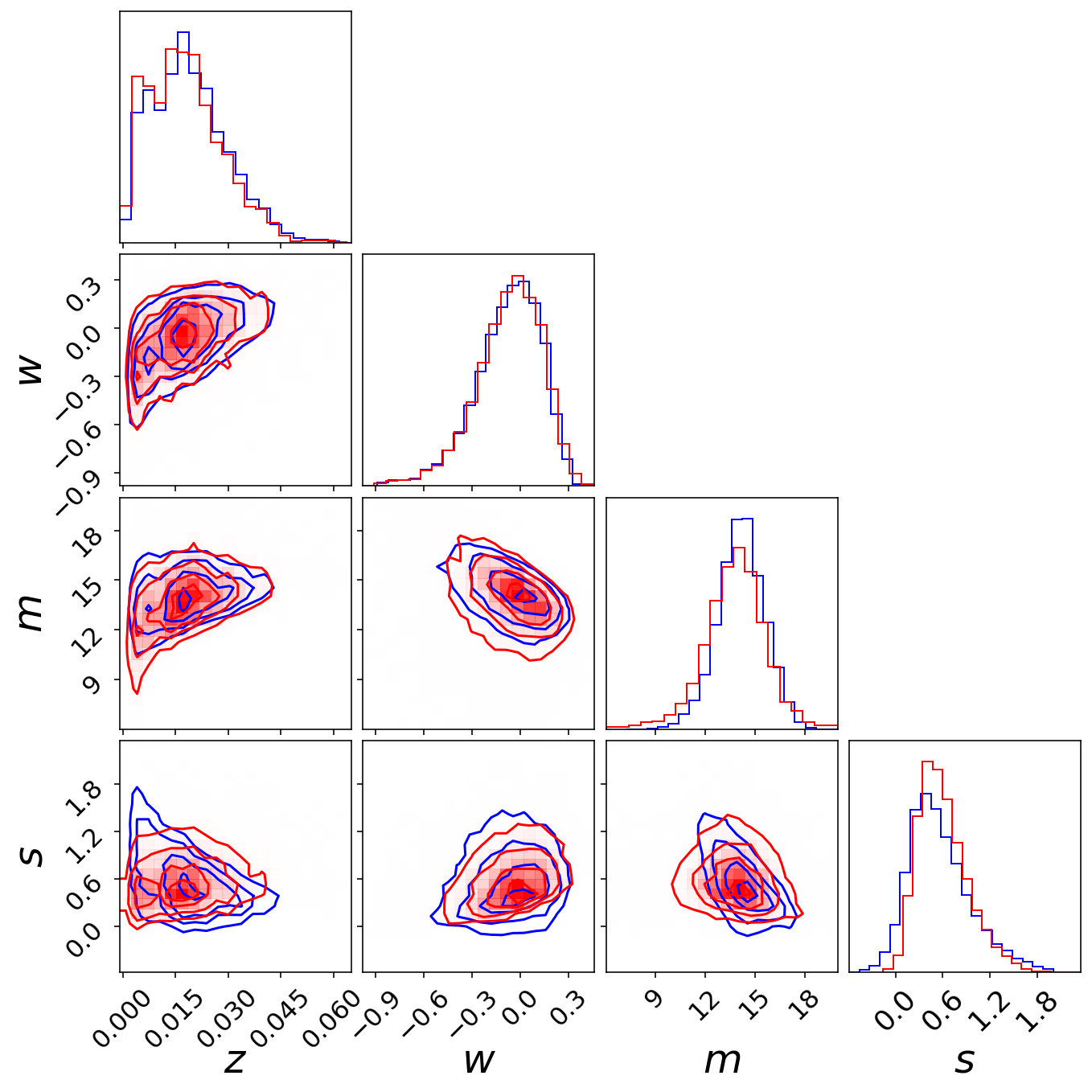}
\caption{Distributions of the observable quantities (redshift, velocity width, apparent magnitude, and flux) for a single mock catalogue (red) compared to the CF4 Tully-Fisher data (blue).}
\label{fig:mockresult2d}
\end{figure}

\begin{figure*}\centering 
\includegraphics[width=1.8\columnwidth]{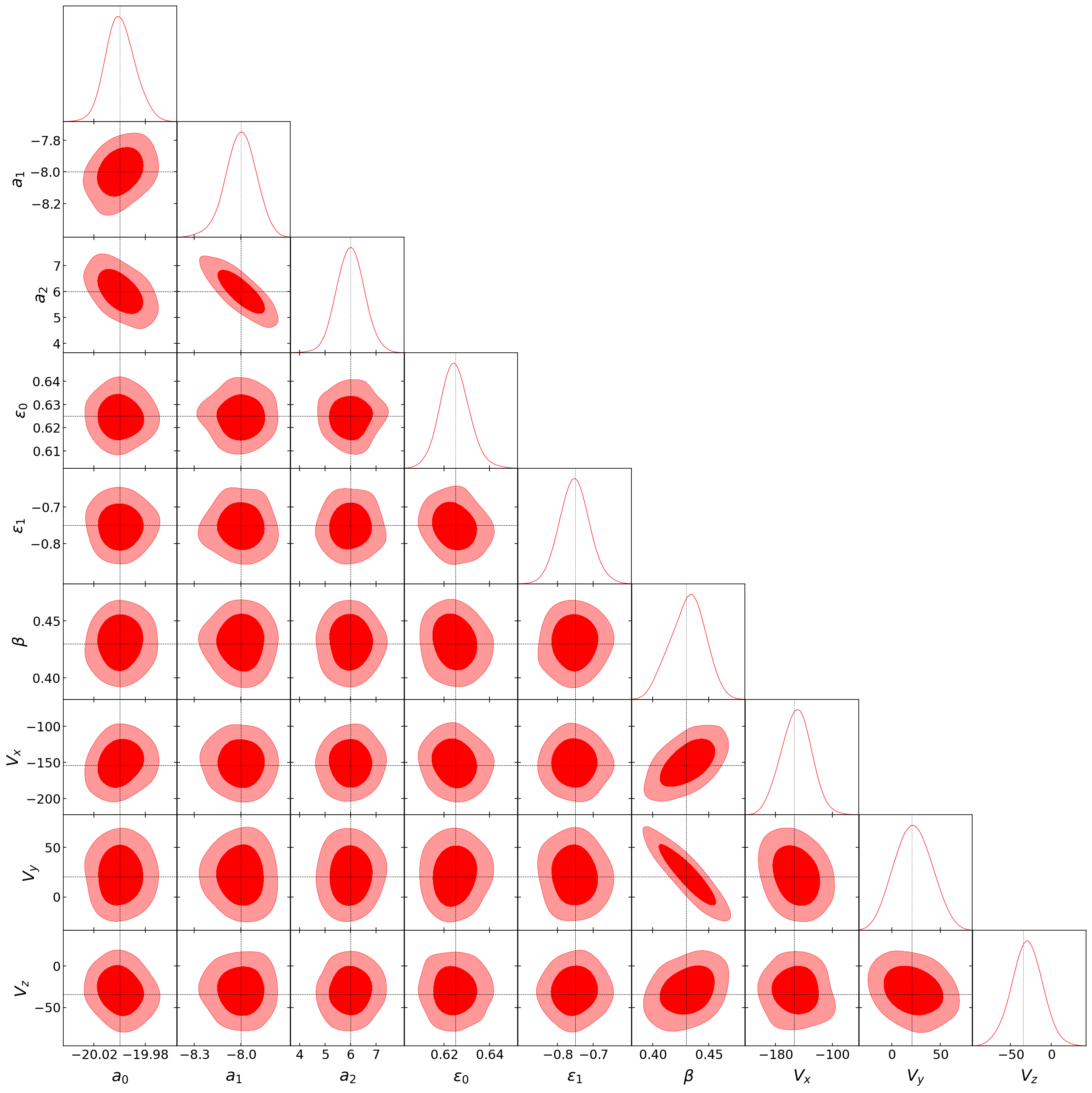}
\caption{Constraints from 1000 mock CF4 samples, with the dark and light red shadings showing 68\% and 95\% confidence regions for the pairwise joint posterior probabilities; input values shown by crosshairs. The parameters $a_0$, $a_1$ and $a_{2}$ are the intercept, slope and curvature coefficients of the Tully-Fisher relation; $\epsilon_{0}$ and $\epsilon_{1}$ are the intercept and slope of the Tully-Fisher scatter model; $\beta$ is the velocity field scaling parameter and ($V_x$,$V_y$,$V_z$) are the residual bulk flow components in Supergalactic Cartesian coordinates in \kms\ \citep[inputs from the 2M++ peculiar velocity model;][]{Carrick_2015}.}
\label{fig:mockres}
\end{figure*}

In order to test the validity of our methodology, check the corrections for selection effects, and test for residual biases in the results, we generate mock data in which the model parameters and selection functions are known and examine the results obtained by applying our method to these mocks.

Our goal is to generate a realistic mock catalogue of redshifts, \HI\ line widths, and apparent magnitudes having the same intrinsic distributions and covariances and the same imposed selection criteria as the actual data. The simulation of CF4 Tully-Fisher observables is complicated by the fact that CF4 is a compilation of different surveys, each with its own selection criteria. For the purpose of generating mock catalogues to test our methodology, we impose the observational selection criteria of the ALFALFA survey, since it accounts for 7341 of the 10737 galaxies with Tully-Fisher peculiar velocities in CF4. To construct a mock sample of `observed' sources drawn from an `intrinsic' source population, we use plausible models (described below in Section~\ref{models}) to simulate the intrinsic population and then impose the ALFALFA selection criteria (described above in Section~\ref{selection}) to obtain the observed sample. However, as discussed in Section~\ref{methodology}, by using conditional probabilities we avoid any negative impact from some of the more uncertain aspects of the simulation. 

\subsection{Modelling the intrinsic population}
\label{models}

The number density of galaxies with \HI\ mass $M_{\textrm{HI}}$ is usually defined as,
\begin{equation}
\phi(M_{\textrm{HI}}) = \frac{\textrm{d}N_{gal}}{\textrm{d}V \textrm{d}\log_{10}(M_{\textrm{HI}})} [\textrm{Mpc}^{-3}\textrm{dex}^{-1}]
\end{equation}
It is observed to be well approximated by a Schechter function \citep{Jones_2018}
\begin{equation}
\phi(M_{\textrm{HI}}) =
 \ln(10) \, \phi_* 
 \exp\left[-\frac{M_{\textrm{HI}}}{M_{\textrm{HI}}^*}\right]
 \left(\frac{M_{\textrm{HI}}}{M_{\textrm{HI}}^*}\right)^{\alpha+1}
\label{eq:HIMF}
\end{equation}
parameterised by faint-end slope $\alpha$+1 and characteristic mass $M_{\textrm{HI}}^*$. We use the \HI\ mass function derived from the complete ALFALFA catalogue by \citet{Jones_2018}; the parameters of this mass function are given in Table~\ref{tab:paramsknown}.

To model the \HI\ fluxes, we use the standard relation, given by \citet{2017meyer}, between flux and \HI\ mass
\begin{equation} 
M_{\rm HI} = 2.35 \times 10^5 \, D_C^2 \, S_{21} ~{\rm M}_{\sun}
\label{eq:massflux}
\end{equation}
where $D_C$ is the comoving distance in units of Mpc. Note that there is no $(1+z)^2$ term because flux density ($S_{21}$) is defined as the flux in units of Jy\kms.

To generate a set of \HI\ velocity widths corresponding to the \HI\ masses, we use a mass-conditional velocity width function, modelled by a Gumbel distribution, taken from \citet{Jones_2015}. The probability of a galaxy of mass $10^m$\,M$_{\sun}$ having a velocity width $10^\omega$\kms\ is given by
\begin{equation}
P(\omega|m) = \frac{1}{b(m)}\frac{e^{(x(m)+e^{-x(m)})}}{e^{-e^{-x_{\rm min}}}-e^{-e^{-x_{\rm max}}}},
\label{eq:MCWF}
\end{equation}
where $x(m) = \frac{a(m)-\omega}{b(m)}$. 

The parameters of this distribution, $a(m)$ and $b(m)$, are taken directly from equations C1--C3 of \citet{Jones_2015}:
\begin{align}
a &= 0.322m - 0.728 \\
b &=
\begin{cases}
-0.0158m + 0.316 & (m \leq 9.83) \\
-0.0578m + 0.729 & (m > 9.83) ~.\\
\end{cases}
\end{align}
Following \citet{Jones_2015}, we adopt $x_{\rm min}$\,=\,1.2 and $x_{\rm max}$\,=\,3.0, spanning the range of velocity widths in the CF4 Tully-Fisher sample.

The correlation between velocity width and absolute magnitude is described by the Tully-Fisher relation. For the mocks, we use the best-fit version of the relation in the form given by Equation~\ref{eq:TFreln} (Section~\ref{nonlinearity}) with intrinsic scatter given by Equation~\ref{eq:TFscatter} (Section~\ref{scattermodel}).

For simplicity, we do not model cosmic structure in simulating the clustering of the mock galaxies; instead we assign redshifts and positions simply by bootstrapping the redshifts and positions of the galaxies in the CF4 Tully-Fisher data, thereby forcing the mock sample to have large-scale structure, redshift distribution, and sky coverage similar to the observed sample. We assign a line-of-sight peculiar velocity to each galaxy based on the peculiar velocity model given by Equation~\ref{eq:velocitymodel}, using the 2M++ velocity model of \citet{Carrick_2015} after subtracting their fitted external dipole and dividing by their fitted velocity scale parameter ($\beta$\,=\,0.43). We use the software\footnote{https://github.com/KSaid-1/pvhub} developed by \citet{Said_2020} to extract the model values at a given location in redshift space. 

For our mocks, all the parameters involved in these models are assumed to be known; some values are taken from previous analyses, while others are fitted in this analysis to match the observed data. The values of the parameters we use in generating the mock samples are listed in Table~\ref{tab:paramsknown}. To assess the effects of the errors associated with the 2M++ reconstruction from \citet{Carrick_2015}, we also generate a suite of mocks for which the values of $\beta$ and ($V_x$,$V_y$,$V_z$) are each drawn from a Gaussian distribution with mean and errors corresponding to the \citet{Carrick_2015} measurements and associated uncertainties: $\beta$\,=\,0.43\,$\pm$\,0.02 and $\mathbf{V}_{\textrm{ext}}$\,=\,(+89\,$\pm$\,21,$-$131\,$\pm$\,23,+17\,$\pm$\,26)\kms in Galactic Cartesian coordinates and $\mathbf{V}_{\textrm{ext}}$\,=\,($-$154\,$\pm$\,23,+21\,$\pm$\,26,$-$35\,$\pm$\,22)\kms in Supergalactic Cartesian coordinates.

\begin{table}
	\centering
	\caption{Parameters of Tully-Fisher model, peculiar velocity model, and HI mass and velocity width distributions used in generating mock samples. $\mathbf{V}_{\textrm{ext}}$ is given in Supergalactic Cartesian coordinates. These are compared to the derived parameters for the Tully-Fisher model and peculiar velocity model.}
	\label{tab:paramsknown}
	\begin{tabular}{lccl}
		\hline
		\hline
		Parameter & Input value & Fitted value & Description \\
		\hline
		\multicolumn{4}{l}{Tully-Fisher model}\\
		\hline
  		$a_0$ & $-$20 & $-$20.00\,$\pm$\,0.01 & intercept\\
		$a_1$ & $-$8 & $-$8.0\,$\pm$\,0.1 & slope\\
		$a_{2}$ & $6$ & 6.0\,$\pm$\,0.5 & curvature coefficient\\
    	$\epsilon_0$ & 0.625 & 0.625\,$\pm$\,0.006 & scatter model intercept\\
		$\epsilon_1$ & $-$0.75 & $-$0.75\,$\pm$\,0.04 & scatter model slope\\
		\hline
		\multicolumn{4}{l}{Peculiar velocity model}\\
		\hline
		$\beta$ & 0.43 & 0.43\,$\pm$\,0.02 & \cite{Carrick_2015}\\
		$V_x$ [\kms] &  $-$154 & $-$151\,$\pm$\,21 & \cite{Carrick_2015}\\
		$V_y$ [\kms] & ~~$+$21 & 22\,$\pm$\,19 & \cite{Carrick_2015}\\
		$V_z$ [\kms] &  ~$-$35 & $-$30\,$\pm$\,19 & \cite{Carrick_2015}\\
		\hline
		\multicolumn{4}{l}{\HI\ mass function}\\
		\hline
		$\alpha_{\textrm{HI}}$ & $-$1.25 & & \citet{Jones_2018} \\
		$\log(M_{\textrm{HI}}^*/M_{\sun})$ & 9.94 & & \citet{Jones_2018} \\
		\hline
	\end{tabular}
\end{table}

\subsection{Procedure for generating mocks}
\label{procedure}

Given the ingredients described in the previous two sections, we here outline the full procedure used for generating the mock sample.
\begin{enumerate}
\item Assign a galaxy position, $(z,\alpha,\delta)$, by random-sampling the positions of the galaxies in the CF4 Tully-Fisher catalogue.
\item Use the position and observed redshift to draw a peculiar velocity from a Gaussian distribution with mean given by the fiducial 2M++ velocity model parameters from \citet{Carrick_2015} (via Equation~\ref{eq:velocitymodel}) and a standard deviation of 250\kms; this yields the cosmological redshift and distance.
\item Assign an \HI\ mass by random-sampling the \HI\ mass function given in Equation~\ref{eq:HIMF}, limited to the range $6 \leq \log(M_{\rm HI}/M_{\sun}) \leq 11$.
\item Compute the \HI\ flux, $S_{21}$, via Equation~\ref{eq:massflux} using the \HI\ mass and comoving distance. 
\item Assign an \HI\ velocity width, $\Wmxc$, by random-sampling the conditional velocity width distribution given by Equation~\ref{eq:MCWF} and the galaxy's \HI\ mass.
\item Assign an inclination $i$ by random-sampling the distribution of observed CF4 inclinations.
\item Compute the observed \HI\ line width as $\Wmx = \Wmxc \sin i$.
\item Calculate the flux corresponding to the ALFALFA 50\% completeness limit, $S_{21,\textrm{lim}}$, from $\Wmx$ using Equation~\ref{eq:complim}. 
\item If $S_{21} \geq S_{21,\textrm{lim}}$, include the galaxy in the mock sample and proceed to step~(x); if not, exclude the galaxy and return to step~(i).
\item Determine the absolute magnitude $M$ corresponding to $\Wmxc$ using the Tully-Fisher relation (Equation~\ref{eq:TFreln}), then add a random error from a Gaussian distribution of width given by the Tully-Fisher scatter model (Equation~\ref{eq:TFscatter}).
\item Compute the apparent magnitude from the absolute magnitude, observed redshift, and comoving distance using Equation~\ref{eq:distancemodulus} (with $H_0$\,=\,100$h$\kms\,Mpc$^{-1}$).
\item Repeat until the mock sample has the desired number of galaxies.
\end{enumerate}

Figure~\ref{fig:mockTFR} shows the Tully-Fisher relation from a single mock catalogue, for comparison with Figure~\ref{fig:cf4data}. Note that the `observed' sample includes galaxies below the subsequently-imposed lower limit to the velocity width. There are no outliers as the generating model is purely Gaussian. Figure~\ref{fig:mockresult2d} shows the distributions of the actual and mock observables for the CF4 Tully-Fisher dataset and their pairwise correlations. It demonstrates that the mocks provide a reasonably realistic representation of the CF4 Tully-Fisher data, including the effective magnitude selection imposed by the flux selection.

We generate 1000 realisations of the mock catalogue using the input parameters in Table~\ref{tab:paramsknown} and apply the method of Section~\ref{methodology}. A velocity width cut is applied at $\log{\Wmxc}$\,=\,2.2, as with the data. The resulting parameter constraints are shown in Figure~\ref{fig:mockres} and listed in Table~\ref{tab:paramsknown} and are consistent with the input values, demonstrating that the selection criteria applied to the sample do not bias the results obtained using this method. 

There is nonetheless a correlation of $\beta$ with $\mathbf{V}_{\textrm{ext}}$; most significantly, an anti-correlation with $V_y$ and, to a lesser extent, correlations with $V_x$ and $V_z$. This is because, in the volume of the CF4 Tully-Fisher dataset, the average bulk flow of the 2M++ peculiar velocity model is partially correlated with $\mathbf{V}_{\textrm{ext}}$. Removing the velocity model's external bulk flow from the predicted velocities of all galaxies in the CF4 Tully-Fisher dataset, we find their mean motion to be ($-$68,+53,$-$73)\kms\ in Supergalactic Cartesian coordinates. This means 80\% of the average model peculiar velocity of the sample is aligned with the external bulk flow, which is $\mathbf{V}_{\textrm{ext}}$\,=\,($-$154\,$\pm$\,23,+21\,$\pm$\,26,$-$35\,$\pm$\,22)\kms\ in Supergalactic Cartesian coordinates. The angle between these vectors is only 38\degree. As a result, the external bulk flow is partially degenerate with the scaling parameter $\beta$. In the much larger volume of the WALLABY survey, this degeneracy disappears (Section~\ref{WALLABY}).

\begin{figure*}\centering 
\includegraphics[width=1.8\columnwidth]{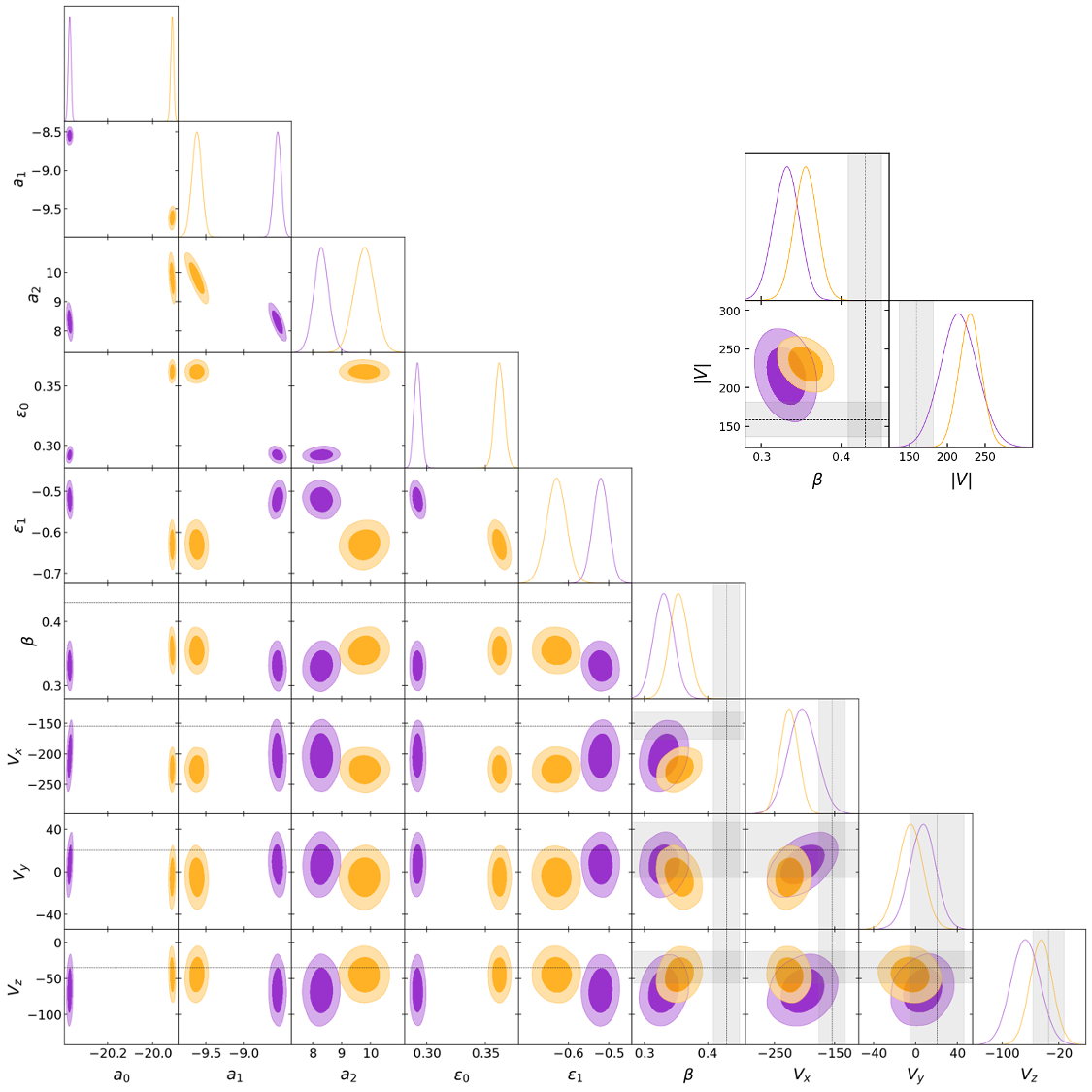}
\caption{Results of our method applied to the CF4 Tully-Fisher dataset using the 2M++ peculiar velocity field model. Purple shows results for the SDSS $i$-band sample and orange for the WISE $W1$-band sample. Dark and light shadings indicate 68\% and 95\% confidence regions for the pairwise joint posterior probabilities. The fitted parameters for the two samples are not expected to be the same: the Tully-Fisher relations in the two bands will differ, as will the velocity field parameters (since the samples cover different volumes with galaxies having different biases). The top-right inset shows the joint constraints of $\beta$ and the amplitude of the residual bulk flow, $|V|$. Grey lines show the values reported by \citet{Carrick_2015} from the 2M++ velocity model, with the grey bands showing the corresponding 1$\sigma$ uncertainty.}
\label{fig:CF4_constraints}
\end{figure*}

\section{Results}
\label{results}

Having validated our methodology on mock data, we proceed to apply it to the CF4 data. Figure~\ref{fig:CF4_constraints} shows the pairwise posterior probability contours for the parameters of both the Tully-Fisher model ($a_0$, $a_1$, $a_{2}$, $\epsilon_0$, $\epsilon_1$) and the peculiar velocity model ($\beta$, $V_x$, $V_y$, $V_z$) obtained using our method applied to the CF4 Tully-Fisher samples, while Table~\ref{tab:paramsfinal} lists the corresponding parameter estimates and uncertainties (the residual bulk flow is given in Supergalactic Cartesian coordinates and the CMB frame).

\begin{table}
\centering
\caption{Parameters of the Tully-Fisher relation and peculiar velocity field fitted to the Cosmicflows-4 SDSS $i$-band and WISE $W1$-band samples.}
	\label{tab:paramsfinal}
	\begin{tabular}{lcc}
        \hline\hline
        Parameter & SDSS $i$-band & WISE $W1$-band\\
		\hline
		$a_0$ & $-$20.367\,$\pm$\,0.006 & $-$19.914\,$\pm$\,0.007 \\
		$a_1$ & $-$8.55\,$\pm$\,0.05 & $-$9.62\,$\pm$\,0.07 \\
		$a_2$ & 8.3\,$\pm$\,0.3 & 9.8\,$\pm$\,0.4 \\
		$\epsilon_0$ & 0.292\,$\pm$\,0.003 & 0.362\,$\pm$\,0.004 \\
		$\epsilon_1$ & $-$0.52\,$\pm$\,0.02 & $-$0.63\,$\pm$\,0.03 \\
		$\beta$ & 0.33\,$\pm$\,0.03 & 0.36\,$\pm$\,0.02 \\
		$V_x$ [\kms] &  ~~$-$203\,$\pm$\,35 &  $-$225\,$\pm$\,15 \\
		$V_y$ [\kms] & ~~7\,$\pm$\,13 &~~~~$-$7\,$\pm$\,13 \\
		$V_z$ [\kms] & ~~$-$65\,$\pm$\,20 & ~~$-$46\,$\pm$\,16 \\
  		\hline
        \multicolumn{3}{l}{$\mathbf{V}_{\textrm{ext}}$ is in Supergalactic Cartesian coordinates and the CMB frame.}	
    \end{tabular}
\end{table}

\begin{figure*}\centering
\includegraphics[width=2\columnwidth]{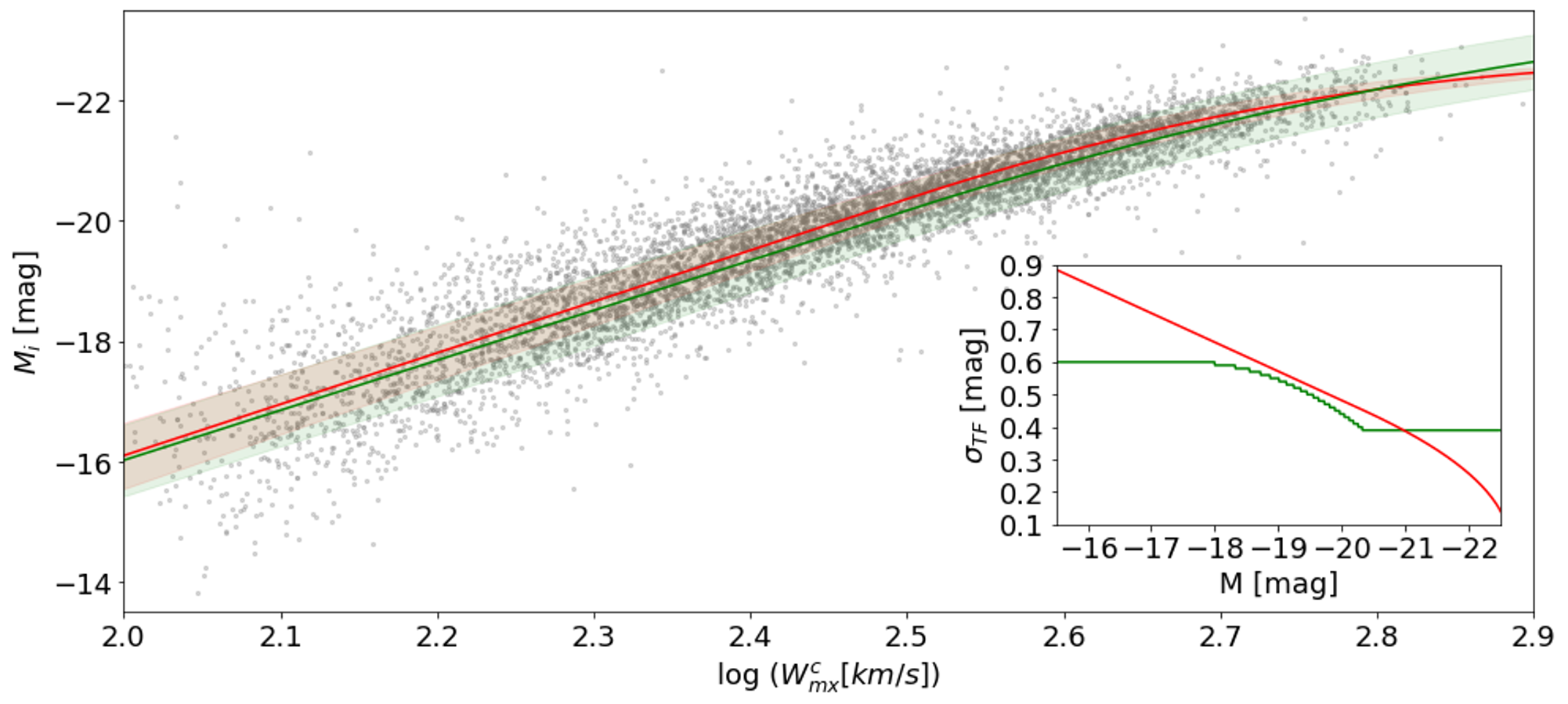}
\caption{A comparison of the Tully-Fisher relation and scatter model fits from this analysis (red) and those of \citet[][green]{K2020}. The Tully-Fisher scatter models are shown as a function of absolute magnitude in the inset; \citet[][Figure~9]{K2020} modelled the scatter as a quadratic function of absolute magnitude at intermediate magnitudes and constants at brighter and fainter magnitudes. Note that the absolute magnitudes assume $H_0$\,=\,100\kms\,Mpc$^{-1}$ and that we have converted the \citet{K2020} Tully-Fisher zero-point to be consistent with this.}
\label{fig:scattermodel_compare}
\end{figure*}

\subsection{Tully-Fisher model parameters}
\label{TFmodelfits}

The fitted Tully-Fisher relation parameters for our final sample of 6,224 galaxies with $i$-band magnitudes are $a_0$\,=\,$-$20.367\,$\pm$\,0.006, $a_1$\,=\,$-$8.55\,$\pm$\,0.05 and $a_{2}$\,=\,8.3\,$\pm$\,0.3; this is a `direct' fit, minimising the residuals in magnitude. \citet[][\!\!b]{K2020} fitted the `inverse' Tully-Fisher relation, minimising the residuals in velocity width, and using a subsample of $\sim$\,600 spirals in clusters to calibrate the Tully-Fisher relation and obtain $H_0$\,=\,74.8\kms\,Mpc$^{-1}$. They report $i$-band Tully-Fisher parameters $a_0$\,=\,$-$20.84\,$\pm$\,0.10 with the zero-point correction applied (corresponding to $a_0$\,=\,$-$20.21 for $H_0$\,=\,100\kms\,Mpc$^{-1}$), $a_1$\,=\,$-$8.32\,$\pm$\,0.13 and $a_{2}$\,=\,5.3\,$\pm$\,0.9 \citep[][Table~2]{2020cf4}. Figure~\ref{fig:scattermodel_compare} compares our Tully-Fisher relation fit (and scatter model) to theirs. Although direct and inverse fits generally give differing results \citep[see, e.g.,][]{Magoulas_2012}, in this case there is a numerous sample, the fitted range spanned is large relative to the scatter, and the two fitting methods yield consistent (at 1.7$\sigma$) linear slopes for the Tully-Fisher relation. There is a marginally significant (3.1$\sigma$) difference in the curvature parameters for the Tully-Fisher relations, which we ascribe primarily to the substantial difference in the scatter models at bright magnitudes and a significant correlation between the slope $a_1$ and curvature $a_2$. The most interesting difference between the two Tully-Fisher relations is the small (0.43\,mag) but significant (4.3$\sigma$) offset in the Tully-Fisher relation zeropoint. 

The scatter about our fit corresponds to an effective 28\% overall distance error for the Cosmicflows-4 Tully-Fisher data set, whereas \citet{2020cf4} report a 22\% overall distance error. This discrepancy is too large to be explained only by differences in the parameters of the Tully-Fisher relations. However the RMS distance errors in \citet{2020cf4} were calculated using the Tully-Fisher intrinsic scatter model determined by \citet{K2020}, based on a parabolic fit to the scatter in absolute magnitude as a function of absolute magnitude. This quadratic scatter model was not extrapolated to galaxies at the bright and faint ends in the full CF4 Tully-Fisher dataset, but instead set to constant values. The inset in Figure~\ref{fig:scattermodel_compare} shows their scatter model compared to the one used in this work. We see that our model predicts higher scatter at the faint end, lower scatter at the bright end, and is roughly consistent in the intermediate regime where \citet{K2020} fit their model. Actual measurements of the scatter in the full CF4 Tully-Fisher data set (see Figure~\ref{fig:mag_scatter}) suggest that our scatter model is a more accurate reflection of the intrinsic scatter than that of \citet{2020cf4}; if anything, we slightly underestimate the scatter at the faint end. For the set of apparent $i$-band magnitudes and velocity widths common to both analyses, the actual scatter about the \citet{2020cf4} Tully-Fisher relation is 0.61\,mag (equivalent to a 28\% distance error), whereas the \citet{2020cf4} Tully-Fisher scatter model would suggest a 0.48\,mag scatter (equivalent to a 22\% distance error). We conclude that the lower estimate of the distance errors in Cosmicflows-4 is due to underestimating the intrinsic scatter in the Tully-Fisher relation at fainter absolute magnitudes, and that, for galaxies fainter than the calibration sample, the Cosmicflows-4 Tully-Fisher distances are not as precise as previously quoted.

\begin{figure*}\centering
\includegraphics[width=\textwidth]{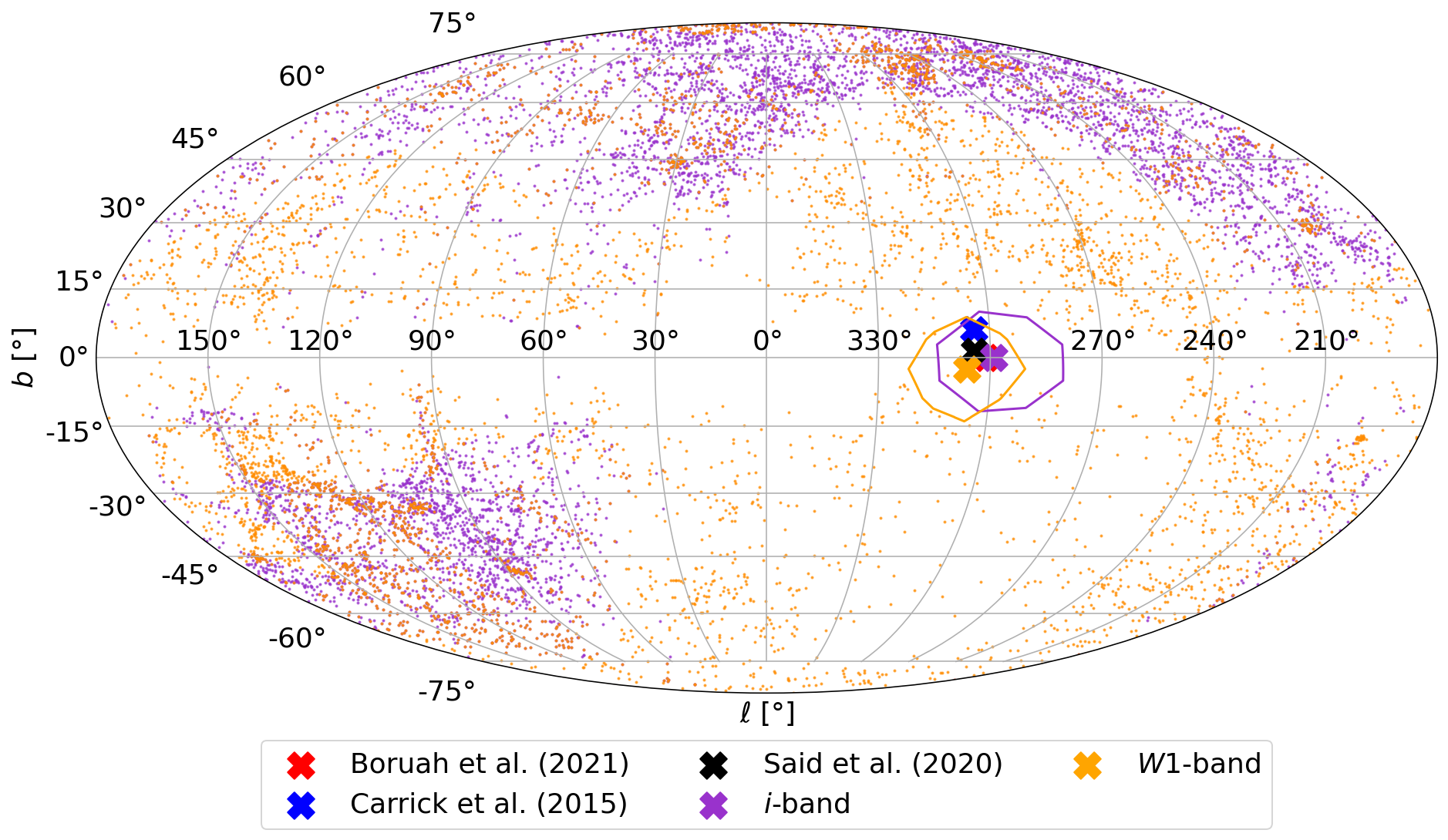}
\caption{The direction of the residual bulk flow in Galactic Cartesian coordinates, as determined here from the SDSS $i$-band data (purple cross and 1$\sigma$ error ellipse) and the WISE $W1$-band data (orange cross and 1$\sigma$ error ellipse), and from previous studies by \citet{Said_2020} (black cross), \citet{Carrick_2015} (blue cross), and \citet{boruah2020peculiar} (red cross). The SDSS and WISE CF4 samples are shown as purple and orange dots.}
\label{fig:bulkflowsky}
\end{figure*}

\begin{figure*}\centering 
\includegraphics[width=2\columnwidth]{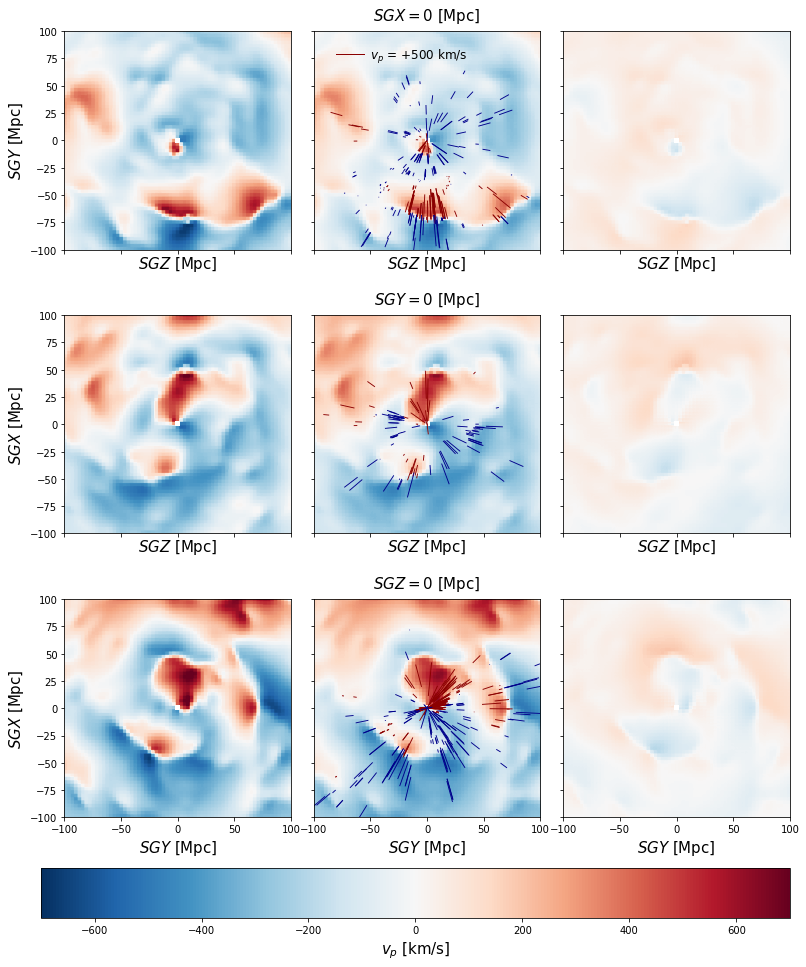}
\caption{A comparison of line-of-sight peculiar velocities. The 2M++ peculiar velocity model as derived by \citet{Carrick_2015} (left column), our fitted velocity field for the WISE sample with parameters listed in Table~\ref{tab:paramsfinal} (middle column), and our fitted velocity field minus the 2M++ velocity field (right column). Peculiar velocities are in \kms\ and plotted in Supergalactic Cartesian coordinates for three orthogonal slices: SGX\,=\,0\,$h^{-1}$\,Mpc (top row), SGY\,=\,0\,$h^{-1}$\,Mpc (middle row), and SGZ\,=\,0\,$h^{-1}$\,Mpc (bottom row). Peculiar velocities of individual galaxies from the CF4 Tully-Fisher sample (see Section~\ref{distpv} estimated from Eq.~\ref{eq:zccondprob}) are shown as vectors. The midpoint of each vector corresponds to the location of the galaxy and the length of the vector is scaled to the magnitude of the line-of-sight peculiar velocity, with a +500\kms\ vector in the top-middle panel for scale. The direction of the vector is represented by its colour: blue for a negative peculiar velocity (inward to the origin) or red for a positive peculiar velocity (outward from the origin). Galaxies included in each row are located in a circular wedge around the central plane with opening angle at the origin of 1.15\,deg, corresponding to a thickness of 4\,$h^{-1}$\,Mpc at 100\,$h^{-1}$\,Mpc radius.}
\label{fig:pvfields}
\end{figure*}

\subsection{Peculiar velocity model parameters}
\label{PVmodelfits}

Figure~\ref{fig:bulkflowsky} shows the SDSS $i$-band and WISE $W1$-band samples from CF4 on the sky, along with the directions of the residual bulk flows derived from fitting these samples. Figure~\ref{fig:pvfields} directly compares the line-of-sight peculiar velocities from our fitted peculiar velocity field (using the parameters of Table~\ref{tab:paramsfinal}) to those predicted from the 2M++ redshift survey by \citet{Carrick_2015}, overlaying the peculiar velocities we derive for the individual CF4 galaxies (see Section~\ref{distpv} below). 

We obtain $\beta$\,=\,0.33\,$\pm$\,0.03 for the SDSS $i$ band sample and $\beta$\,=\,0.36\,$\pm$\,0.02 for the WISE $W1$ band sample, giving a weighted mean of $\beta$\,=\,0.35\,$\pm$\,0.02. Despite there being 30\% more galaxies with $i$-band magnitudes than $W1$-band magnitudes ($N$\,=\,4,723 for the $W1$ band and $N$\,=\,6,224 for the $i$ band), the $W1$-band uncertainty for $\beta$ is actually about 50\% smaller. The uncertainty in velocity field parameters is linked to sky coverage as well as sample size, so the fact WISE covers the whole sky likely explains why it yields a more stringent constraint for $\beta$ compared to SDSS. \citet{Carrick_2015} reported $\beta$\,=\,0.431\,$\pm$\,0.021, 3$\sigma$ higher than our result, but the corresponding $f\sigma_8$ measurements are still consistent; there is also reason to believe that the \citet{Carrick_2015} estimate is biased high (see Section~\ref{growthrate}). 

The CMB frame residual bulk flow of our $i$-band fit is ($-$203\,$\pm$\,25,7\,$\pm$\,13,$-$65\,$\pm$\,20)\kms\ in Supergalactic Cartesian coordinates, (+104\,$\pm$\,21,$-$187\,$\pm$\,25,$-$1\,$\pm$\,14)\kms\ in Galactic Cartesian coordinates, and $|V|$\,=\,215\,$\pm$\,25\kms, $(l,b)$\,=\,(299$^\circ$,0$^\circ$) in Galactic polar coordinates. The residual bulk flow of our $W1$-band fit is ($-$225\,$\pm$\,15,$-$7\,$\pm$\,13,$-$46\,$\pm$\,16)\kms\ in Supergalactic Cartesian coordinates, (+135\,$\pm$\,16,$-$186\,$\pm$\,15,$-$12\,$\pm$\,13)\kms\ in Galactic Cartesian coordinates, and $|V|$\,=\,230\,$\pm$\,12\kms, $(l,b)$\,=\,(306$^\circ$,$-$3$^\circ$) in Galactic polar coordinates. The weighted mean of these two estimates is $|V|$\,=\,227\,$\pm$\,11\kms, $(l,b)$\,=\,(303$^\circ$,$-$1$^\circ$) in Galactic polar coordinates.

We can compute the $\chi^2$ by comparing the peculiar velocities predicted by this best-fitting model to the peculiar velocities predicted only using each galaxy's offset from the best-fitting Tully-Fisher relation and the intrinsic scatter from Section~\ref{TFmodelfits}. This results in a reduced $\chi^2$ of 2.6 for the $i$-band with 6,215 degrees of freedom and 7.6 in the $W1$-band with 4,714 degrees of freedom. The $i$-band galaxies are better fit by the combined Tully-Fisher and peculiar velocity model.

These results agree within 3$\sigma$ with each other and with earlier CMB frame results: \citet{Carrick_2015} found (+89\,$\pm$\,21,$-$131\,$\pm$\,23,+17\,$\pm$\,26)\kms\ in Galactic Cartesian coordinates and ($-$154\,$\pm$\,23,21\,$\pm$\,26,$-$35\,$\pm$\,22)\kms\ in Supergalactic Cartesian coordinates for the 2M++ volume, \citet{boruah2020peculiar} found (+88\,$\pm$\,13,$-$146\,$\pm$\,11,+0\,$\pm$\,9)\kms\ in Galactic Cartesian coordinates, and \citet{Said_2020} found $\mathbf{V}_{\textrm{ext}}$\,=\,(+94\,$\pm$\,10,$-$138\,$\pm$\,12,+4\,$\pm$\,12)\kms\ in Galactic Cartesian coordinates. The directions of these various residual bulk flow measurements are shown in Figure~\ref{fig:bulkflowsky} with their 1$\sigma$ error ellipses. The directions are all consistent, implying that there is a net dipole induced by the mass distribution outside the volume covered by the 2M++ survey ($z<0.07$). In general, the amplitude of our residual bulk flow is higher than these previous estimates. This could partly be explained by the different volumes sampled. Although each of these estimates are confined to the 2M++ volume, any differences in sampling would result in slightly different residual bulk flows. In addition, this higher $|V|$ could be linked with our lower estimate of $\beta$, as their joint constraint in Figure~\ref{fig:CF4_constraints} shows a modest anti-correlation between these two parameters. 

We can also compute the predicted total bulk flow of our sample volume by averaging the peculiar velocities from the model with our best-fit $\beta$ and $\mathbf{V}_{\textrm{ext}}$. Doing this, we obtain $\mathbf{B}$\,=\,($-$359\,$\pm$\,4,\,$-$9\,$\pm$\,4,\,$-$129\,$\pm$\,4)\kms\ for the SDSS sample and $\mathbf{B}$\,=\,($-$286\,$\pm$\,4,\,52\,$\pm$\,4,\,$-$119\,$\pm$\,4)\kms\ for the WISE sample, corresponding to amplitudes of 382\,$\pm$\,5\kms\ and 314\,$\pm$\,6\kms\ respectively. We note that these uncertainties are only statistical and that the systematic uncertainties have not been characterised. \citet{whitford_2023} measured the bulk flow for the full CF4 sample and obtained $\mathbf{B}$\,=\,($-$391\,$\pm$\,104,\,$-$119\,$\pm$\,93,\,$-$126\,$\pm$\,122)\kms\ or $\mathbf{B}$\,=\,($-$382\,$\pm$\,165,\,48\,$\pm$\,149,\,$-$135\,$\pm$\,154)\kms\ using two different estimators, corresponding to amplitudes of 428\,$\pm$\,108\kms\ and 408\,$\pm$\,165\kms\ respectively. These measurements are all higher than the $\Lambda$CDM prediction of 194\,$\pm$\,86\kms\ \citep{whitford_2023} for the bulk flow amplitude for the full CF4 volume, but all by less than 2$\sigma$ except one (the SDSS sample estimate is within 3$\sigma$).

\subsection{Growth rate}
\label{growthrate}

Having obtained an estimate for the $\beta$ velocity field scaling parameter, we can also derive an estimate for the growth rate of cosmic structure, $f\!\sigma_8$, using $f\!\sigma_8$\,=\,$\beta\sigma_{8,g}$, where $\sigma_{8,g}$ is the RMS fluctuation in galaxy number in spheres of radius 8\,$h^{-1}$\,Mpc. This enters through Equation~\ref{eq:velocity_beta} and relates to the galaxies used to determine the density field, not the galaxies for which we measure peculiar velocities.

The uncertainty in our estimate of $f\!\sigma_8$ arises in part from the measurement error for $\beta$ from our samplev and in part from sampling variance (cosmic variance) due to the finite volume of the 2M++ redshift survey used to predict the velocity field and the finite volume of the peculiar velocity sample. 

For individual galaxies the measurement error is dominated by the intrinsic scatter in the distance indicator relation; contributions from uncertainties in the predicted velocity field are smaller by about an order of magnitude. While our method accounts for these through $\sigma_{\rm TF}$ and $\sigma_v$, it also assumes that both errors are uncorrelated between galaxies. This is true for the intrinsic scatter in the Tully-Fisher relation, but does not hold for the velocity field, as galaxies over relatively large volumes will have correlated velocities. Accounting for such correlated errors requires the computation of a full velocity covariance matrix and will be addressed in future work.

The second source of uncertainty, cosmic variance, was discussed in \citet{Hollinger_2021} and \citet{Hollinger_2023}. They used mocks of the 2M++ survey in combination with various Tully-Fisher peculiar velocity datasets (all smaller than CF4) to estimate that the cosmic variance due to differences in local mean density and in density variance ($\sigma_8$) results in a $\sim$5\% scatter on $f\!\sigma_8$. We adopt this estimate and combine it in quadrature with our measurement uncertainty to obtain our overall uncertainty in $f\!\sigma_8$. This accounts for sample variance due to the finite volume of the density field, but not for any additional sample variance that might result from the smaller volume enclosed by the peculiar velocity sample.

So, our quoted uncertainty for $f\!\sigma_8$ does not account for two possible sources of additional uncertainty: correlated errors in the velocity field and variance due to sampling peculiar velocities within the 2M++ volume. With that caveat, we proceed to compare our measurement with other results and with various model predictions.

For the 2M++ galaxy redshift compilation, \citet{Carrick_2015} report $\sigma_{8,g}$\,=\,0.99, which results in a growth rate $f\!\sigma_8$\,=\,0.33$\,\pm\,$0.04 from the measurement of $\beta$ in the $i$-band and $f\!\sigma_8$\,=\,0.36\,$\pm$\,0.03 from the $W1$-band. These measurements are consistent with those of other estimates at similar effective redshifts ($z$\,$\approx$\,0.02): \citet{Carrick_2015} measured $f\!\sigma_8$\,=\,0.401\,$\pm$\,0.024, \citet{boruah2020peculiar} measured $f\!\sigma_8$\,=\,0.401\,$\pm$\,0.017, and \citet{Davis_2011} measured $f\!\sigma_8$\,=\,0.31\,$\pm$\,0.06. \citet{Hollinger_2023} analysed the biases affecting the 2M++ velocity field reconstruction and found that the result of \citet{Carrick_2015} may be biased high. They reported a value of $f\!\sigma_8$\,=\,0.362\,$\pm$\,0.023. This corrected value slightly decreases the difference between our estimates and that of \citet{Carrick_2015}.

\begin{figure*}\centering 
\includegraphics[width=2\columnwidth]{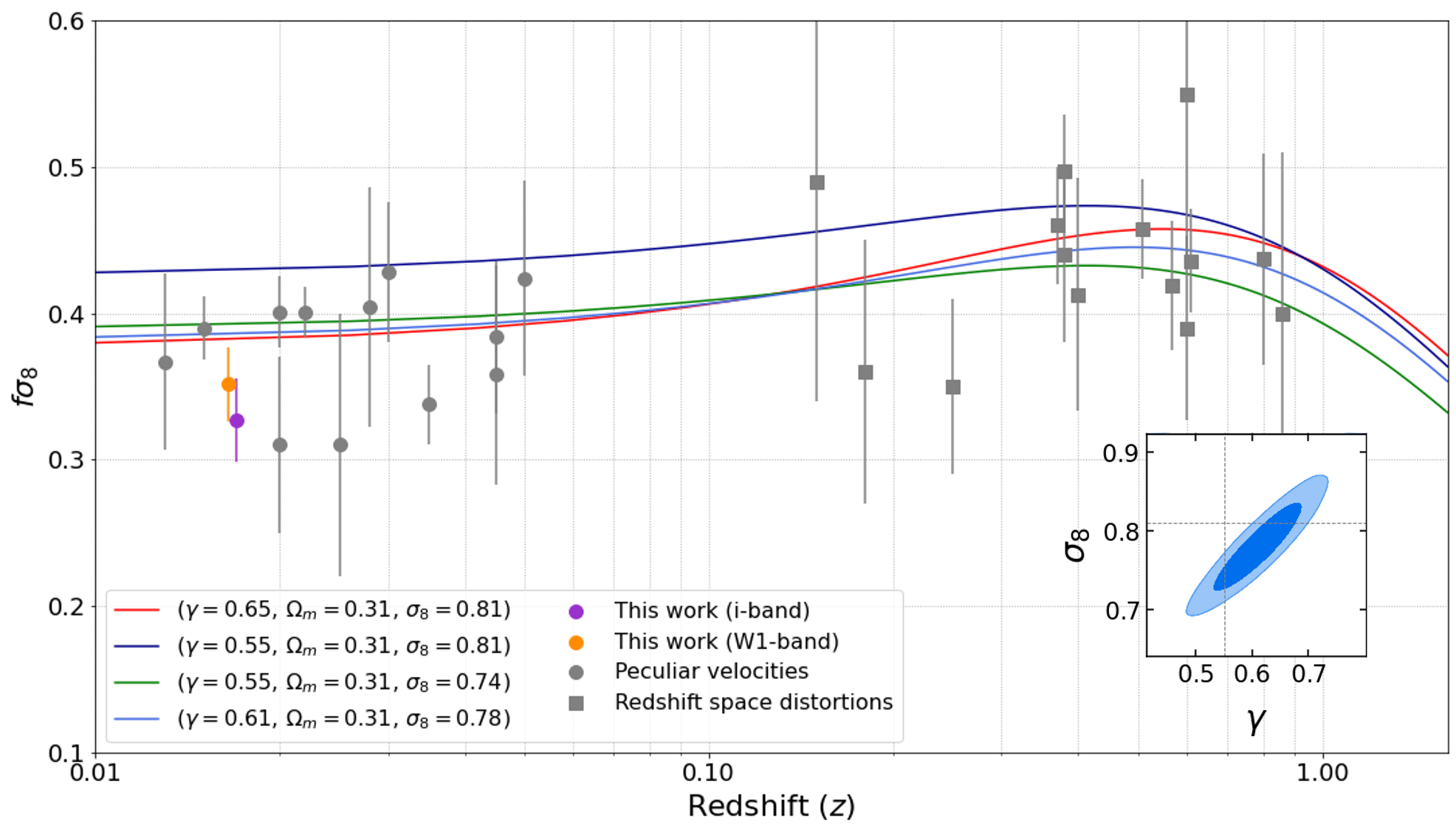}
\caption{Comparison of $f\!\sigma_8$ constraints from various measurements at different redshifts (see Table~\ref{tab:fcompare}). The curves show models for the evolution of $f\!\sigma_8$ with redshift for four different cosmologies. Dark blue: Planck parameters \citep{2018planck} with the GR value of $\gamma=0.55$. The red, green, and light blue curves are naive fits to the data neglecting correlations between data points and degeneracies with the Alcock-Paczynski effect. Red: if $\Omega_{m}$ and $\sigma_{8}$ are fixed at the Planck values, the best fit is $\gamma=0.65\pm0.02$. Green: if $\Omega_{m}$ and $\gamma$ are fixed at the Planck/GR values, the best fit is $\sigma_{8}=0.74\pm0.01$. Light blue: if $\Omega_{m}$ is fixed at the Planck value, the best fit is $\sigma_{8}=0.78\pm0.04$ and $\gamma=0.61\pm0.05$; the inset shows the joint probability for these two parameters; the crosshairs correspond to the baseline Planck and GR values.}
\label{fig:fcompare}
\end{figure*}

More broadly, Table~\ref{tab:fcompare} and Figure~\ref{fig:fcompare} show that our measurements of $f\!\sigma_8$ from the CF4 Tully-Fisher sample are consistent with those of other peculiar velocity analyses at low redshifts and with measurements at higher redshifts obtained from redshift space distortions. We note that the uncertainties on growth rate measurements that used a density reconstruction to predict peculiar velocities are generally significantly smaller than those obtained from other methods. This may be a result of correlated velocity errors that are not accounted for by our method, or possibly from higher-point statistics that enter into, and inform, the reconstruction \citep[see][]{Schmittfull_2015}. The dark blue curve in Figure~\ref{fig:fcompare} shows the prediction for the evolution of the growth rate assuming General Relativity \citep[for GR, $\gamma$\,=\,6/11\,$\approx$\,0.55;][]{Wang_1998,Linder_2005} and the baseline Planck $\Lambda$CDM parameters \citep[$\Omega_m$\,=\,0.31, $\sigma_8$\,=\,0.81;][]{2020_Planck}. Taking the quoted uncertainties at face value, both the $i$-band and the $W1$-band measurements of $f\!\sigma_8$ are marginally consistent with this model at 3$\sigma$. Given that we have not accounted for some potential sources of uncertainty, this result cannot be interpreted as a significant discrepancy with the Planck CMB predictions based on a standard $\Lambda$CDM+GR cosmology. The fact that all measurements of $f\!\sigma_8$ based on peculiar velocities lie below this prediction (mostly by <2$\sigma$, although the \citet{Said_2020} measurement is >3$\sigma$ below) might be ascribed to a combination of under-estimated errors and common sample volumes.

Since $f\!\sigma_8$\,=\,$\Omega_m(z)^\gamma\sigma_8$, we can try to improve the agreement between measurements and model by allowing the cosmological parameters $\gamma$, $\Omega_{m}$ and $\sigma_{8}$ to vary. First, we allow $\gamma$ to be a free parameter while fixing $\Omega_{m}$ and $\sigma_{8}$ to the baseline Planck values, and find the best fit is $\gamma$\,=\,0.65\,$\pm$\,0.02 (red curve). These observations thus tend to favour theories of gravity with higher values of $\gamma$, such as Dvali-Gabadadze-Porrati gravity \citep[for DGP gravity, $\gamma$\,=\,11/16\,$\approx$\,0.69;][]{Dvali_2000,Linder_2007}. Second, we fit $\sigma_{8}$ while setting $\gamma$\,=\,0.55 and $\Omega_{m}$\,=\,0.31, and find that a lower fluctuation amplitude of $\sigma_{8}$\,=\,0.74\,$\pm$\,0.01 is favoured (green curve). Lastly, we allow both $\gamma$ and $\sigma_{8}$ to be free and obtain $\sigma_{8}$\,=\,0.78\,$\pm$\,0.04 and $\gamma$\,=\,0.61\,$\pm$\,0.05 (light blue curve); the inset shows the joint posterior probability distribution. We emphasise that this is a naive analysis: it assumes the measurements are independent, when in fact they are not; a more rigorous analysis would account for the correlations between the measurements and likely result in less stringent constraints. In addition, for $z \gtrsim 0.1$, $f\!\sigma_8$ is degenerate with the redshift-distance relation due to the Alcock-Paczynski effect \citep{Alcock_1979}. Therefore, in order to perform a proper fit using the higher-redshift RSD measurements, a correction needs to be made \citep[see][]{Hudson_2012}, but this requires the appropriate covariance matrices. Nonetheless, the fact remains that existing measurements of $f\!\sigma_8$ generally lie below the prediction of $\Lambda$CDM and GR, particularly at lower redshifts where the Alcock-Paczynski effect is negligible.

\begin{table}
	\centering
	\caption{Comparison of $f\!\sigma_{8}$ constraints from this work and various literature measurements based on peculiar velocities or redshift space distortions at different effective redshifts $z_{\rm eff}$.}
	\label{tab:fcompare}
	\begin{tabular}{lcl}
		\hline
		\hline
		$z_{\rm eff}$ & $f\!\sigma_{8}$ & Reference \\
		\hline
		0.013 & 0.367\,$\pm$\,0.06 & \citet{Lilow_2021} \\
		0.015 & 0.39\,$\pm$\,0.022 & \citet{Stahl_2021} \\
  		0.017 & 0.36\,$\pm$\,0.03 & This work \\
    	0.017 & 0.33\,$\pm$\,0.04 & This work \\
		0.02 & 0.31\,$\pm$\,0.06 & \citet{Davis_2011} \\
		0.02 & 0.401\,$\pm$\,0.024 & \citet{Carrick_2015} \\
        0.02 & 0.362\,$\pm$\,0.023 & \citet{Hollinger_2023} \\
		0.022 & 0.401\,$\pm$\,0.017 & \citet{boruah2020peculiar} \\
		0.025 & 0.31\,$\pm$\,0.09 & \citet{Branchini_2012} \\
		0.028 & 0.404\,$\pm$\,0.082 & \citet{Qin_2019} \\
		0.03 & 0.428\,$\pm$\,0.048 & \citet{Huterer_2017} \\
		0.035 & 0.338\,$\pm$\,0.027 & \citet{Said_2020} \\
		0.045 & 0.384\,$\pm$\,0.052 & \citet{Adams_2020} \\
		0.045 & 0.358\,$\pm$\,0.075 & \citet{Turner_2023} \\
		0.05 & 0.424\,$\pm$\,0.067 & \citet{Adams_2017} \\
		0.067 & 0.423\,$\pm$\,0.055 & \citet{Beutler_2012} \\
		0.15 & 0.49\,$\pm$\,0.15 & \citet{Howlett_2016} \\
		0.18 & 0.36\,$\pm$\,0.09 & \citet{Blake_2013} \\
		0.25 & 0.35\,$\pm$\,0.06 & \citet{Samushia_2012} \\
		0.37 & 0.46\,$\pm$\,0.04 & \citet{Samushia_2012} \\
		0.38 & 0.497\,$\pm$\,0.039 & \citet{Alam_2017} \\
		0.4 & 0.413\,$\pm$\,0.08 & \citet{Blake_2012} \\
		0.51 & 0.458\,$\pm$\,0.035 & \citet{Alam_2017} \\
		0.57 & 0.419\,$\pm$\,0.044 & \citet{Beutler_2014} \\
		0.6 & 0.55\,$\pm$\,0.12 & \citet{Pezzotta_2017} \\
		0.61 & 0.436\,$\pm$\,0.034 & \citet{Alam_2017} \\
		0.8 & 0.437\,$\pm$\,0.072 & \citet{Blake_2012} \\
		0.86 & 0.4\,$\pm$\,0.11 & \citet{Pezzotta_2017} \\
		\hline
	\end{tabular}
\end{table}

\subsection{Distances and peculiar velocities}
\label{distpv}

We report two estimates of the comoving distances and the corresponding peculiar velocities: (i)~an estimate derived from each galaxy's offset with respect to the best-fit Tully-Fisher relation by solving $m(z_c)$\,=\,$m$ (see Equation~\ref{eq:mpred}), with a random error given by the Tully-Fisher model scatter computed from the galaxy's velocity width; and (ii)~an estimate obtained from the full posterior distribution derived from maximising the conditional probability of the cosmological redshift given the observables and the best-fit model parameters (the equivalent of Equation~\ref{eq:zccondprob}, but using the conditional magnitude distribution given by combining Equations~\ref{eq:mcondprob} and~\ref{eq:maglimsoft}).

A peculiar velocity, $v_{\!p}=cz_p$, can be computed from a comoving distance estimate and an observed redshift by first numerically inverting the redshift-distance relation to convert from log-distance ratio to cosmological redshift, then using Equation~\ref{eq:redshifts} to obtain the peculiar velocity. However, this non-linear transformation means the posterior distribution of the peculiar velocity is no longer Gaussian (it is close to log-normal) and its uncertainties are asymmetric. It can be represented as a non-Gaussian distribution by taking into account the Jacobian of the transformation \citep[e.g.,][]{Johnson_2014}, by providing a fitting formula \citep[e.g.,][]{Springob_2014} or an approximation \citep[e.g.,][]{Scrimgeour_2016}, or by correcting in some other manner \citep[e.g.,][]{Hoffman_2021,Qin_2021}. 

\begin{figure}
\centering 
\includegraphics[width=1\columnwidth]{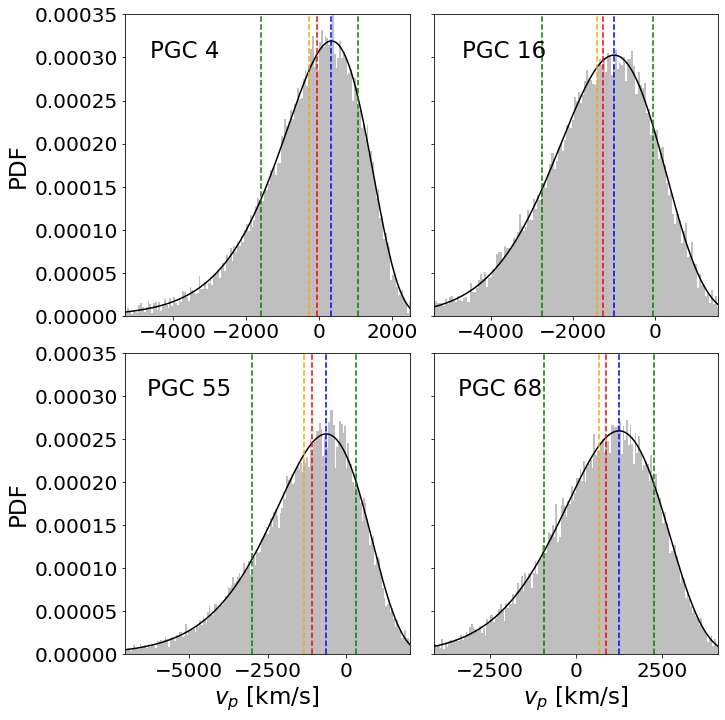}
\caption{Log-normal fits (black lines) to the probability distributions obtained from the MCMC chains of estimated peculiar velocities $v_{\!p}$ from $i$-band Tully-Fisher offsets (grey histograms) for three galaxies in the sample (PGC = 4, 12, 16, and 55; see Table~\ref{tab:preddist} for values of $\overline{\eta}$ and $v_{p,\textrm{median}}$). Dashed blue, red, and orange lines shows the modes, medians, and means computed using Equation~\ref{eq:lognorm_mmm}; dashed green lines are the 16\% and 84\% percentiles of the distribution given by Equation~\ref{eq:quantile}.}
\label{fig:vp_dist}
\end{figure}

Here we fit a log-normal probability density function to the estimated peculiar velocity distribution for each galaxy $i$,
\begin{equation}
P_i(v_{\!p}) = e^{-\frac{1}{2}\left( \frac{\eta - \overline{\eta}_i}{\sigma_{\eta i}} \right)^{2}} 10^{\eta}
\label{eq:lognorm}
\end{equation}
where $\eta$ is the log-distance ratio, which is assumed to have a Gaussian distribution with mean $\overline{\eta}_i=\log{[D_C(z_i)/D_C(z_{ci})]}$ and standard deviation $\sigma_{\eta i}$. The mode, mean, and median of the peculiar velocity distribution correspond to values of $\eta$ given by 
\begin{align}
\eta_{\textrm{mode},i} &= \overline{\eta}_i + \sigma_{\eta i}^{2}\ln{10} \nonumber \\
\eta_{\textrm{median},i} &= \overline{\eta}_i \nonumber \\ 
\eta_{\textrm{mean},i} &= \overline{\eta}_i - \frac{1}{2}\sigma_{\eta i}^{2}\ln{10} ~. 
\label{eq:lognorm_mmm}
\end{align}
The quantiles of $\eta$ are
\begin{equation}
\eta_p = \overline{\eta}_i + \sqrt{2}\sigma_{\eta i}\,\textrm{erf}^{-1}(2p-1)
\label{eq:quantile}
\end{equation}
where $\textrm{erf}^{-1}$ is the inverse error function and $p$ is the percentile. 
The peculiar velocity $v_{\!p}$ corresponding to each $\eta$ is given by 
\begin{equation}
v_{\!p} = c\left( \frac{z_i - z_c(\eta, z_i)}{1+z_c(\eta,z_i)} \right) ~.
\label{eq:eta_to_vp}
\end{equation}
Examples of fits to the distributions of estimated peculiar velocities for four galaxies are shown in Figure~\ref{fig:vp_dist}.

A convenient alternative to specifying fitting functions for the non-Gaussian posterior distributions is to use an approximate peculiar velocity estimator that has Gaussian errors because it is proportional to the log-distance ratio, as suggested by \citet[][hereafter WF15]{Watkins_2015}. This is appealing for many applications, so we also report a variant of the WF15 approximate peculiar velocity estimator with Gaussian errors, namely
\begin{equation}
\widetilde{v}_{\!p} \equiv \frac{cz}{1+\epsilon z} \, \eta \ln10
\label{eqn:PVapprox}
\end{equation}
with $\epsilon = 0.75$ and $\eta=\log{[D_C(z)/D_C(z_{c})]}$. This estimator, which is derived and compared to the WF15 estimator in Appendix~\ref{PVest}, has multiple advantages in addition to Gaussian errors: (i)~it uses only the log-distance ratio and the observed redshift (unlike the WF15 estimator, which uses a modified redshift that is dependent on the cosmological model); (ii)~it gives similar or smaller fractional systematic errors for the estimated peculiar velocity (relative to the WF15 estimator) at all redshifts; (iii)~it makes the amplitudes of the systematic fractional errors for positive and negative peculiar velocities almost the same (although, as for the WF15 estimator, these errors are not symmetric); and (iv)~it makes the fractional systematic errors asymptote around zero at higher redshifts (whereas those for the WF15 estimator diverge).

Distance and peculiar velocity measurements for the galaxies in our sample are reported in the table provided in the supplementary online material, with example data shown in Appendix~\ref{resultstable}. The columns of this table provide the following information: 
(1) common name; 
(2) ID number of galaxy in the Principal Galaxy Catalog (PGC);
(3) observed redshift in the CMB frame;
(4) right ascension in degrees;
(5) declination in degrees;
(6-7) logarithm of the inclination-corrected \HI\ linewidth, and its uncertainty;
(8) $i$-band magnitude;
(9) $W1$-band magnitude;
(10-11) cosmological redshift, $z_c$, computed from the full posterior distribution derived by maximising the conditional probability for $z_c$ given the observables and the best-fit model parameters (the equivalent of Equation~\ref{eq:zccondprob}, but using the conditional magnitude distribution given by combining Equations~\ref{eq:mcondprob} and~\ref{eq:maglimsoft});
(12-13) comoving distance in $h^{-1}$\,Mpc corresponding to $z_c$ in column~10, and its 1$\sigma$ uncertainty;
(14-16) line-of-sight peculiar velocity in \kms\ corresponding to $z_c$ in column~10, approximated by the median of its log-normal probability distribution (see Equations~\ref{eq:lognorm} and~\ref{eq:lognorm_mmm}), and its asymmetric 16--84\% confidence interval (see Equation~\ref{eq:quantile});
(17-18) logarithm of the distance ratio (log-distance ratio), $\eta=\log{[D_C(z)/D_C(z_{c})]}$, and its 1$\sigma$ uncertainty;
(19-20) approximate peculiar velocity, $\widetilde{v}_{\!p}$ (see Equation~\ref{eqn:PVapprox}) and its 1$\sigma$ uncertainty;
(21-31) equivalent quantities to columns 10-20, but using the $W1$-band Tully-Fisher parameters;
(32-53) equivalent quantities to columns 10-31, where $z_c$ is computed from the galaxy's offset with respect to the best-fit Tully-Fisher relation by solving $m(z_c)$\,=\,$m$ (see Equation~\ref{eq:mpred}).

\section{Forecasts for WALLABY}\label{WALLABY}

The Wide-field ASKAP L-band Legacy All-sky Blind surveY \citep[WALLABY;][]{Koribalski_2020, Westmeier_2022} started observations in 2022 and over 5 years is expected to detect 21\,cm emission from around 210,000 galaxies at low redshifts ($z < 0.1$ and a median redshift $z \approx 0.05$). For about 40\% of these galaxies, the signal-to-noise ratio is expected to be sufficiently high, and the inclination sufficiently edge-on, to provide redshift-independent distances via the Tully-Fisher relation \citep{Courtois_2022}, corresponding to about 84,000 late-type galaxies with Tully-Fisher distances. By 2027, WALLABY is expected to have measured these Tully-Fisher distances with $\sim$20\% precision \citep{Courtois_2022} over 14000\,deg$^2$ (1.4$\pi$ sr) \citep{Westmeier_2022} of the southern sky. The WALLABY sample will provide an independent probe of peculiar velocities within a volume similar to the 6dFGS survey. It will be the largest Tully-Fisher peculiar velocity sample to date, substantially extending the Cosmicflows-4 Tully-Fisher dataset and complementing the DESI \citep{DESI_2023} and 4MOST Hemisphere Survey \citep{Taylor_2023}, which will measure Fundamental Plane peculiar velocities over, respectively, most of the northern and southern hemispheres. 

A mock catalogue for an ideal WALLABY survey was developed by \citet{Koribalski_2020} using the SHARK model for galaxy evolution \citep{Lagos_2018} and the SURFS simulations of cosmic structure \citep{Elahi_2018}. Two simulation boxes were used: MICRO-SURFS and MEDI-SURFS. MEDI covers a volume of $210^{3}\,h^{-1}$\,Mpc, while MICRO covers the nearby Universe with a volume of only $40^{3}\,h^{-1}$\,Mpc but much higher resolution. For each galaxy in this WALLABY reference simulation, realistic \HI\ line profiles were simulated. 

\begin{figure*}\centering 
\includegraphics[width=\textwidth]{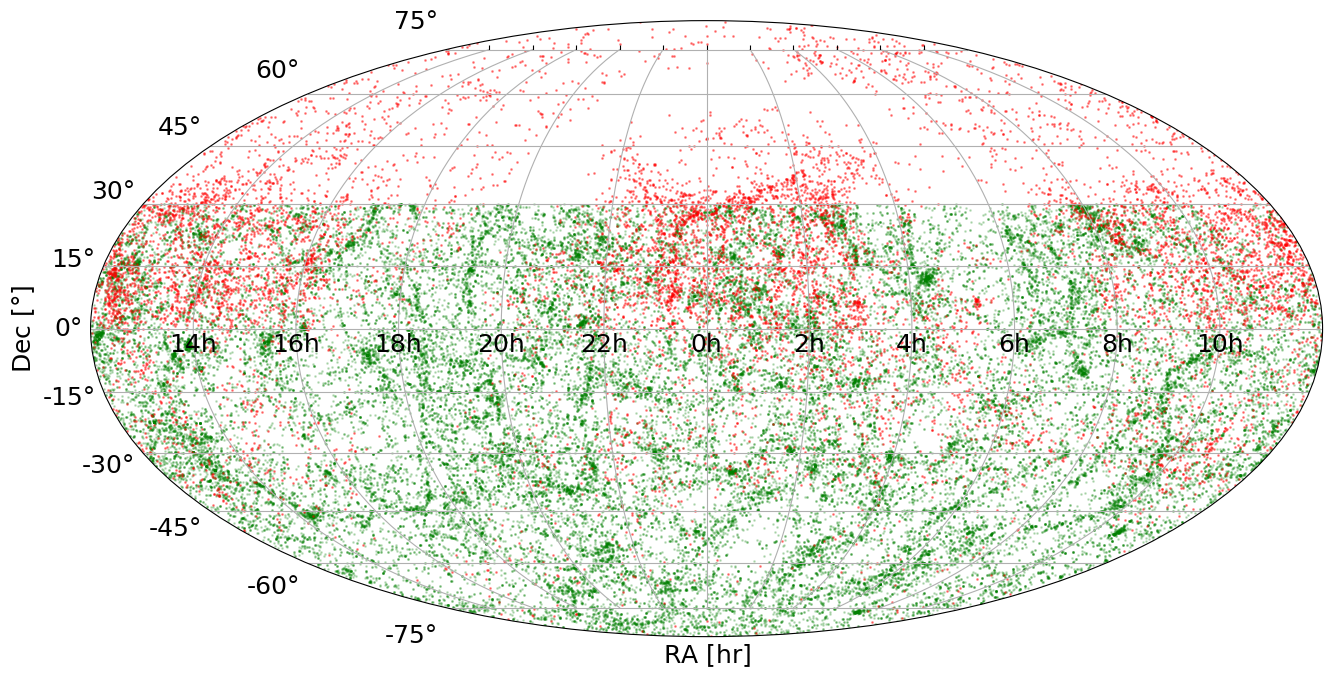}
\caption{Sky coverage of the full CF4 Tully-Fisher dataset (red) and the WALLABY reference simulation for 84,000 galaxies (green).}
\label{fig:wallabyfootprint}
\end{figure*}

\begin{figure}\centering 
\includegraphics[width=0.99\columnwidth]{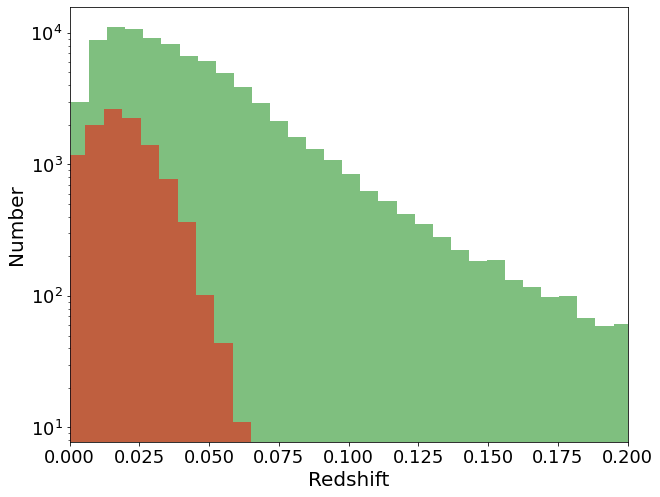}
\caption{Redshift distributions of the full CF4 Tully-Fisher sample (red) and the WALLABY reference simulation for 84,000 galaxies (green).}
\label{fig:wallabyredshift}
\end{figure}

While there are over 400,000 galaxies in the full simulation, we reduced this to a more realistic number, given the actual time allocated to the survey and the fact that not all measurements will be suitable for a Tully-Fisher analysis. The main requirements are a sufficiently high signal-to-noise \HI\ measurement (all galaxies in the simulation already pass this test) and inclinations greater than 45$\degree$. As stated above, approximately 84,000 Tully-Fisher distances are expected to be derived at $z<0.1$ in the full WALLABY survey. Selecting only galaxies with inclinations greater than 45$\degree$, we take as our sample the 84,000 galaxies with the highest \HI\ fluxes in the simulation. Figure~\ref{fig:wallabyfootprint} shows the sky coverage of the full CF4 data compared to this trimmed-down WALLABY Tully-Fisher reference simulation, while Figure~\ref{fig:wallabyredshift} shows a comparison of the redshift distributions. We use these simulated \HI\ linewidths, galaxy redshifts, inclinations, and spatial distributions to create realistic mocks of the WALLABY Tully-Fisher dataset. 

Peculiar velocities are generated from the 2M++ model and absolute magnitudes are generated using an imposed Tully-Fisher relation. We again use a curved linear relation and a linear scatter model. In the CF4 Tully-Fisher data the mean distance error is approximately 28\%, while for the SFI++ Tully-Fisher survey \citep{Masters_2006} put the mean distance error at around 18\%. We expect the typical distance errors for the WALLABY sample to be within this range, although we do not yet have a reliable estimate. We therefore generate two WALLABY mocks with approximately 30\% and 20\% mean distance errors by appropriately scaling the CF4 scatter model parameters, $\epsilon_{0}$ and $\epsilon_{1}$; scaling by 0.9 results in distance errors comparable to those of CF4 (29\%) and scaling by 0.65 results in distance errors comparable to those of SFI++ (21\%).

\begin{figure}\centering 
\includegraphics[width=\columnwidth]{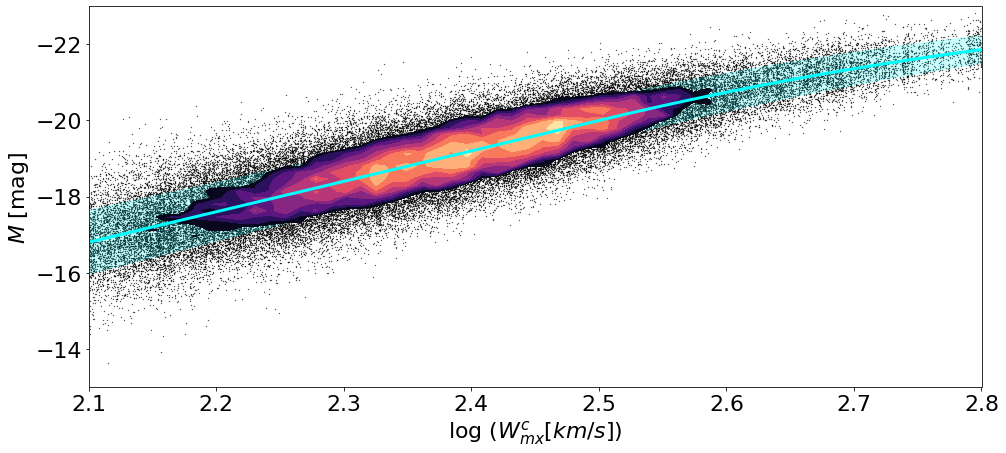}
\includegraphics[width=\columnwidth]{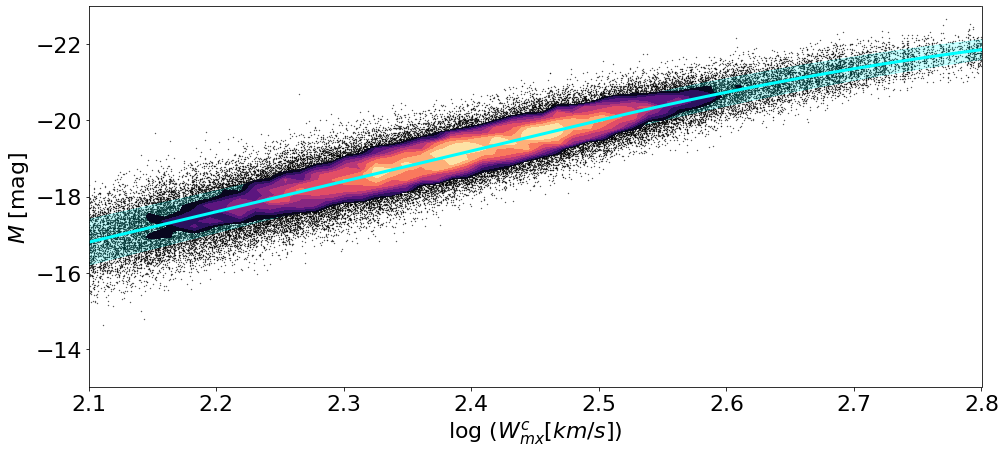}
\caption{Tully-Fisher relations for the WALLABY mock surveys with 29\% distance errors (top) and 21\% distance errors (bottom). Cyan lines and shading show the fits to the Tully-Fisher relations and scatter models.}
\label{fig:wallabyTFR}
\end{figure}

\begin{figure*}\centering 
\includegraphics[width=2\columnwidth]{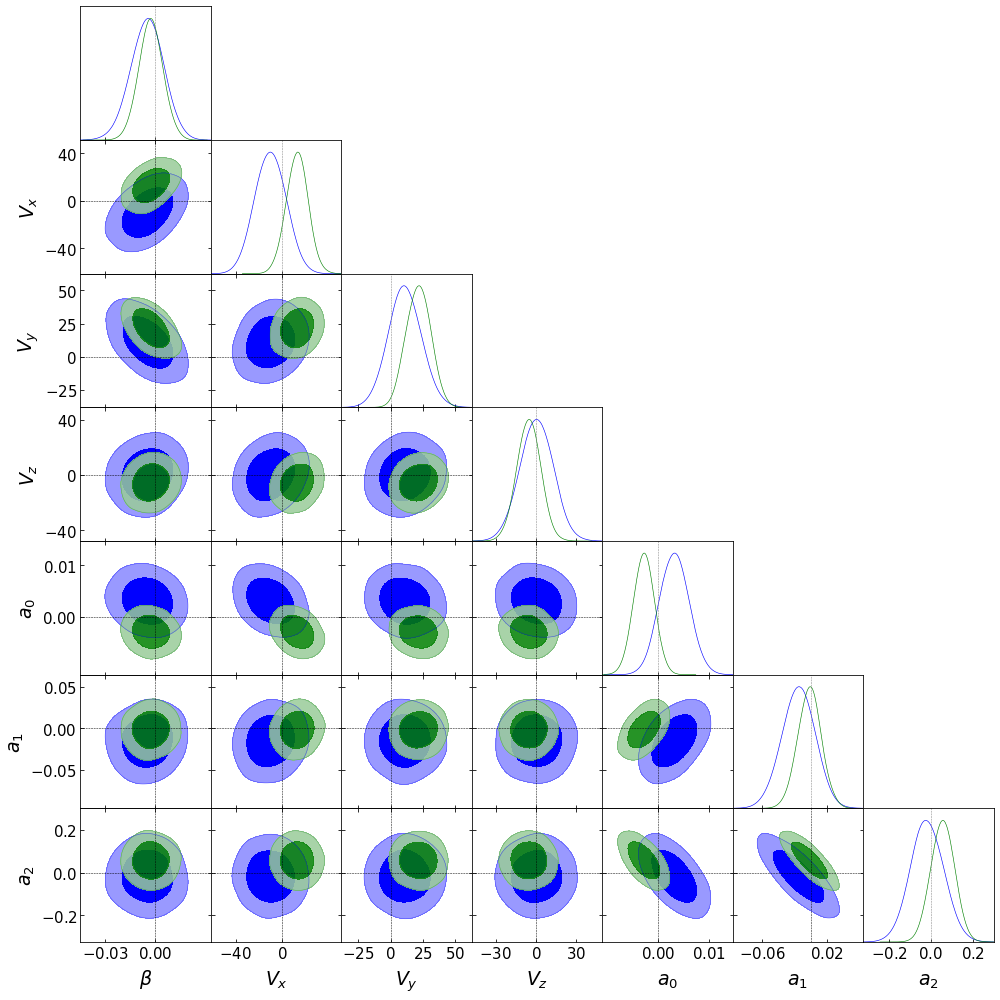}
\caption{Peculiar velocity and Tully-Fisher parameter constraints obtained from the WALLABY Tully-Fisher mocks. The constraints are all relative to the expected values (if known) or the best fit (if not). Green and blue contours show the optical mocks with distance errors of 21\% and 29\%, respectively. We omit the parameters of the Tully-Fisher scatter models.}
\label{fig:wallabycorner}
\end{figure*}

We have applied our method to both WALLABY Tully-Fisher mocks. Starting with the sample of 84,000 galaxies, then applying a cut at $\log{\Wmxc}=2.2$ and removing outliers, leaves us with 70,024 galaxies. Figure~\ref{fig:wallabyTFR} shows the fitted Tully-Fisher models for both mocks. The best-fit parameter values (and uncertainties) for the Tully-Fisher and peculiar velocity models are listed in Table~\ref{tab:paramswallaby}; the pairwise constraints on these parameters are shown in Figure~\ref{fig:wallabycorner}. We recover $\beta$\,=\,0.427\,$\pm$\,0.007 and $\mathbf{V}_{\textrm{ext}}$\,=\,($-$140\,$\pm$\,10,+42\,$\pm$\,9,$-$40\,$\pm$\,9)\kms\ for the mock with 21\% distance errors and $\beta$\,=\,0.425\,$\pm$\,0.010 and $\mathbf{V}_{\textrm{ext}}$\,=\,($-$164\,$\pm$\,14,+32\,$\pm$\,13,$-$34\,$\pm$\,12)\kms\ for the mock with 29\% distance errors, consistent at around 1$\sigma$ with the expected value of $\beta$\,=\,0.43 and the residual bulk motion in the 2M++ model of \citet{Carrick_2015}. 

\begin{figure}\centering 
\includegraphics[width=1\columnwidth]{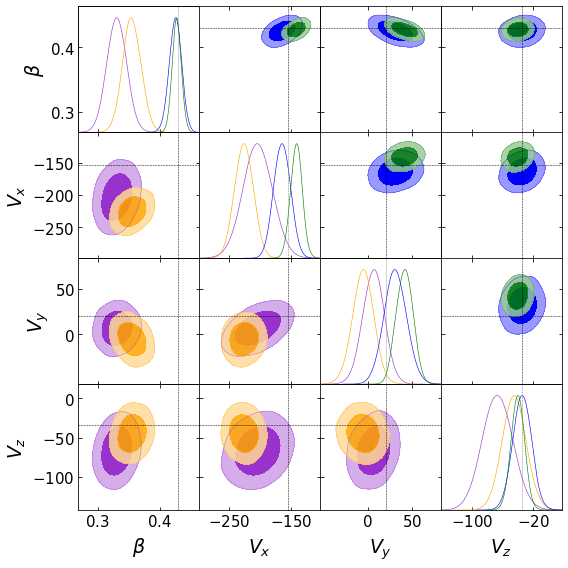}
\caption{Comparison of cosmological parameter constraints from this analysis using Cosmicflows-4 $i$-band and $W1$-band data (purple and orange respectively) and forecast constraints from WALLABY optical mocks with 21\% and 29\% distance errors (green and blue  respectively). Crosshairs correspond to the input parameters of the 2M++ peculiar velocity model.}
\label{fig:wallabycf4}
\end{figure}

Figure~\ref{fig:wallabycf4} compares the constraints on cosmological parameters obtained from the CF4 analysis (6,224 SDSS galaxies) and the WALLABY mocks (70,024 galaxies). This comparison suggests WALLABY will yield a factor of 2--3 improvement over Cosmicflows-4 in constraints on $\beta$. The uncertainty in $f\!\sigma_8$ will be improved by a slightly larger factor, since cosmic variance will be reduced with the larger sample size and volume of WALLABY. The improvement here is not as large as might be expected when compared to our WISE sample results, but the improvement due to the larger sample size of WALLABY is partially negated by its more limited sky coverage. Indeed, in versions of WALLABY mocks where sky positions were randomised, the uncertainty on $\beta$ decreased by almost a factor of 2. Ideally, therefore, WALLABY-derived peculiar velocities should be combined with surveys of the northern sky (such as SDSS and DESI) to maximise the constraining power of this methodology. Ultimately, these high-precision measurements of $f\!\sigma_8$ derived from WALLABY and other peculiar velocity surveys will bolster our ability to constrain $\sigma_8$ and $\gamma$. 

\begin{table}
	\centering
	\caption{Parameter constraints for the WALLABY mocks.}
	\label{tab:paramswallaby}
    \begingroup   
    \setlength{\tabcolsep}{4pt}
	\begin{tabular}{lccc}
		\hline
		\hline
		Parameter & 29\% scatter & 21\% scatter \\
		\hline
		$a_0$ & $-$20.000\,$\pm$\,0.003& $-$20.003\,$\pm$\, 0.002\\
		$a_1$ & $-$8.02\,$\pm$\,0.02& $-$8.00\,$\pm$\,0.01\\
  		$a_2$ &5.98\,$\pm$\,0.08 & 6.06\,$\pm$\,0.06\\
		$\epsilon_0$ & 0.563\,$\pm$\,0.002 & 0.406\,$\pm$\,0.001\\
		$\epsilon_1$ & $-$0.665\,$\pm$\,0.009& $-$0.499\,$\pm$\,0.006\\
		$\beta$ & 0.425\,$\pm$\,0.010 &0.427\,$\pm$\,0.007\\
		$V_x$ [\kms] & $-$164\,$\pm$\,14 &$-$140\,$\pm$\,10 \\
		$V_y$ [\kms] &~~$+$32\,$\pm$\,13 &~~$+$42\,$\pm$\,9 \\
		$V_z$ [\kms] &~~$-$34\,$\pm$\,12 &~~$-$40\,$\pm$\,9 \\
		\hline
	\end{tabular}
    \endgroup
\end{table}

\section{Conclusions}
\label{conclusions}

We present a method for simultaneously fitting models for the Tully-Fisher relation and the peculiar velocity field. The method uses the conditional probability for the observed magnitude as a function of the observed HI velocity width; it is thus a `direct' fit of the Tully-Fisher relation between magnitude and velocity width. It uses the cosmological redshift for each galaxy (corresponding to its comoving distance) predicted by a peculiar velocity model at the observed redshift, together with the galaxy's observed \HI\ velocity width, to give the expected apparent magnitude based on a model for the Tully-Fisher relation; it is thus a redshift-space method that minimises Malmquist bias. The conditional probability for each galaxy is computed from the predicted and observed magnitudes using a model for the scatter about the Tully-Fisher relation. The best-fit parameters of the models for the Tully-Fisher relation and its scatter, and for the peculiar velocity model, are estimated by maximising the likelihood obtained as the product over the sample of each galaxy's conditional probability distribution. We then estimate the cosmological redshift of each galaxy (and so its comoving distance and peculiar velocity) using the best-fitting models for the peculiar velocity field (and its scatter) and the Tully-Fisher relation (and its scatter). We are thus able to provide two distance and peculiar velocity estimates for each galaxy, one based on the measured offset from the Tully-Fisher relation and the other the posterior prediction combining the Tully-Fisher offset and the best-fit peculiar velocity model.

We first apply this method to the Cosmicflows-4 (CF4) catalogue of Tully-Fisher measurements. We modify the conventional linear Tully-Fisher relation to account for the observed curvature at the bright end of the relation and for the varying scatter along the relation. For the peculiar velocity model, we adopt the relative peculiar velocities derived from the density field reconstruction based on the 2M++ redshift survey by \citet{Carrick_2015}, but leaving free the velocity scaling parameter $\beta$ and a residual bulk motion $\mathbf{V}_{\textrm{ext}}$  that approximates the effect of the mass distribution external to the 2M++ volume. The full model for the Tully-Fisher relation and the peculiar velocity field thus involves a total of nine parameters: five Tully-Fisher parameters (two for the linear relation, one for the curvature at the bright end, and two for the linear change in scatter along the relation) and four cosmological parameters (the scaling parameter $\beta$ and the three components of the residual bulk motion).

This method is tested on simulated CF4 Tully-Fisher data, which confirms that it recovers the model parameters accurately (i.e.\ without significant bias). For the observed CF4 Tully-Fisher data, we trim the CF4 samples to remove galaxies with velocity widths below $\log \Wmxc$\,=\,2.2 and outliers more than 3$\sigma$ from the Tully-Fisher relation, leaving final samples of 6,224 galaxies with SDSS $i$-band magnitudes and 4,723 galaxies with WISE $W1$-band magnitudes. We find the velocity field scaling parameter to be $\beta=0.33\pm0.03$ for the SDSS sample and $\beta=0.36\pm0.02$ for the WISE sample. The residual bulk flow in Supergalactic Cartesian coordinates and the CMB frame is $\mathbf{V}_{\textrm{ext}}$\,=\,($-$203\,$\pm$\,35,7\,$\pm$\,13,$-$65\,$\pm$\,20)\kms\ for the SDSS sample and $\mathbf{V}_{\textrm{ext}}$\,=\,($-$225\,$\pm$\,15,$-$7\,$\pm$\,13,$-$46\,$\pm$\,16)\kms\ for the WISE sample. These results are consistent at <3$\sigma$ with previous determinations of $\beta$ and the amplitude and direction of $\mathbf{V}_{\textrm{ext}}$ based on fits to the 2M++ predicted velocity field using largely independent datasets \citep[][see Figure~\ref{fig:bulkflowsky}]{Carrick_2015,Said_2020,boruah2020peculiar}.

Since the bias of the galaxies in the 2M++ sample is $b=0.99$ \citep{Carrick_2015}, our measurements of $\beta$ correspond to growth rates of $f\!\sigma_8$\,=\,0.33\,$\pm$\,0.04 for the SDSS sample and $f\!\sigma_8$\,=\,0.36\,$\pm$\,0.03 for the WISE sample, where we have included a 5\% sample variance in the uncertainties as derived by \citet{Hollinger_2023}. These are consistent with other estimates of the growth rate at similar effective redshifts ($z$\,=\,0.017), listed in Table~\ref{tab:fcompare}. The relatively high precision of the WISE measurement showcases the constraining power of an all-sky dataset. These results reinforce the tendency for growth rate estimates from peculiar velocities or redshift-space distortions to fall below the predictions from the Planck CMB measurements \citep{2018planck} based on a standard $\Lambda$CDM+GR cosmology (see Figure~\ref{fig:fcompare}). However, we do note that uncertainties may be too optimistic for some measurements due to neglecting or underestimating sample variance. 

The analysis developed here can also be applied to new \HI\ Tully-Fisher peculiar velocity surveys such as WALLABY. We apply our method to existing WALLABY simulations \citep{Koribalski_2020} and show that the simulated WALLABY data can be well described with modest modifications to the Tully-Fisher model developed for CF4. We find that WALLABY on its own could provide a factor 2--3 improvement in the precision with which $\beta$ can be measured and will yield a valuable new low-redshift constraint on $f\!\sigma_8$. By supplementing WALLABY with peculiar velocities in the northern hemisphere from other surveys, even greater precision can be achieved.

\section*{Acknowledgements}

We thank Danail Obreschkow for providing the WALLABY reference simulation used in this work. MMC acknowledges support from a Royal Society Wolfson Visiting Fellowship (RSWVF{\textbackslash}R3{\textbackslash}223005) while on sabbatical at the University of Oxford. KS acknowledges support from the Australian Government through the Australian Research Council’s Laureate Fellowship funding scheme (project FL180100168). We acknowledge the use of the following analysis packages: Astropy \citep{astropy}, GetDist \citep{getdist}, emcee \citep{emcee}, and Matplotlib \citep{matplotlib}.

\section*{Data Availability}

This work uses previously published data as referenced and described in text. Details of the analysis and code can be made available upon reasonable request to the corresponding author.



\bibliographystyle{mnras}
\bibliography{sources}

\appendix

\section{Peculiar velocity estimators}
\label{PVest}

\begin{figure}\centering 
\includegraphics[width=\columnwidth]{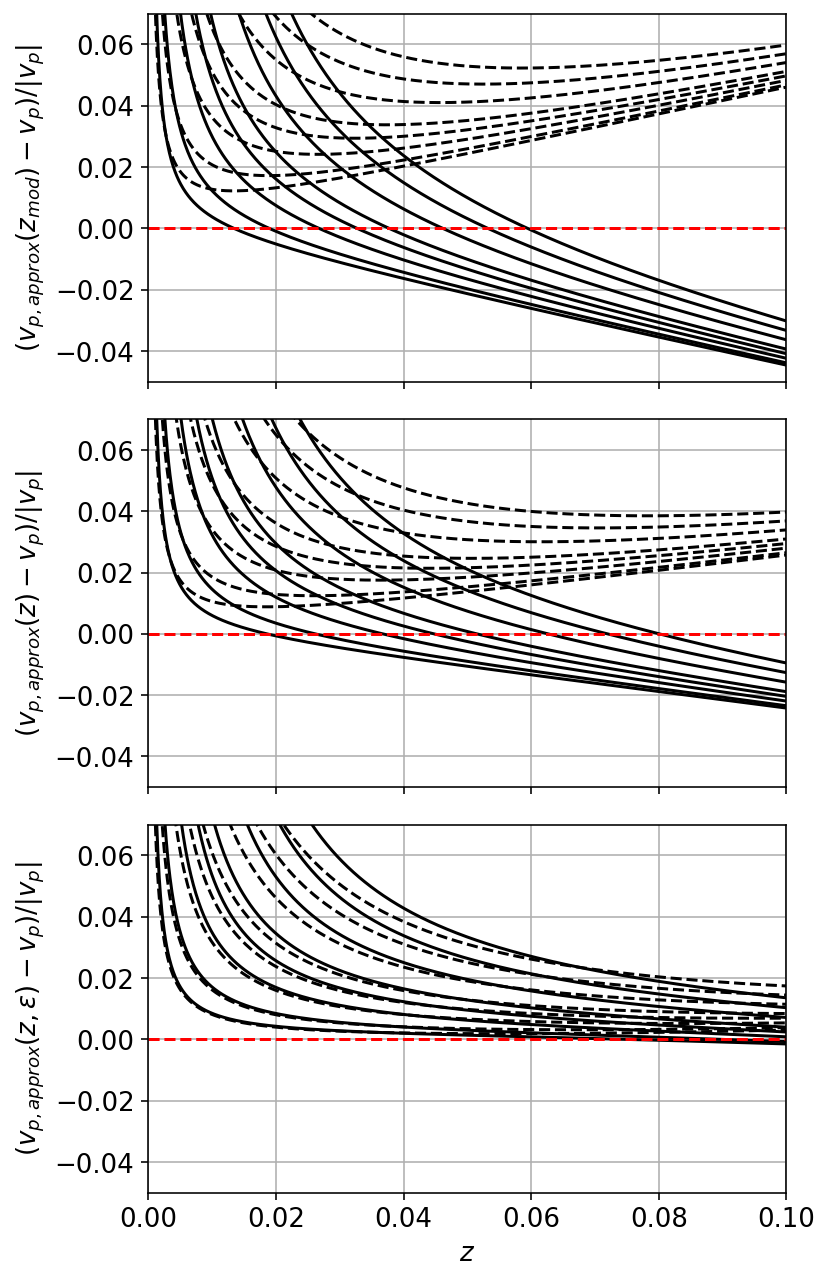}
\caption{Comparison of approximations for the peculiar velocity. The three panels show the fractional errors, as a function of redshift and peculiar velocity, that result from the peculiar velocity approximations given by, top to bottom, Equations~\ref{eqn:PVapprox-zmod}, \ref{eqn:PVapprox-zobs} and~\ref{eqn:PVapprox-zobseps} (with $\epsilon=0.75$ in the latter case). The curves show the fractional errors for true peculiar velocities of $\pm$50, $\pm$100, $\pm$200, $\pm$300, $\pm$400, $\pm$600, $\pm$800 and $\pm$1000\kms, with the solid and dashed lines corresponding to positive and negative peculiar velocities respectively, and peculiar velocity amplitude increasing from the bottom-left curve to the top-right curve.}
\label{fig:vp_approx_compare}
\end{figure}

\citet{Watkins_2015} propose an approximate peculiar velocity estimator that is proportional to the log-distance ratio and therefore inherits its Gaussian error distribution. Their estimator is based on an approximate linear relation between a modified redshift, $z_{\rm mod}$, and luminosity distance  
\begin{equation}
D_L(z) \approx c z_{\rm mod}/H_0
\label{eq:DLzmod}
\end{equation}
where, for a flat $\Lambda$CDM cosmology, \citet{Chiba_1998} and \citet{Visser_2004} show that
\begin{equation}
z_{\rm mod} \approx z[1 + z(1-q_0)/2 - z^2(2-q_0-3q_0^2)/6] ~.
\label{eq:zmod}
\end{equation}
In this expression, $q_0 = 3\Omega_m/2 -1$ is the present-day deceleration parameter; if $\Omega_m$\,=\,0.315 then $q_0$\,=\,$-$0.5275. On dividing by $1+z$, one obtains an equivalent expression for the comoving distance \citep{Davis_2014},
\begin{equation}
D_C(z) \approx cz_\ast/H_0 \approx \frac{cz_{\rm mod}}{H_0(1+z)} \approx \frac{cz_{\rm mod}}{H_0(1+z_{\rm mod})} ~.
\label{eq:DCzmod}
\end{equation}

An approximate relation between peculiar velocity and log-distance ratio can be derived by noting that the peculiar velocity is the difference between the recession velocity inferred from the observed redshift, $z$, and the recession velocity inferred from the cosmological redshift, $z_c$. Thus
\begin{align} 
v_{\!p} &= H_0 D_C(z) - H_0 D_C(z_c) \nonumber \\
    &\approx cz_\ast - H_0 D_C(z_c)
        & &[{\rm using}~cz_\ast \approx H_0 D_C(z)] \nonumber \\
    &\approx -cz_\ast \ln \left( \frac{H_0 D_C(z_c)}{cz_\ast} \right)
        & &[{\rm using}~\ln x \approx x -1~{\rm for}~x \approx 1] \nonumber \\
    &\approx cz_\ast \ln \left( \frac{D_C(z)}{D_C(z_c)} \right) 
        & &[{\rm using}~cz_\ast \approx H_0 D_C(z)] \nonumber \\
    &\approx \frac{cz_{\rm mod}}{(1+z_{\rm mod})} \, \eta \, \ln10 
        & &[{\rm using~the~definition~of}~z_\ast]
\label{eqn:PVapprox-zmod}
\end{align}
which is the expression obtained by \citet{Watkins_2015}.

A related peculiar velocity approximation can be derived starting from the redshift product rule, $(1+z)= (1+z_c)(1+z_p)$, so that
\begin{align} 
v_{\!p} &= c \left( \frac{z-z_c}{1+z_c} \right) \nonumber \\
    &\approx \frac{cz_c}{1+z_c} \ln \left( \frac{z}{z_c} \right) 
        & &[{\rm using}~\ln x \approx x -1~{\rm for}~x \approx 1] \nonumber \\
    &\approx \frac{cz_c}{1+z_c} \ln \left( \frac{D_C(z)}{D_C(z_c)} \right) \ln10 
        & &[{\rm using}~cz \approx H_0 D_C(z)] \nonumber \\
    &\approx \frac{cz}{1+z} \, \eta \, \ln10
        & &[{\rm using}~z \approx z_c]
\label{eqn:PVapprox-zobs}
\end{align}
This relation has the advantage of simplicity (it uses only $z$ and not $z_{\rm mod}$, so is independent of the cosmological model), but makes the apparently crude approximations $cz \approx H_0 D_C(z)$ (although this is buried within the log-distance ratio) and subsequently $z_c \approx z$.

These two approximate estimators for $v_{\!p}$ have very similar form, differing only in whether we use the observed redshift or the modified redshift in the pre-factor. Because they are linear in $\eta$, they inherit the same error distribution as the log-distance ratio (the errors in $z$ are negligible). \citet{Howlett_2022} noted that these expressions should be accurate as long as the true peculiar velocity (not necessarily the measured value) satisfies $v_{\!p} \ll cz$.

We can test these approximate relations for a range of redshifts and peculiar velocities to see how they perform. The left two panels of Figure~\ref{fig:vp_approx_compare} show, as a function of $z$ and for a set of $v_{\!p}$ values, the fractional error in $v_{\!p}$ that results from each approximation. Note that there are two regimes: (i)~at lower redshifts, below the elbow of the curve, the fractional error is large because the amplitude (absolute value) of $v_{\!p}$ is no longer much less than $cz$; and (ii)~at higher redshifts, above the elbow of the curve, the results asymptote to an approximately linear relation. As the amplitude of $v_{\!p}$ increases, the elbow occurs at a higher redshift and a higher fractional error amplitude, and is less sharply bent. It is important to recognise that the errors are not symmetric (and do not necessarily have opposite signs) for positive and negative peculiar velocities: in general, $v_{p,{\rm approx}} - v_{\!p} > 0$; this is always true at low $z$ (for any $v_{\!p}$) and for positive $v_{\!p}$ (at any $z$); it is only not true for negative $v_{\!p}$ at higher redshifts. Interestingly, the simpler approximation using $z$ is slightly better at all redshifts than the approximation using $z_{\rm mod}$. 

A concern for both approximations is that the amplitudes of the fractional errors are increasing at higher redshifts. However, it turns out that this can be remedied by a further slight empirical modification of the approximation. The third panel of Figure~\ref{fig:vp_approx_compare} shows the fractional errors resulting from an approximation of the form
\begin{equation}
v_{\!p} \approx \frac{cz}{1+\epsilon z} \, \eta \, \ln10
\label{eqn:PVapprox-zobseps}
\end{equation}
with $\epsilon = 0.75$. This approximation has multiple advantages: (i)~it uses $z$ rather than $z_{\rm mod}$, so is simpler and independent of the cosmological model; (ii)~it gives smaller fractional errors at all redshifts around and above the knee of the relation; (iii)~it makes the amplitudes of the fractional errors for positive and negative peculiar velocities almost the same over the redshift range of interest (though both errors are positive, so still not symmetric); and (iv)~it makes the fractional errors asymptote around zero at higher redshifts. We therefore use Equation~\ref{eqn:PVapprox-zobseps} in this work to provide an approximate peculiar velocity estimator with Gaussian errors.

\begin{table*}
\label{tab:preddist}
\raggedright

\section{Distances and peculiar velocities}
\label{resultstable}

The first five rows of the table are show below as an example; the complete version of this table is available online.

\centering
\begin{tabular}{ccccrccccccccc}
\vspace{1\baselineskip}\\
\vspace{1\baselineskip}\\
\multicolumn{8}{c}{CF4 observables}\\ \hline
Name& PGC  & $z$& $\alpha$($\degree$) & \multicolumn{1}{c}{$\delta$($\degree$)} & $\log{\Wmxc}$(\kms)  & $m_{i}$(mag) & $m_{W1}$(mag) \\
(1)&(2)&(3)&(4)&\multicolumn{1}{c}{(5)}&(6-7)&(8)&(9) \\ \hline
UGC12889   &  2 & 0.01575 & 0.00684 & $+$47.27460 & 2.744\,$\pm$\,0.029 &  ---  & 11.89 \\
PGC000004  &  4 & 0.01369 & 0.01462 & $+$23.08753 & 2.189\,$\pm$\,0.014 & 15.12 & 16.07 \\
PGC000012  & 12 & 0.02064 & 0.03605 &  $-$6.37400 & 2.606\,$\pm$\,0.021 &  ---  & 13.59 \\
PGC000016  & 16 & 0.01770 & 0.04699 &  $-$5.15886 & 2.515\,$\pm$\,0.025 & 13.58 & 13.99 \\
UGC12898   & 55 & 0.01485 & 0.15600 & $+$33.60058 & 2.260\,$\pm$\,0.025 & 15.22 & 16.20 \\ 
\hline
\end{tabular} 
\vspace{1\baselineskip}\\
\begin{tabular}{ccccc}
\multicolumn{5}{c}{Distances predicted by the combined peculiar velocity and Tully-Fisher model in the $i$-band}\\ \hline
$z_{c}$ & $D_{C}$($h^{-1}$Mpc)& $v_{\!p}$(\kms)&$\eta$&$\widetilde{v}_{\!p}$(\kms)\\
(10-11)&(12-13)&(14-16)&(17-18)&(19-20)\\ \hline
--- & --- & --- & --- & ---\\
0.0137\,$\pm$\,0.0008 & 41\,$\pm$\,2 & ~5 $\substack{-249 \\ ~+235}$ & 0.001\,$\pm$\,0.026 & ~9\,$\pm$\,234 \\
--- & --- & --- & --- & --- \\
0.0185\,$\pm$\,0.0008 & 55\,$\pm$\,2 & ~$-$217 $\substack{-233 \\ ~+224}$ & $-$0.018\,$\pm$\,0.018 & $-$217\,$\pm$\,217 \\
0.015\,$\pm$\,0.0008 & 45\,$\pm$\,2 & ~$-$51 $\substack{-239 \\ ~+227}$ & $-$0.005\,$\pm$\,0.023 & ~$-$51\,$\pm$\,235 \\
\hline
\end{tabular}
\vspace{1\baselineskip}\\
\begin{tabular}{ccccc}
\multicolumn{5}{c}{Distances predicted by the combined peculiar velocity and Tully-Fisher model in the $W1$-band}\\ \hline
$z_{c}$ & $D_{C}$($h^{-1}$Mpc)& $v_{\!p}$(\kms)&$\eta$&$\widetilde{v}_{\!p}$(\kms)\\
(21-22)&(23-24)&(25-27)&(28-29)&(30-31)\\ \hline
0.0164\,$\pm$\,0.0007 & 49\,$\pm$\,2 & ~$-$196 $\substack{-222 \\ ~+212}$ & $-$0.018\,$\pm$\,0.02 & ~$-$193\,$\pm$\,214 \\
0.0138\,$\pm$\,0.0008 & 41\,$\pm$\,2 & ~$-$20 $\substack{-248 \\ ~+233}$ & $-$0.002\,$\pm$\,0.026 & ~$-$19\,$\pm$\,247 \\
0.0216\,$\pm$\,0.0008 & 64\,$\pm$\,2 & ~$-$270 $\substack{-232 \\ ~+225}$ & $-$0.019\,$\pm$\,0.016 & ~$-$267\,$\pm$\,225 \\
0.0183\,$\pm$\,0.0008 & 55\,$\pm$\,2 & ~$-$175 $\substack{-238 \\ ~+228}$ & $-$0.014\,$\pm$\,0.019 & ~$-$169\,$\pm$\,229 \\
0.015\,$\pm$\,0.0008 & 45\,$\pm$\,2 & ~$-$49 $\substack{-247 \\ ~+234}$ & $-$0.005\,$\pm$\,0.024 & ~$-$51\,$\pm$\,245 \\
\hline
\end{tabular}
\vspace{1\baselineskip}\\
\begin{tabular}{ccccc}
\multicolumn{5}{c}{Distances predicted by the Tully-Fisher model in the $i$-band }\\ \hline
$z_{c}$ & $D_{C}$($h^{-1}$Mpc)& $v_{\!p}$(\kms)&$\eta$&$\widetilde{v}_{\!p}$(\kms)\\
(32-33)&(34-35)&(36-38)&(39-40)&(41-42)\\ \hline
--- & --- & --- & --- & ---\\
0.013\,$\pm$\,0.003 & 39\,$\pm$\,8 & ~276 $\substack{-898 \\ ~+728}$ & 0.03\,$\pm$\,0.09 & ~281\,$\pm$\,843 \\
--- & --- & --- & --- & ---\\
0.022\,$\pm$\,0.003 & 66\,$\pm$\,9 & ~$-$1219 $\substack{-915 \\ ~+804}$ & $-$0.09\,$\pm$\,0.06 & ~$-$1085\,$\pm$\,723 \\
0.018\,$\pm$\,0.004 & 53\,$\pm$\,11 & ~$-$774 $\substack{-1108 \\ ~+915}$ & $-$0.07\,$\pm$\,0.09 & ~$-$710\,$\pm$\,913 \\
\hline
\end{tabular}
\vspace{1\baselineskip}\\
\begin{tabular}{ccccc}
\multicolumn{5}{c}{Distances predicted by the Tully-Fisher model in the $W1$-band}\\ \hline
$z_{c}$ & $D_{C}$($h^{-1}$Mpc)& $v_{\!p}$(\kms)&$\eta$&$\widetilde{v}_{\!p}$(\kms)\\
(43-44)&(45-46)&(47-49)&(50-51)&(52-53)\\ \hline
0.017\,$\pm$\,0.002 & 52\,$\pm$\,5 & ~$-$426 $\substack{-521 \\ ~+473}$ & $-$0.04\,$\pm$\,0.04 & ~$-$430\,$\pm$\,430 \\
0.015\,$\pm$\,0.004 & 44\,$\pm$\,11 & ~$-$131 $\substack{-1228 \\ ~+953}$ & $-$0.01\,$\pm$\,0.11 & ~$-$94\,$\pm$\,1034 \\
0.026\,$\pm$\,0.004 & 77\,$\pm$\,11 & ~$-$1423 $\substack{-1095 \\ ~+959}$ & $-$0.09\,$\pm$\,0.06 & ~$-$1263\,$\pm$\,842 \\
0.022\,$\pm$\,0.004 & 66\,$\pm$\,11 & ~$-$1175 $\substack{-1112 \\ ~+951}$ & $-$0.09\,$\pm$\,0.07 & ~$-$1085\,$\pm$\,844 \\
0.021\,$\pm$\,0.005 & 62\,$\pm$\,15 & ~$-$1542 $\substack{-1545 \\ ~+1233}$ & $-$0.13\,$\pm$\,0.1 & ~$-$1318\,$\pm$\,1014 \\
\hline
\end{tabular}
\end{table*}


\bsp	
\label{lastpage}
\end{document}